\documentclass[prd,nofootinbib,showpacs,preprint]{revtex4}
\usepackage{graphicx}
\usepackage{epsfig}
\usepackage{amsmath}
\usepackage{amsfonts}
\usepackage{amssymb}
\usepackage{url}
\usepackage{hyperref}
\usepackage{subfigure}
\usepackage{cancel}
\newcommand{\beqa}{\begin{eqnarray}}
\newcommand{\eeqa}{\end{eqnarray}}
\newcommand{\beq}{\begin{equation}}
\newcommand{\eeq}{\end{equation}}

\newcommand{\nn}{\nonumber}
\newcommand{\bmt}{\begin{pmatrix}}
\newcommand{\emt}{\end{pmatrix}}
\usepackage[toc,page]{appendix}
\usepackage{comment}
\usepackage{xcolor}
\newcommand{\be}{\begin{equation}}
\newcommand{\ee}{\end{equation}}
\newcommand{\bea}{\begin{eqnarray}}
\newcommand{\eea}{\end{eqnarray}}

\begin{document}
\title{Explaining the $R_{K}$ and $R_{D^{(*)}}$  anomalies with vector  leptoquarks}
\author{Suchismita Sahoo$^a$, Rukmani Mohanta$^a$, Anjan K. Giri$^b$ }
%\email{}
\affiliation{$^a$School of Physics, University of Hyderabad,
              Hyderabad - 500046, India\\
              $^b$ Physics Department, IIT Hyderabad, Kandi - 502285, India  }
%%%%%%%%%%%%%%%%%%%%%%%%%%%%%%%%%%%%%%%%%%%%%%%%%%%%%%%%%%%%
\begin{abstract}
%%%%%%%%%%%%%%%%%%%%%%%%%%%%%%%%%%%%%%%%%%%%%%%%%%%%%%%%%

Recently the $B$ factories BaBar and Belle as well as the LHCb experiment  have reported several anomalies  in the semileptonic $B$ meson decays such as $R_{K}$ and  $R_{D^{(*)}}$ etc.  We investigate these deviations by considering the vector leptoquarks relevant for both $b \to s l^+ l^-$ and $b \to c l \bar \nu_l$ transitions. The leptoquark parameter space is constrained  by using the experimentally measured branching ratios of $B_s \to l^+ l^-$, $\bar B \to X_s l^+ l^- (\nu \bar \nu)$ and $B_u^+ \to l^+ \nu_l$ processes. Using the constrained leptoquark couplings, we compute the branching ratios, forward-backward asymmetries, $\tau$ and $D^*$ polarization parameters in the $\bar B \to D^{(*)} l \bar \nu_l$ processes. We find that the vector leptoquarks can explain both  $R_{D^{(*)}}$ and $R_K$ anomalies simultaneously. Furthermore, we study the rare leptonic $B_{u, c}^* \to l \bar\nu$ decay processes in this  model.
\end{abstract}
\maketitle
%%%%%%%%%%%%%%%%%%%%%%%%%%%%%%%%%%%%%%%%%%%%%%%%%%%%%%%%%%%%%%%%%%%%%%%%%%%%%%%%
\section{Introduction}
%%%%%%%%%%%%%%%%%%%%%%%%%%%%%%%%%%%%%%%%%%%%%%%%%%%%%%%%

The standard model (SM) of particle physics explains almost all the  experimental data observed so far, to  a very good level of accuracy. But it is unable to account for some of the fundamental problems of nature, such as the hierarchy in fermion masses,  matter dominance of the universe and dark matter content etc. Therefore, we strongly   believe that there exists some kind of new physics at high scale and the low-energy version of the same could be the SM. The study of nuclear beta decay has set the $V-A$ current structure of the weak interactions which describes various charged current interactions in all the generation of quarks and leptons to a high precision.  However, the recently measured experimental data indicate that the processes involving third generation of fermions in both the initial and final states are comparably  less precise than the first two generations. The couplings of third generation fermions to the electroweak gauge sector is comparatively stronger due to their larger masses and thus sensitive to new physics which could modify the $V-A$ structure of the SM.   In this context, the study of $B^{(*)}_c \to \tau \bar{\nu}_l$ and $B \to D^{(*)} \tau \bar{\nu}_l$ charge current processes, involving the quark level transition $b \to c$  are captivating. 
 Recently BaBar\cite{BaBar1,BaBar2} and Belle \cite{Belle1, Belle2} have measured the  ratio of branching fractions of $\bar B \to  D \tau \bar \nu_\tau$  over $\bar B \to  D l \bar \nu_l$, where $l=e$, $\mu$  and the current experimental average  \cite{hfag} is  
\bea
R_{D}=\frac{{\rm Br}\left(\bar{B} \to D \tau \bar \nu_\tau \right)}{{\rm Br}\left(\bar{B} \to D l \bar \nu_l \right)} = 0.397 \pm 0.040 \pm 0.028, 
\eea
which has  $1.9\sigma$ deviation from its SM result 
$R_D^{\rm SM} = 0.300 \pm 0.008$ \cite{RD-SM}.
In addition, both the $B$ factories and LHCb \cite{Lhc1} have reported  $3.3\sigma$ discrepancy \cite{hfag}  in the measurement of $R_{D^*}$ 
\bea
R_{D^*}=\frac{{\rm Br}\left(\bar{B} \to D^* \tau \bar \nu_\tau \right)}{{\rm Br}\left(\bar{B} \to D^* l \bar \nu_l \right)} = 0.316 \pm 0.016 \pm 0.010,
\eea
from its SM prediction $R_{D^*}^{\rm SM} = 0.252 \pm 0.003$ \cite{Fajfer}. 
These observations may be considered as  the smoking gun signals for the violation of lepton flavour universality (LFU). The dominant theoretical uncertainties are reduced in these observables, as the hadronic uncertainties cancel out to a large extent in these ratios.  
The branching ratio of semileptonic $b \to c l \bar \nu_l$ process  can be computed precisely due to the  light mass  of leptons in the final state, thus the deviation in $R_{D^{(*)}}$  could be from  new physics affecting $\bar B \to  D^{(*)} \tau \bar \nu_\tau$ processes. 
Since these decays occur at tree level in the SM,  new physics models with mass of the new particles near the TeV scale would be required to explain the $R_{D^{(*)}}$ anomalies. The branching ratios of $\bar B \to \bar D^{(*)} \tau \bar \nu_\tau$ processes  and the associated  $R_{D^{(*)}}$ anomalies have been investigated in the literature both in the SM as well as  in various new physics models \cite{Bauer, sakaki, fazio, RD-star, Tanaka, Ruckl, RD-star-LQ, kosnik}.

 Another interesting observable  is the lepton non-universality parameter $(R_K)$ in  $B^+ \to K^+ l^+ l^-$ process,  defined as \cite{hiller} 
\bea
R_K=\frac{{\rm Br}\left(B^+ \to K^+ \mu^+ \mu^- \right)}{{\rm Br}\left(B^+ \to K^+  e^+ e^- \right)}.
\eea
This parameter has recently been measured at   LHCb  with the value 
$R_K= 0.745^{+0.090}_{-0.074} \pm 0.036$ \cite{RK},
which has $2.6\sigma$ deviation from its SM value $R_K = 1.0003 \pm 0.0001$  in the dilepton invariant mass squared bin $\left( 1 \leq q^2 \leq 6 \right) {\rm GeV^2}$. The deviation in the ratios of  branching fractions of other exclusive and inclusive $b \to s$  semileptonic decays \cite{pdg} into dimuon over the dielectron is a  compelling reason to infer possible violation of lepton universality. Various new physics models have been  considered in the literature \cite{RK-TH} to explain the lepton non-universality $(R_K)$ parameter.   The decay rate \cite{Kstar-decayrate} of $B \to K^* \mu^+ \mu^-$ process and the famous $P_5^\prime$ angular observable \cite{p5p}   also have   $\sim 3\sigma$ deviation \cite{p5prime} from the corresponding  SM predictions. Furthermore, the discrepancy of $3.3\sigma$ is  found in the decay rate of  $B_s \to \phi \mu^+ \mu^-$ process in the low $q^2$ region \cite{phi-decayrate}.

In this paper, we pursue the analysis of  semileptonic decays of $B$ meson mediated through  charged-current  $b \to c l \bar \nu_l$ and FCNC $b \to s l^+ l^-$ transitions in the vector leptoquark (LQ) model.
 In most of the studies in the  literature, the authors  have discussed either $R_K$ or $R_D^{(*)}$ anomaly, but not both on the same footing. In the Ref. \cite{RK-n-RD}, both the $R_{D^{(*)}}$ and $R_{K}$ anomalies have been investigated in the $(3,2,1/6)$ scalar LQ model. According to the scenario presented in  \cite{Bauer}, the extension of SM with the $SU(2)_L$ singlet scalar LQ can  accommodate $R_K$ through a loop correction and $R_{D^{(*)}}$ via the tree level LQ contribution.  However, in Ref. \cite{funchal}, it has been argued that a simultaneous explanation of $R_K$ and $R_{D^{(*)}}$ is not realistic and  would imply serious phenomenological problems elsewhere.  In this work, we would like to focus on  both the anomalies $R_{D^{(*)}}$  and $R_K$ as well as some other observables in the $b \to c l \bar{\nu}$ decay processes.   We calculate the branching ratios,  forward-backward asymmetries, the $\tau$ and $D^*$ polarizations of $B \to D^{(*)} \tau \bar{\nu}$ processes in the vector LQ model. We also estimate the branching ratios of the rare leptonic $B_{u, c}^* \to \tau \bar{\nu}$ decay processes. LQs can couple or decay   to a quark and a lepton simultaneously and carry both  baryon number $(B)$ and lepton number $(L)$. They can have spin $0$ (scalar) or spin $1$ (vector) and can be characterized by their fractional electric  charge $(Q)$ and fermion number $(F=3B+L)$. $|F|$ can be either $0$ or $2$ depending on the coupling of LQ to the fermion-antifermion pair or fermion-fermion pair. Such LQs exist in some extended SM theories \cite{georgi} such as grand unified theories based on $SU(5)$, $SO(10)$ etc. \cite{georgi, georgi2}, Pati-Salam model, technicolor model \cite{schrempp} and composite model \cite{kaplan}. To avoid rapid proton decay, we consider  the LQ which  does not couple to diquarks and therefore conserve baryon and lepton numbers. The  LQ model in the context of $B$-physics anomalies has been  studied in the literature \cite{RD-star-LQ, mohanta, mohanta1, mohanta2, davidson, kosnik, Bauer, sakaki, RK-n-RD}.

The outline of this paper is as follows. In section II, we describe the effective Hamiltonian involving $b \to c \tau \bar \nu$ and $b \to s l^+ l^-$ quark level transition in the SM. We also discuss the  relevant  vector LQ contributions to $b \to c l \bar{\nu}_l$ and $b \to s l^+ l^-$ processes. In section III, we compute the constraint on LQ parameter space by using  the recently measured branching ratios of $B_q \to l^+ l^-$,  $\bar B \to X_s l^+ l^-(\nu \bar \nu)$ and $B_{u}^+ \to l^+ \nu$ processes, where $l=e, \mu, \tau$.  The branching ratios, forward-backward asymmetries, $\tau$ and $D^{(*)}$ polarization in $B \to D^{(*)} \tau \bar{\nu}$ processes are presented in section IV. We also describe the deviation in lepton non-universality, $R_{D^{(*)}}$ and $R_{K^{(*)}}$ in this. We work out the branching ratios of the rare $B_{u, c}^* \to l \nu$ decay processes in section V and section  VI contains the summary and  conclusion.
%%%%%%%%%%%%%%%%%%%%%%%%%%%%%%%%%%%%%%%%%%%%%%%%%%%%%%%%%%%%%%%%%%%%%%%%%%%%%%%%%%%%%
\section{Effective Hamiltonian for $b \to c \tau \bar{\nu}_l$ and $b \to s l^+ l^-$ processes}
%%%%%%%%%%%%%%%%%%%%%%%%%%%%%%%%%%%%%%%%%%%%%%%%%%%%%%%%%%%%%

In the SM, the effective Hamiltonian mediating the  semileptonic decays $b \to c \tau \bar{\nu}_l$, considering neutrinos  only to be  left handed, is given as  \cite{sakaki}
\bea
\mathcal{H}_{eff}=\frac{4G_F}{\sqrt{2}} V_{cb} \Big [ \left(\delta_{l\tau} + C_{V_1}^l \right) \mathcal{O}_{V_1}^l + C_{V_2}^l \mathcal{O}_{V_2}^l +  C_{S_1}^l \mathcal{O}_{S_1}^l +  C_{S_2}^l \mathcal{O}_{S_2}^l+ C_{T}^l \mathcal{O}_{T}^l \Big ],
\eea
where $G_F$ is the Fermi constant, $V_{cb}$ is the Cabibbo-Kobayashi-Maskawa (CKM) matrix element and the index $l$ stands for neutrino flavour, $l=e, \mu, \tau$. The $C_X^l$ coefficients, with $X=V_{1,2}, S_{1,2}, T$ are the Wilson coefficients and the corresponding current-current operators are
\bea
&&\mathcal{O}_{V_1}^l = \left(\bar{c}_L \gamma^\mu b_L \right) \left(\bar{\tau}_L \gamma_\mu \nu_{lL} \right), \nn \\
&&\mathcal{O}_{V_2}^l = \left(\bar{c}_R \gamma^\mu b_R \right) \left(\bar{\tau}_L \gamma_\mu \nu_{lL} \right), \nn \\
&&\mathcal{O}_{S_1}^l = \left(\bar{c}_L  b_R \right) \left(\bar{\tau}_R \nu_{l L} \right), \nn \\
&&\mathcal{O}_{S_2}^l = \left(\bar{c}_R b_L \right) \left(\bar{\tau}_R \nu_{l L} \right), \nn \\
&&\mathcal{O}_{T}^l = \left(\bar {c}_R \sigma^{\mu \nu}  b_L \right) \left(\bar{\tau}_R \sigma_{\mu \nu} \nu_{lL} \right),
\eea
where $q_{L(R)} = L(R)q$ are the chiral quark fields with $L(R)=(1\mp \gamma_5)/2$ as the projection operators. Since the flavour of neutrino is not observed at $B$-factories all generations of neutrinos can be taken into  account to reveal the signature of new physics (NP). In the standard model, the contribution to the $b \to c \tau \bar{\nu}_\tau$ process is indicated as $\delta_{l \tau}$ and  the Wilson coefficients ($C_X^l$) are  zero. These coefficients  can only be generated in new physics models.
% This new couplings can be bound experimentally, so that the effects of the new operators can be scrutinised in physical observables.

The effective Hamiltonian describing the processes induced by $b \to s l^+ l^-$ transitions in the SM is given by  \cite{b-s-Hamiltonian}
\bea
{\cal H}_{eff} &=& - \frac{ 4 G_F}{\sqrt 2} V_{tb} V_{ts}^* \Bigg[\sum_{i=1}^6 C_i(\mu) O_i +\sum_{i=7,9,10,S, P} \Big ( C_i(\mu) O_i
+ C_i'(\mu) O_i' \Big )
\Bigg]\;,\label{ham}
\eea
%\bea
%{\cal H}_{eff} &=& - \frac{ 4 G_F}{\sqrt 2} V_{tb} V_{ts}^* \Bigg[\sum_{i=1}^6 C_i(\mu) O_i +C_7 \frac{e}{16 \pi^2} \Big(\bar s \sigma_{\mu \nu}
%(m_s P_L + m_b P_R ) b\Big) F^{\mu \nu} \nn\\
%&&+C_9^{eff} \frac{\alpha}{4 \pi} (\bar s \gamma^\mu P_L b) \bar l \gamma_\mu l + C_{10} \frac{\alpha}{4 \pi} (\bar s \gamma^\mu P_L b)
%\bar l \gamma_\mu \gamma_5 l\Bigg]\;,\label{ham}
%\eea
where  $V_{tb}V_{ts}^*$ is the product of CKM matrix elements and  $C_{i}$'s are the Wilson coefficients evaluated at the renormalization  scale $\mu=m_b$ \cite{b-s-Wilson}. The corresponding
effective operators are given as 
\bea
O_7^{(\prime)} &=&\frac{e}{16 \pi^2} \Big(\bar s \sigma_{\mu \nu}
\big (m_s L(R) + m_b R(L) \big ) b\Big) F^{\mu \nu}, \nn\\
O_9^{(\prime)}&=& \frac{\alpha}{4 \pi} \big(\bar s \gamma^\mu L(R) b\big)(\bar l \gamma_\mu l)\;,~~~~~~~ O_{10}^{(\prime)}= \frac{\alpha}{4 \pi} \big(\bar s \gamma^\mu 
L(R) b \big)(\bar l \gamma_\mu \gamma_5 l),\;\nn \\
O_S^{(\prime)}&=& \frac{\alpha}{4 \pi} \big (\bar s  L(R) b \big )(\bar l  l)\;,~~~~~~~~~~~~~ O_{P}^{(\prime)}= \frac{\alpha}{4 \pi} \big (\bar s  
L(R) b\big )(\bar l  \gamma_5 l)\;,
\eea
where $\alpha$ is the fine structure constant. There is no  contribution of primed Wilson coefficient as well as (pseudo)scalar coefficients in the SM and they arise only in the physics  beyond SM.
 In the following subsections, we will discuss the possible LQ bosons relevant for the $b \to c l \bar{\nu}_l$ and $b \to s l^+  l^-$ quark level transitions.
%%%%%%%%%%%%%%%%%%%%%%%%%%%%%%%%%%%%%%%%%%%%%%%%%%%%%%%%%%%%%%%%%%%%%%%%%%%%%%%%%%%%
\subsection{New physics contribution  due to the exchange of vector leptoquark}
%%%%%%%%%%%%%%%%%%%%%%%%%%%%%%%%%%%%%%%%%%%%%%%%%%%%%%%%%%%%

In the leptoquark model,  the new particles, i.e., leptoquarks,    interact with quarks and leptons simultaneously  and carry both baryon and lepton numbers.
Leptoquarks have ten different multiplets \cite{Ruckl}  under the  $SU(3)_C \times SU(2)_L \times U(1)_Y$ SM gauge symmetries, with flavour non-diagonal couplings.  Out of these,  half are scalars  and the rest  have vectorial nature under the Lorentz transformation. The scalar (vector) LQs have spin $0~(1)$ and could potentially contribute to the FCNC processes involving the quark level transitions $b \to s l^+ l^-$ and $b \to c l^- \bar{\nu}$.   Out of all possible LQ multiplets, six  LQ bosons  are relevant for the $b \to c l \bar{\nu}$ processes whose quantum numbers are presented in Table I. Here $S_{1, 3}$ and $R_2$ are the scalar LQ bosons, $U_{1, 3}^\mu$ and $V_2^\mu$ are the vector LQs. 
In this work, we investigate  the $U_1^\mu=(3, 1, 2/3)$ and $U_3^\mu=(3, 3, 2/3)$ vector LQs, which have $Y=2/3$, $F=0$ and can mediate  both   $b \to s l^+ l^-$ and $b \to c l^- \bar{\nu}$ quark level transitions.
The charge of LQ is related to hypercharge and weak isospin ($T_3$) through $Q=T_3+Y$.
In order to avoid rapid proton decay we do not consider  diquark interactions, as the presence of both LQ and diquark interactions will violate baryon and lepton number.
The interaction Lagrangian of $U_{1, 3}^\mu$ LQs with the SM fermion bilinear  is given as \cite{sakaki, Ruckl}
\bea
\mathcal{L}^{LQ}&=&\left(h_{1L}^{ij}\bar{Q}_{iL} \gamma^\mu L_{jL} + h_{1R}^{ij}\bar{d}_{iR} \gamma^\mu l_{jR} \right) U_{1\mu} + h_{3L}^{ij}\bar{Q}_{iL} \pmb{\sigma}  \gamma^\mu L_{jL} {\bf U}_{3\mu}\;, \label{Lagrangian}
\eea
where $Q_L (L_L)$ is the left handed quark (lepton) doublet, $u_R (d_R)$ and $l_R$ are the right-handed  up (down) quark and charged-lepton singlet respectively and $\pmb{\sigma} $ represents the Pauli matrices. Here   the LQ couplings are represented by  $h^{ij}$, where $i, j$ are the generation indices of quarks and leptons respectively.

 The fermion fields in Eqn. (\ref{Lagrangian}) are represented in the gauge eigen basis in which  Yukawa couplings of the up type quarks and the charged leptons are diagonal, whereas the down type quark fields are rotated into the mass eigenstate basis by the CKM matrix. Now performing the Fierz transformation, we obtain additional Wilson coefficients to the $b \to c \tau \bar{\nu}_l$ process as \cite{Ruckl},
 \begin{subequations}
\bea
&&C_{V_1}^l=\frac{1}{2\sqrt{2}G_F V_{cb}}\sum_{k=1}^3 V_{k3}\Bigg [ \frac{h_{1L}^{2l}{h_{1L}^{k3}}^*}{M^2_{U_1^{2/3}}} - \frac{h_{3L}^{2l}{h_{3L}^{k3}}^*}{M^2_{U_3^{2/3}}} \Bigg ], \label{CV1} \\
&& C_{V_2}^l=0,  \label{CV2} \\
&& C_{S_1}^l = -\frac{1}{2\sqrt{2}G_F V_{cb}}\sum_{k=1}^3 V_{k3} \frac{2 h_{1L}^{2l}{h_{1R}^{k3}}^*}{M^2_{U_1^{2/3}}} ,
\label{CS1}
\eea
\end{subequations}
where $V_{k3}$ denotes the CKM matrix element, $M_{U_{1(3)}^{2/3}}$ is the mass of the leptoquark and  the superscript denotes the charge of $U_{1(3)}$. 
%%%%%%%%%%%%%%%%%%%%%%%%%%%%%%%%%%%%%%%%%%%%%%%%%%%%%%%%%%%%%%%%%%%%%%%%%%%%%%%%%%%%%%
\begin{table}[h]
\caption{Possible relevant scalar and vector leptoquarks invariant under  $SU(3)_C \times SU(2)_L \times U(1)_Y $  SM gauge group.}
\begin{center}
\begin{tabular}{| c | c | c | c |}
\hline
~Leptoquarks~ & ~Spin~ &~ $F=3B+L$~ & ~$\left( SU(3)_C, SU(2)_L, U(1)_Y\right) $~ \\
\hline
\hline
 $S_1$ & $0$ & $-2$ & $\left( 3^*, 1, 1/3 \right)$ \\
  $S_3$ & $0$ & $-2$ & $\left( 3^*, 3, 1/3 \right)$ \\
  $R_2$ & $0$ & $0$ & $\left( 3, 2, 7/6 \right)$ \\
  $U_1$ & $1$ & $0$ & $\left( 3, 1, 2/3 \right)$ \\
   $U_3$ & $1$ & $0$ & $\left( 3, 3, 2/3 \right)$ \\
   $V_2$ & $1$ & $-2$ & $\left( 3^*, 2, 5/6 \right)$ \\
 \hline
\end{tabular}
\end{center}
\end{table}
%%%%%%%%%%%%%%%%%%%%%%%%%%%%%%%%%%%%%%%%%%%%%%%%%%%%%%%%

After expanding the $SU(2)$ indices of Eqn. (\ref{Lagrangian}), one can notice that  $U_{1, 3}$ vector LQs give additional  contributions   to the Wilson coefficients of $b \to s l_i^+ l_j^-$ processes as 
\begin{subequations}
\bea
  C_{9}^{\rm NP} &=& -C_{10}^{\rm NP}  = \frac{\pi}{\sqrt{2} G_F V_{tb}V_{ts}^* \alpha} 
\Big [  \frac{h_{1L}^{2l}{h_{1L}^{k3}}^*}{M^2_{U_1^{2/3}}} +
 \frac{h_{3L}^{2l}{h_{3L}^{k3}}^*}{M^2_{U_3^{2/3}}} \Big ] \,, \label{C9-bs} \\
 C_9^{\prime \rm NP} &= &C_{10}^{\prime \rm NP} = \frac{\pi}{\sqrt{2} G_F V_{tb}V_{ts}^* \alpha}   \frac{h_{1R}^{2l}{h_{1R}^{k3}}^*}{M^2_{U_1^{2/3}}}\,, \label{c9p-bs}\\
  -C_P^{\rm NP} &= &C_{S}^{\rm NP} = \frac{\sqrt{2} \pi}{ G_F V_{tb}V_{ts}^* \alpha} 
   \frac{h_{1L}^{2l}{h_{1R}^{k3}}^*}{M^2_{U_1^{2/3}}}\,, \label{cs-bs}\\
  C_P^{\prime \rm NP} &=& C_{S}^{\prime \rm NP}= \frac{\sqrt{2} \pi}{ G_F V_{tb}V_{ts}^* \alpha}  \frac{h_{1R}^{2l}{h_{1L}^{k3}}^*}{M^2_{U_1^{2/3}}}\,, \label{csp-bs}
\eea
\end{subequations}
where $l,k$ are the generation indices and $C_{9,10, S, P}^{(\prime)\rm NP}$ are the new  Wilson coefficients which arise due to the exchange of vector LQs  associated with  their respective  operators $\mathcal{O}_{9, 10, S, P}^{(\prime)}$.

%%%%%%%%%%%%%%%%%%%%%%%%%%%%%%%%%%%%%%%%%%%%%%%%%%%%%%%%%%%%%%%%%%%%%%%%%%%%%%%%%%%%%%%%%
\section{Constraint on leptoquark couplings from rare decay processes of $B$ meson}
%%%%%%%%%%%%%%%%%%%%%%%%%%%%%%%%%%%%%%%%%%%%%%%%%%%%%%%%%%

After knowing all the possible  vector LQs suitable for $B \to D^{(*)} l \bar{\nu}_l$ and $B \to K^{(*)} l^+ l^-$  processes and the contribution of  additional new Wilson coefficients to the SM, we now proceed to constrain the new LQ parameter space.   The  relevant leptoquark couplings   can be constrained using both $b \to s l^+ l^-$ and $b \to c l^- \nu_l$ processes.  In this analysis, we obtain the constraints on  various LQ couplings by comparing the theoretical and experimental branching ratio of $B_s \to l^+ l^-$,  $B \to X_s l^+ l^-$ and 
$B \to X_s \nu \bar \nu$ processes,  considering the LQ mass as $M_{\rm LQ} =1$ TeV.    Using the constrained  LQ couplings one can  study  the processes mediated by  $b \to s l^+ l^-$ and $b \to c l \bar{\nu}$ transitions. The new LQ parameter space contributing to $b \to u l \nu_l$ transition is constrained by $B_u \to l \nu_l$ processes.

%%%%%%%%%%%%%%%%%%%%%%%%%%%%%%%%%%%%%%%%%%%%%%%%%%%%
\subsection{$B_s \to l^+ l^-$ processes}
%%%%%%%%%%%%%%%%%%%%%%%%%%%%%%%%%%%%%%%%%%%%%%%%%

The rare leptonic $B_s \to l^+ l^-$ processes, where $l=e, \mu, \tau$, mediated by $b \to s l^+ l^-$ transitions are highly suppressed in the SM and occur via electroweak penguin and box diagrams. These processes are theoretically very clean and the only hadronic parameter involved is the decay constant of $B$ meson, hence well suited for constraining the LQ parameters. The branching ratio of $B_s \to l^+ l^-$ process in the SM is given by \cite{Buras}
\bea
{\rm Br}(B_s \to l^+ l^-) = \frac{G_F^2}{16 \pi^3} \tau_{B_s} \alpha^2 f_{B_s}^2 |C_{10}^{\rm SM}|^2 M_{B_s} m_{l}^2   |V_{tb} V_{ts}^*|^2
 \sqrt{1- \frac{4 m_l^2}{M_{B_s}^2}}\times \left( \left |P \right |^2 + \left |S \right |^2 \right),
\eea
where $P$ and $S$ are defined as
\bea
&&P \equiv \frac{C_{10}^{\rm SM}+C_{10}^{\rm NP}-C_{10}^{\rm '  NP }}{C_{10}^{\rm SM}} + \frac{M_{B_s}^2}{2m_l}  \frac{m_b}{m_b + m_s} \Big( \frac{C_P^{\rm NP}-C_P^{\rm ' NP }}{C_{10}^{\rm SM}} \Big) , \nn \\
&&S \equiv \sqrt{1- \frac{4 m_l^2}{M_{B_s}^2}} \frac{M_{B_s}^2}{2m_l}  \frac{m_b}{m_b + m_s} \Big( \frac{C_S^{\rm NP}-C_S^{\rm '  NP}}{C_{10}^{\rm SM}} \Big).\label{np-wilson}
\eea
Here $C_{10, S, P}^{(') \rm NP}$  are the new Wilson coefficients  arising due to the exchange of vector LQ, which are negligible in the SM.
The theoretical predictions \cite{Bobeth} and the average experimental values of CMS and LHCb \cite{mu-br, CDF, tau-br} for the branching ratios of $B$ meson decaying to all charged leptonic modes are given as
\bea
&&{\rm Br}(B_s \to ee)^{\rm SM} = (8.54 \pm 0.55)\times10^{-14}~ [38], 
~~{\rm Br}(B_s \to ee)^{\rm expt}<2.8\times10^{-7} ~ [39],  \nonumber\\ 
&&{\rm Br}(B_s \to \mu \mu)^{\rm SM}= (3.65 \pm 0.23)\times10^{-9}~  [38], ~~
{\rm Br}(B_s \to \mu \mu)^{\rm expt}=(2.8^{+0.7}_{-0.6} )\times10^{-9} ~[40],  \nonumber\\
&&{\rm Br}(B_s \to \tau \tau)^{\rm SM}  =   (7.73 \pm 0.49) \times 10^{-7}~ [38],~~~
{\rm Br}(B_s \to \tau \tau)^{\rm expt}  <  ~3.0 \times 10^{-3} ~[41]. 
\eea
 If we consider  the LQ couplings as chiral, then only   $C_{10}^{\rm NP}$ Wilson coefficient will  give additional contributions. Now comparing the theoretical  value of   branching ratio of $B_s \to l^+ l^-$ processes with the $1\sigma$ range of the experimental data,  the allowed region of real and imaginary parts of the LQ couplings  are shown in Fig. 1,   for $B_s \to e^+ e^-$ (top left panel), $B_s \to \mu^+ \mu^-$ (top right panel) and $B_s \to \tau^+ \tau^-$ (bottom panel) processes. The constrained values of real and imaginary parts of LQ couplings are given in Table II.

As seen from Eqn. (\ref{np-wilson}), the scalar and pseudoscalar Wilson coefficients are dominated by the $M_B^2/m_l$ multiplication factor, therefore the new physics contribution to the $C_{10}$ Wilson coefficient can be neglected.  Now considering  only the $C_{S, P}^{(') \rm NP}$ new Wilson coefficients, the allowed region  on real and imaginary parts of LQ  couplings  for  $B_s \to e^+ e^-$ (top left panel), $B_s \to \mu^+ \mu^-$ (top right panel) and $B_s \to \tau^+ \tau^-$ (bottom panel) processes are shown in Fig. 2 and the allowed range of LQ couplings are presented in Table II.

%%%%%%%%%%%%%%%%%%%%%%%%%%%%%%%%%%%%%%%%%%%%%%%%%%%%%%%%%%%%%%%%%%%%%%%%%%%%%%
\begin{figure}[h]
\centering
\includegraphics[scale=0.6]{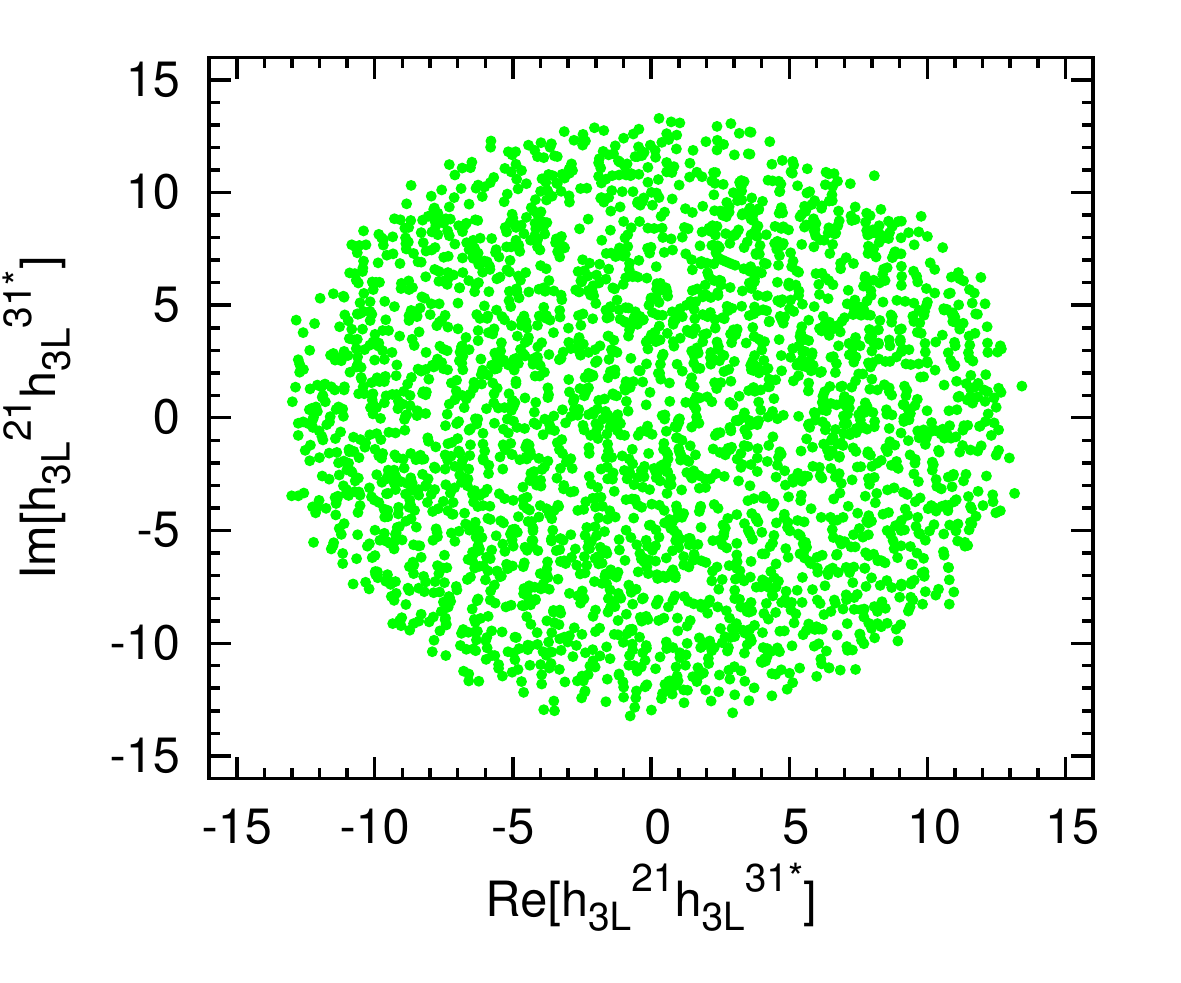}
\quad
\includegraphics[scale=0.6]{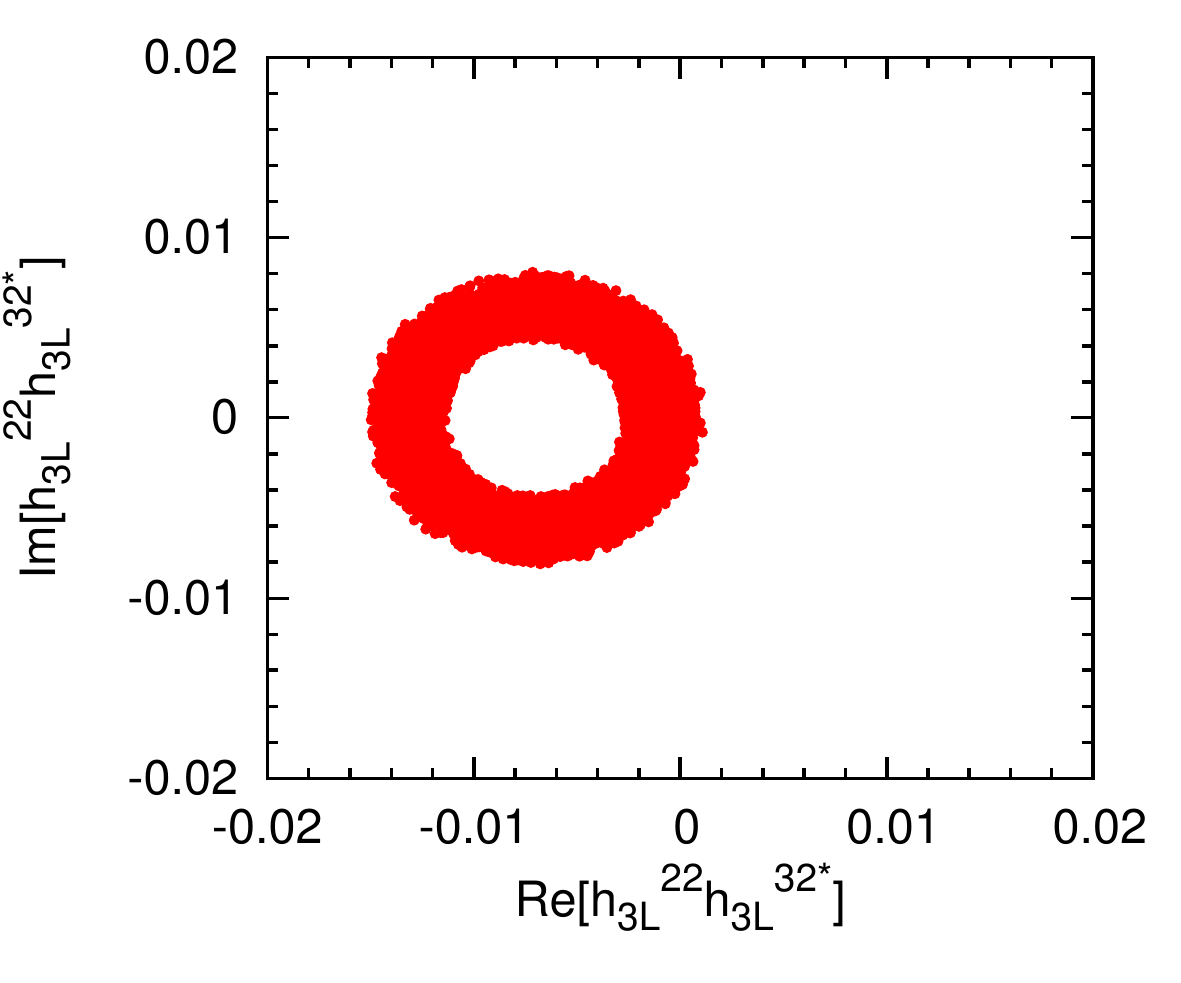}
\quad
\includegraphics[scale=0.6]{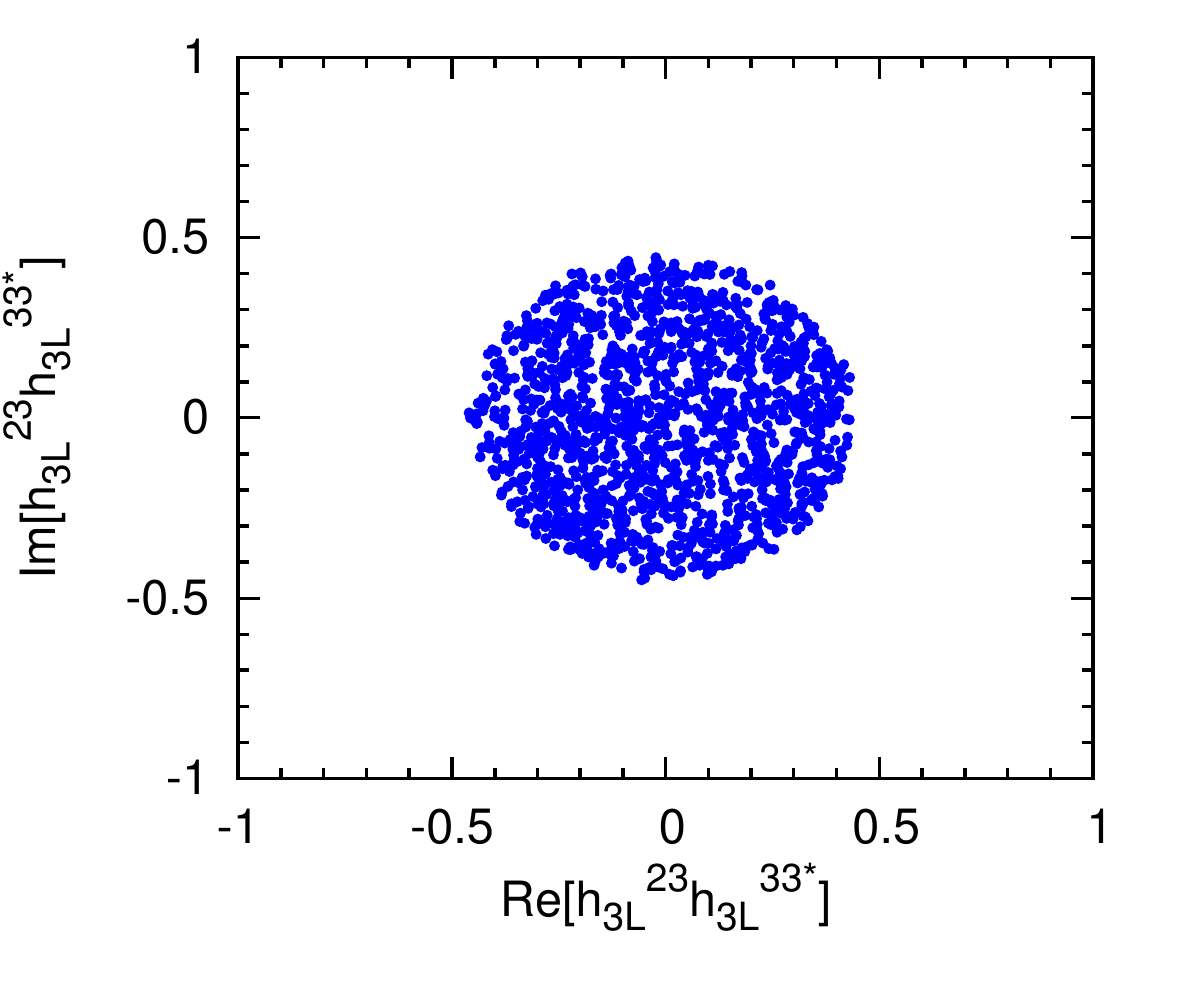}
\caption{Constraints on the real and imaginary parts of the leptoquark couplings from $B_s \to e^+ e^-$  (top left panel), $B_s \to \mu^+ \mu^-$ (top right panel) and $B_s \to \tau^+ \tau^-$ (bottom panel) processes in $U(3,3,2/3)$ leptoquark model.}
\end{figure}
%%%%%
%%%%%%%%%%%%%%%%%%%%%%%%%%%%%%%%%%%%%%%%%%%%%%%%%%%%%%%%%%%%%%%%%%%%%%%%%%%%%%
\begin{figure}[h]
\centering
\includegraphics[scale=0.6]{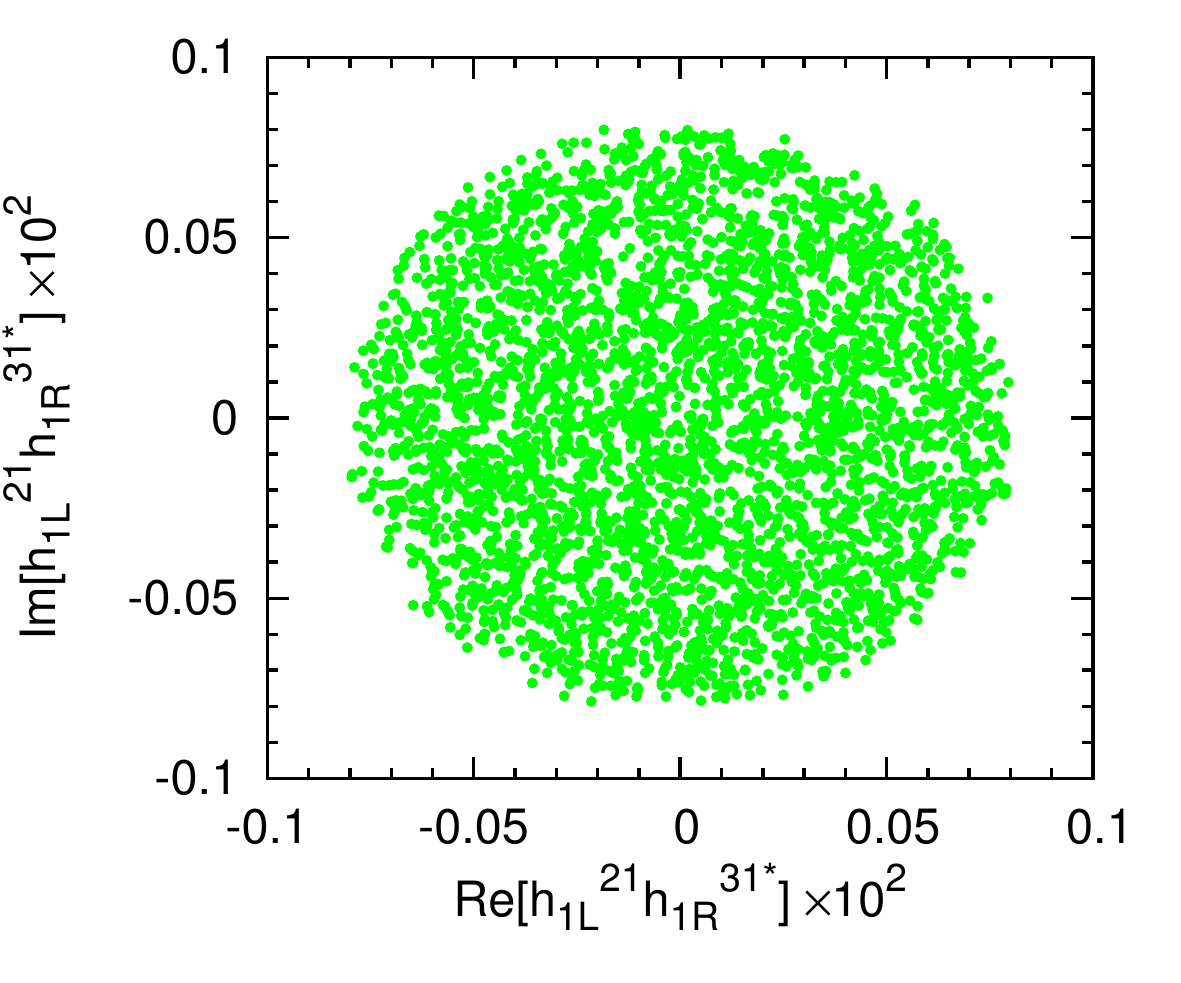}
\quad
\includegraphics[scale=0.6]{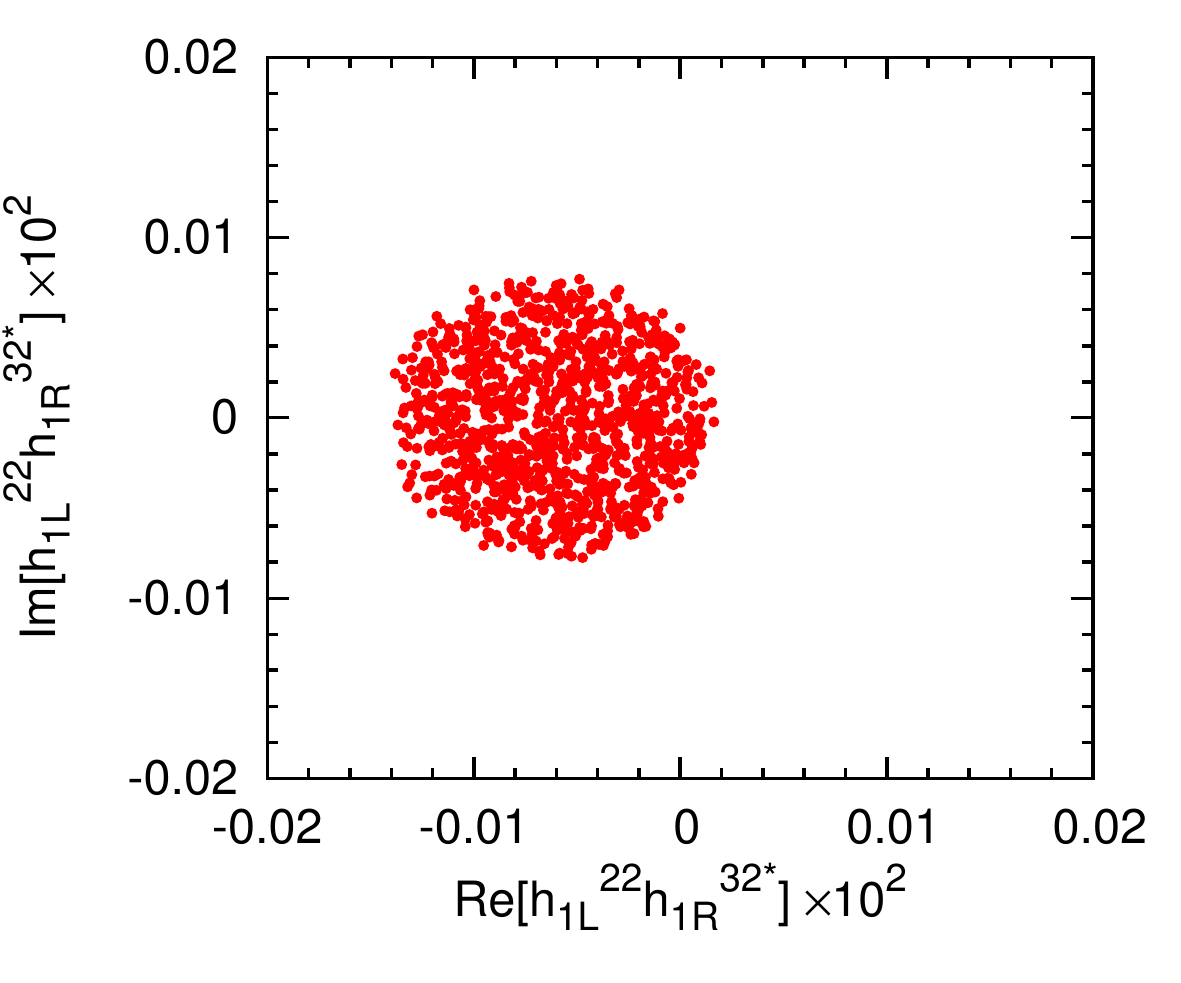}
\quad
\includegraphics[scale=0.6]{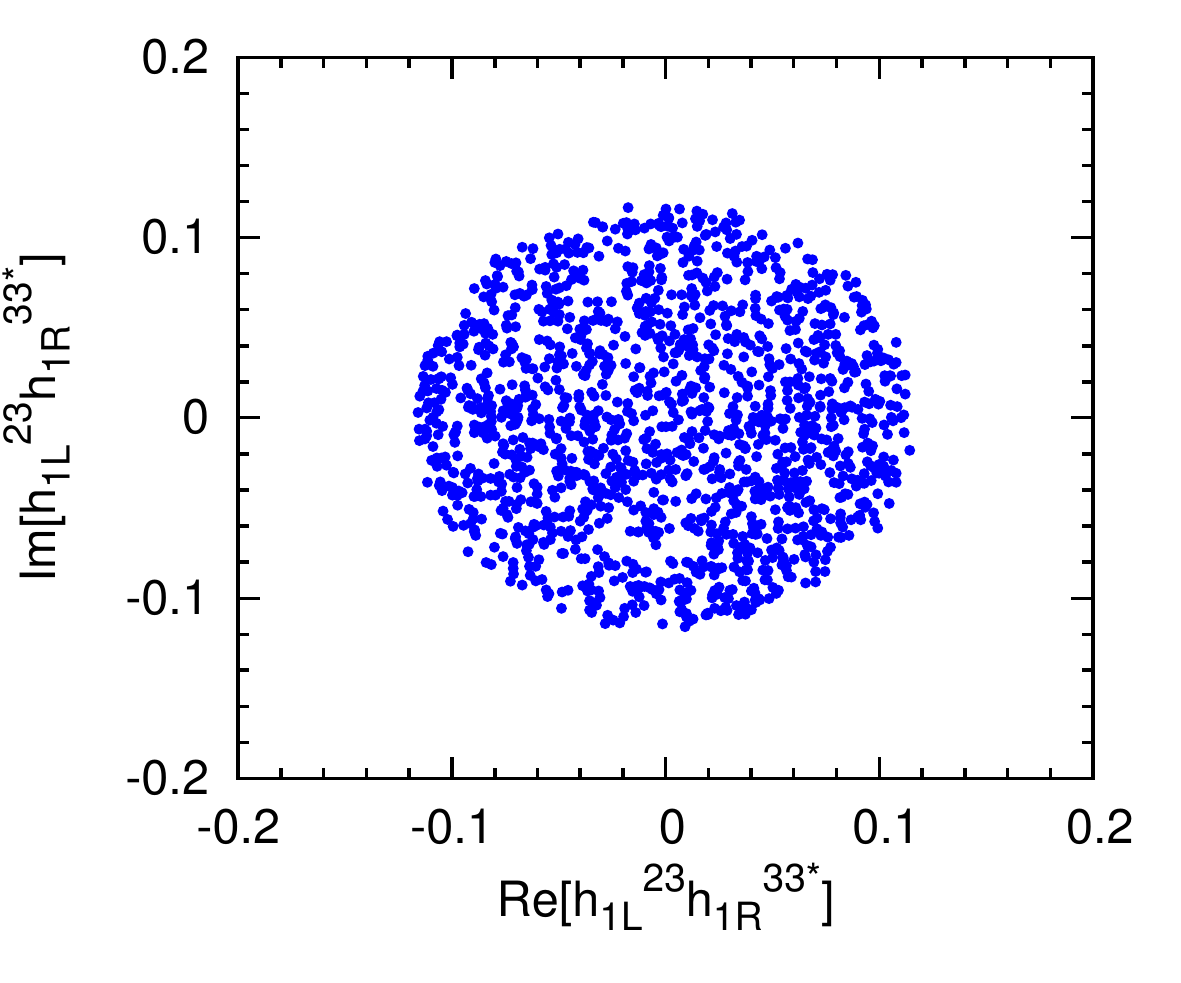}
\caption{Constraints on the real and imaginary parts of the leptoquark couplings from $B_s \to e^+ e^-$  (top left panel), $B_s \to \mu^+ \mu^-$ (top right panel) and $B_s \to \tau^+ \tau^-$ (bottom panel) processes in $U(3,1,2/3)$ leptoquark model.}
\end{figure}
%%%%%%%%%%%%%%%%%%%%%%%%%%%%%%%%%%%%%%%%%%%%%%%%%%%%%%%%%%%%%%%%%%%%%%%%%%%%%%

\begin{table}[h]
\caption{Constraints  on the real and imaginary parts of the leptoquark couplings from $B_s \to l^+ l^-$ processes, where $l=e, \mu, \tau$.}
\begin{center}
\begin{tabular}{| c | c | c |}
\hline
~Leptoquark Couplings~ &~ Real part~ & ~Imaginary Part~  \\
\hline
\hline
$h_{1(3)L}^{21} {h_{1(3)L}^{{31}^*}}$ & $-13.0 \to 13.0$ & $-13 \to 13$ \\
$h_{1(3)L}^{22} {h_{1(3)L}^{{32}^*}}$   &   $-0.016 \to 0.0$ &$-0.008 \to 0.008$  \\
$h_{1(3)L}^{23} {h_{1(3)L}^{{33}^*}}$   &   $-0.4 \to 0.4$ &$-0.4 \to 0.4$  \\
   \hline
 $h_{1L}^{21} {h_{1R}^{31}}^*$ & $(-0.8 \to 0.8) \times 10^{-3}$ & $(-0.8 \to 0.8) \times 10^{-3}$ \\
$h_{1L}^{22} {h_{1R}^{32}}^*$   &   $-0.016 \times 10^{-2} \to 0.0$ &$(-0.8 \to 0.8) \times 10^{-4}$  \\
$h_{1L}^{23} {h_{1R}^{33}}^*$   &   $-0.1 \to 0.1$ &$-0.1 \to 0.1$  \\
 \hline
\end{tabular}
\end{center}
\end{table}
%%%%%%%%%%%%%%%%%%%%%%%%%%%%%%%%%%%%%%%%%%%%%%%%%%%%%%%%%%%%%%%%%%%%%%%%%%%%%%%%%%%%%%
\subsection{$\bar B \to X_s l^+ l^-$ processes}
%%%%%%%%%%%%%%%%%%%%%%%%%%%%%%%%%%%%%%%%%%%%%%%%%%%%%%%%
In this subsection, we discuss the constraint on LQ couplings from the branching ratio of inclusive  $\bar B \to X_s l^+ l^-$  decay process mediated via $b \to s l^+ l^-$ transitions. 
The branching ratio for  this  process in the SM is given by \cite{mohanta, Alok}
\bea
\frac{d{\rm Br}}{ds_1}\biggr|_{\rm SM} &= & B_0 \frac{8}{3} (1-s_1)^2\sqrt{1- \frac{4t^2}{s_1}} \times \biggr[(2s_1+1)
\left(\frac{2t^2}{s_1} +1 \right ) |C_9^{eff}|^2\nn\\
&+& \left (\frac{2(1-4 s_1)t^2}{s_1} +(2 s_1+1)\right )  |C_{10}  |^2
+4 \left (\frac{2}{s_1}+1 \right )\left (\frac{2 t^2}{s_1}+1 \right )\left |C_7 \right |^2\nn\\
&+& 12 \left (\frac{2 t^2}{s_1}+ 1\right ){\rm Re}(C_7 C_9^{eff *} ) \biggr]\;,
\eea
where $t= m_l/m_b^{pole}$,  $s_1=q^2/(m_b^{pole})^2$ and $B_0$ is the normalization constant related to ${\rm Br}(\bar B \to X_c e \bar \nu_e)$ process as
\bea
B_0 = \frac{3 \alpha^2 {\rm Br} (\bar B \to X_c e \bar \nu_e)}{32 \pi^2 f(\hat m_c)\kappa(\hat m_c)}
\frac{|V_{tb} V_{ts}^{*}|^2}{|V_{cb}|^2}\;.
\eea
Here  $\hat m_c = m_c^{pole}/m_b^{pole}$ and the functions $f(\hat m_c)$ and $\kappa(\hat m_c)$ are defined in Ref. \cite{mohanta, Alok}. For the numerical estimation, we use the numerical parameters as $\hat m_c = 0.29 \pm 0.02$ \cite{Ali} and Br$(\bar B \to X_c e \bar \nu_e) = (10.1 \pm 0.4) \%$ \cite{pdg}. For the CKM matrix elements we use the Wolfenstein  parameters with values $A=0.814 ^{+0.023}_{-0.024}$, $\lambda = 0.22537 \pm 0.00061$, $\bar \rho = 0.117 \pm  0.021$ and $\bar \eta = 0.353 \pm 0.013$  \cite{pdg}.  Now using  these parameters, the branching ratios of $\bar B \to X_s l^+ l^-$ processes  in the SM for the low $q^2 \in [1, 6]~ {\rm GeV}^2$ region  are found as
\bea
&&{\rm Br} (\bar B \to X_s e^+ e^-)|_{q^2 \in [1, 6]~{\rm GeV}^2} = (1.67 \pm 0.06) \times 10^{-6}, \\
&&{\rm Br} (\bar B \to X_s \mu^+ \mu^-)|_{q^2 \in [1, 6]~{\rm GeV}^2} = (1.6 \pm 0.61) \times 10^{-6},
\eea
and the  predicted branching ratios in the high $q^2~ (\geq 14.2 ~ {\rm GeV}^2)$ region are given as
\bea
&&{\rm Br} (\bar B \to X_s e^+ e^-)|_{q^2 \geq 14.2~{\rm GeV}^2} = (3.9 \pm 0.15) \times 10^{-7}, \\
&&{\rm Br} (\bar B \to X_s \mu^+ \mu^-)|_{q^2 \geq 14.2~{\rm GeV}^2} = (3.8 \pm 0.25) \times 10^{-7}, \\
&&{\rm Br} (\bar B \to X_s \tau^+ \tau^-)|_{q^2 \geq 14.2~{\rm GeV}^2} = (1.78 \pm 0.29) \times 10^{-7}.
\eea
The corresponding experimental results \cite{Lees} for both low and high $q^2$ regions are given by 
\bea
{\rm Br} (\bar B \to X_s e^+ e^-) &=& (1.93^{+0.47~+0.21}_{-0.45~-0.16} \pm 0.18) \times 10^{-6} ~~~{\rm for ~low}~ q^2, \\ &=& (0.56^{+0.19~+0.03}_{-0.18~-0.03} \pm 0.00) \times 10^{-6}  ~~~{\rm for~ high}~ q^2, \\
{\rm Br} (\bar B \to X_s \mu^+ \mu^-) &=& (0.66^{+0.82~+0.30}_{-0.76~-0.24} \pm 0.07) \times 10^{-6} ~~~{\rm for ~low}~ q^2, \\ &=& (0.60^{+0.31~+0.05}_{-0.29~-0.04} \pm 0.00) \times 10^{-6}  ~~~{\rm for~ high}~ q^2,
%{\rm Br} (B_d \to X_s \tau^+ \tau^-) &\leq& 1\%.
\eea
where the first uncertainties are statistical, the second experimental systematics and the third model-dependent systematics.
Since there is no experimental measurement for the   branching ratio of $\bar B \to X_s \tau^+ \tau^-$ process, we consider the limit as $\sim 1\%$ in our analysis.
 Including the new physics contribution, the total branching ratio of $\bar B \to X_s l^+ l^-$  process is given by \cite{mohanta, Alok}
\bea
\left (\frac{d {\rm Br}}{d s_1 }\right )_{\rm Total}&=&\left (\frac{d {\rm Br}}{d s_1 }\right )_{\rm SM}
+B_0 \Big[\frac{16}{3} (1-s_1)^2 (1+2 s_1)[{\rm Re}(C_9^{eff} C_9^{NP *})+{\rm Re}(C_{10} C_{10}^{NP *})]\nn\\
&+& \frac{8}{3}(1-s_1)^2(1+2 s_1)\left [|C_9^{NP}|^2 +|C_{10}^{NP}|^2 +|C_9^{'NP}|^2+|C_{10}^{'NP}|^2\right ]
\nn\\
&+& 32 (1-s_1)^2 ~{\rm Re}(C_7 C_{10}^{NP *})  \Big]\;,\label{br-1}
\eea
where $C_{9, 10}^{(')NP}$ are the new Wilson coefficients. 
The particle masses and the lifetime of $B$ meson are taken from \cite{pdg}. Now comparing the  theoretical  and experimental  branching ratios, we show the constraints on $U(3,3,2/3)$ LQ couplings from $\bar B \to X_s e^+ e^- (\mu^+ \mu^-)$ process for low $q^2$ (left panel) and high $q^2$ (right panel) in Fig. 3 (Fig. 4) respectively. Similarly in Fig. 5, we show  the allowed region from $\bar B \to X_s \tau^+ \tau^-$ process in  high $q^2$ region. From these figures, the allowed range  of real and imaginary parts of LQ parameter space in the low and high $q^2$ regime are presented in Table III.
%%%%%%%%%%%%%%%%%%%%
%%%%%%%%%%%%%%%%%%%%%%%%%%%%%%%%%%%%%%%%%%%%%%%%%%%%%%%%%%%%%%%%%%%%%%%%%%%%%%
\begin{figure}[h]
\centering
\includegraphics[scale=0.55]{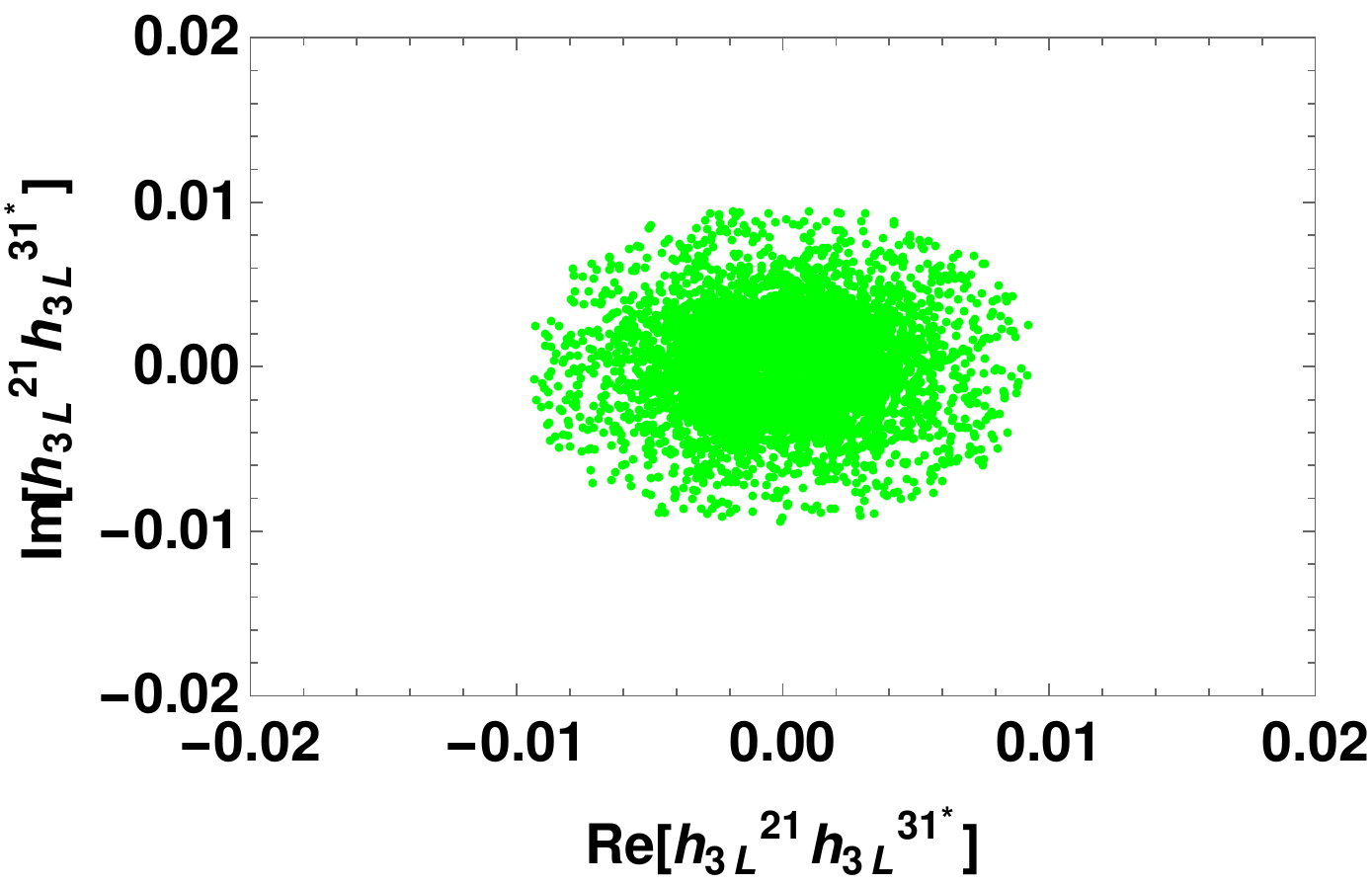}
\quad
\includegraphics[scale=0.55]{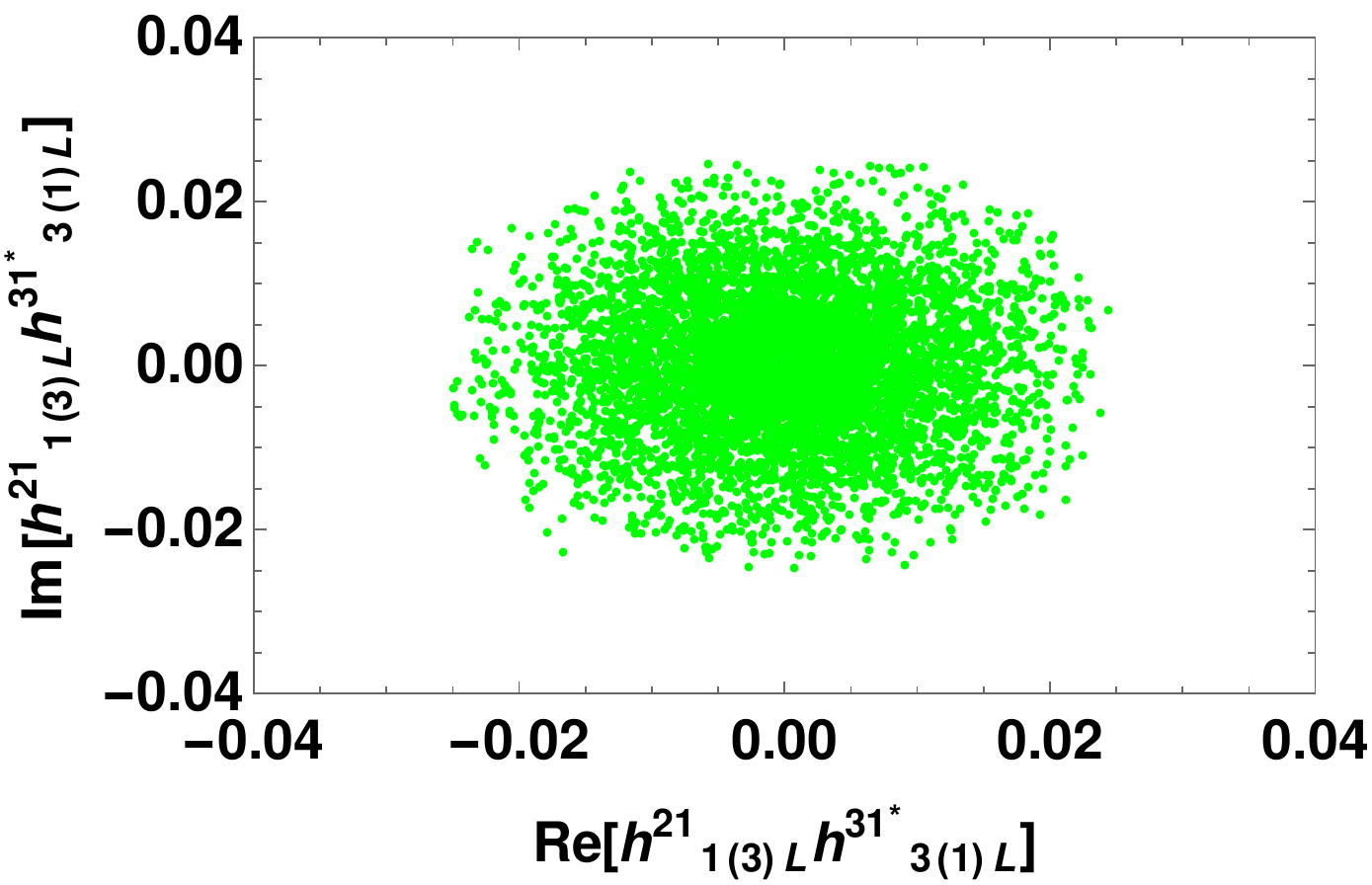}
\caption{Constraints on the real and imaginary parts of the leptoquark couplings from $\bar B \to X_s e^+ e^-$ process in low $q^2$  (left panel) and high $q^2$ region (right panel) in the $U(3,3,2/3)$ leptoquark model.}
\end{figure}
%%%%%%%%%%%%%%%%%%%%%%%%%%%%%%%%%%%%%%%%%%%%%%%%%%%%%%%%%%%%%%%%%%%%%%%%%%%%%%
\begin{figure}[h]
\centering
\includegraphics[scale=0.55]{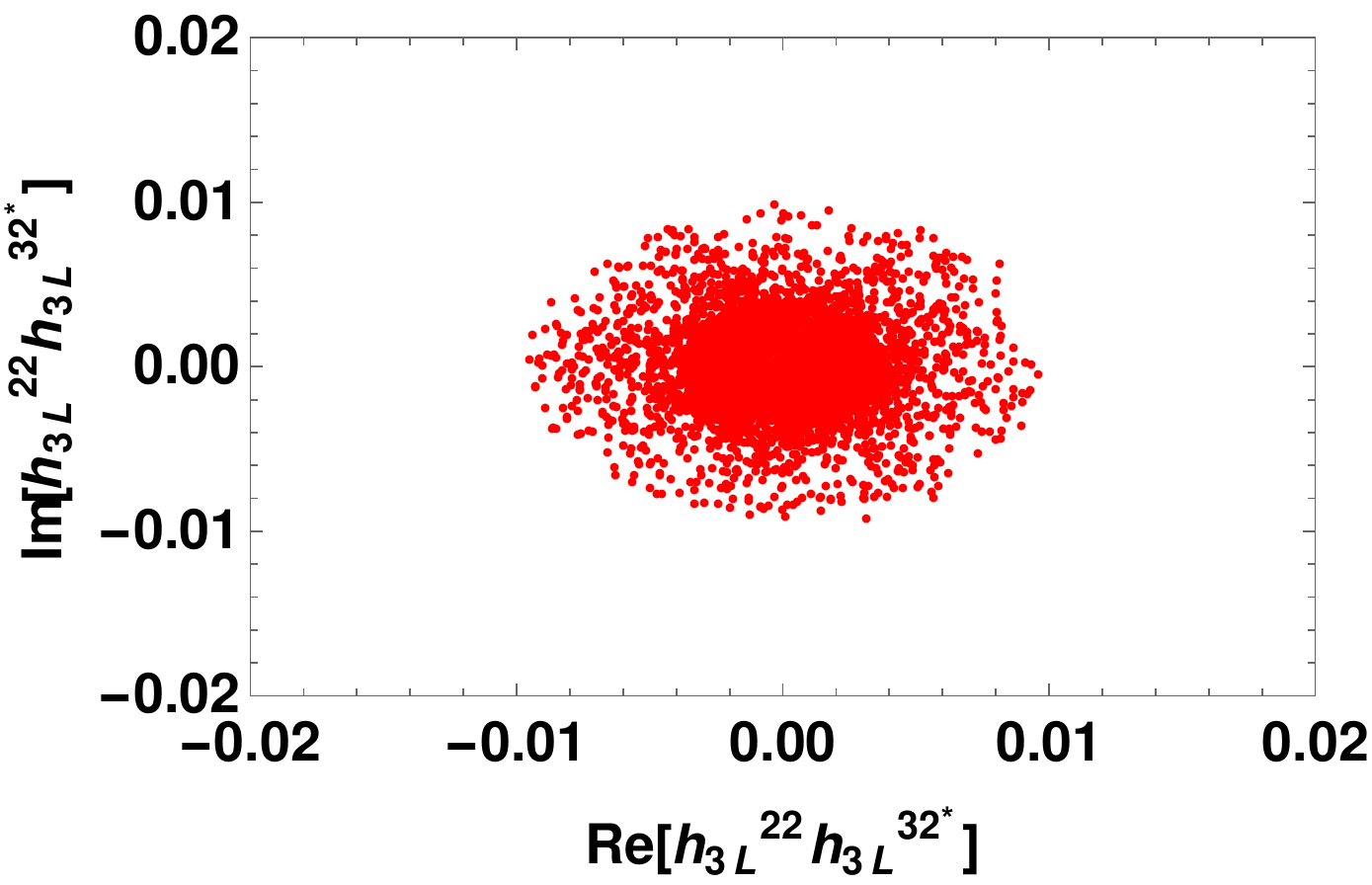}
\quad
\includegraphics[scale=0.55]{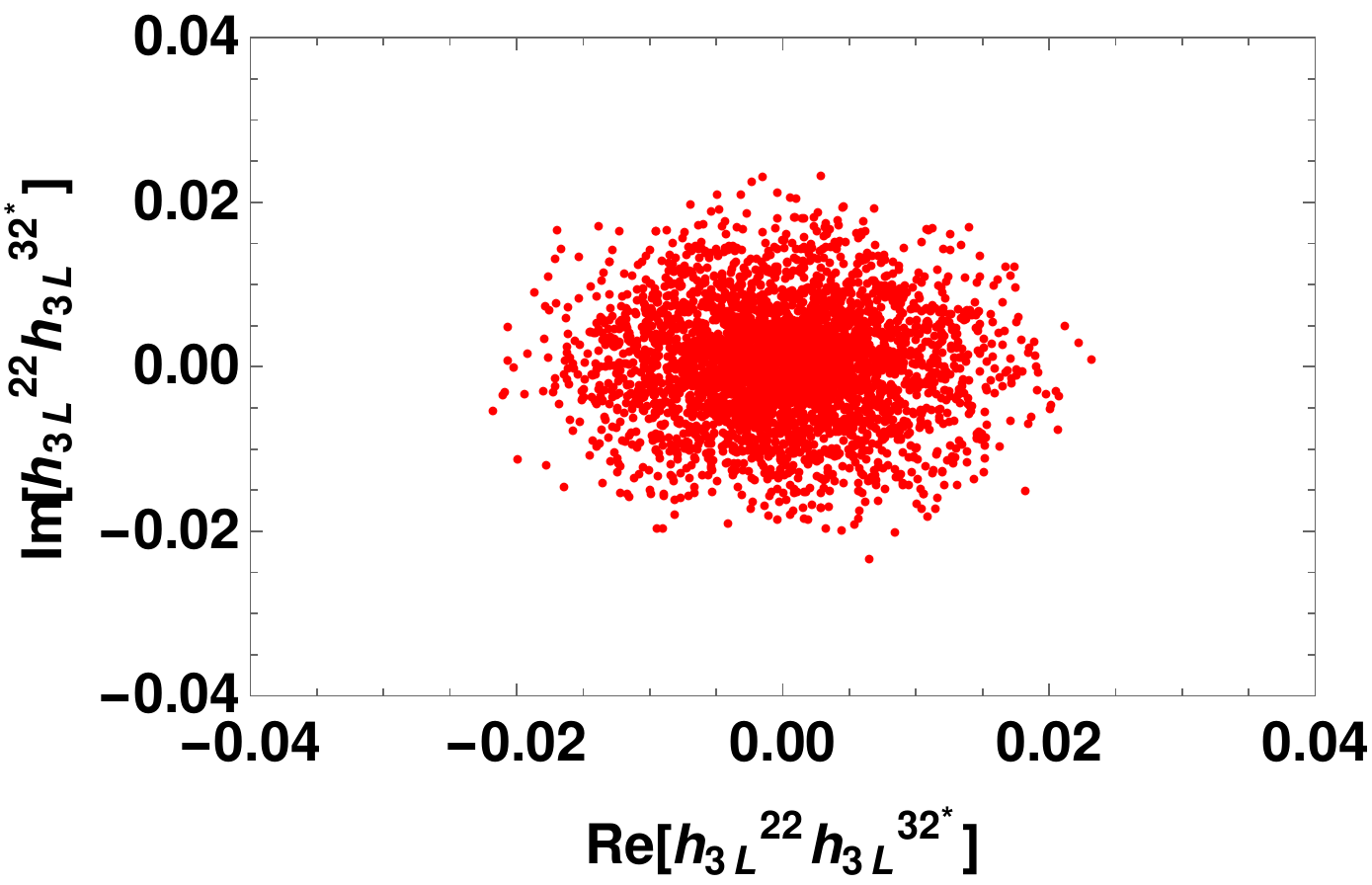}
\caption{Constraints on the real and imaginary parts of the leptoquark couplings from $\bar B \to X_s \mu^+ \mu^-$  process in low $q^2$  (left panel) and high $q^2$ region (right panel) in the $U(3,3,2/3)$ leptoquark model.}
\end{figure}
%%%%%%%%%%%%%%%%%%%%%%%%%%%%%%%%%%%%%%%%%%%%%%%%%%%%%%%%%%%%%%%%%%%%%%%%%%%%%%
\begin{figure}[h]
\centering
\includegraphics[scale=0.6]{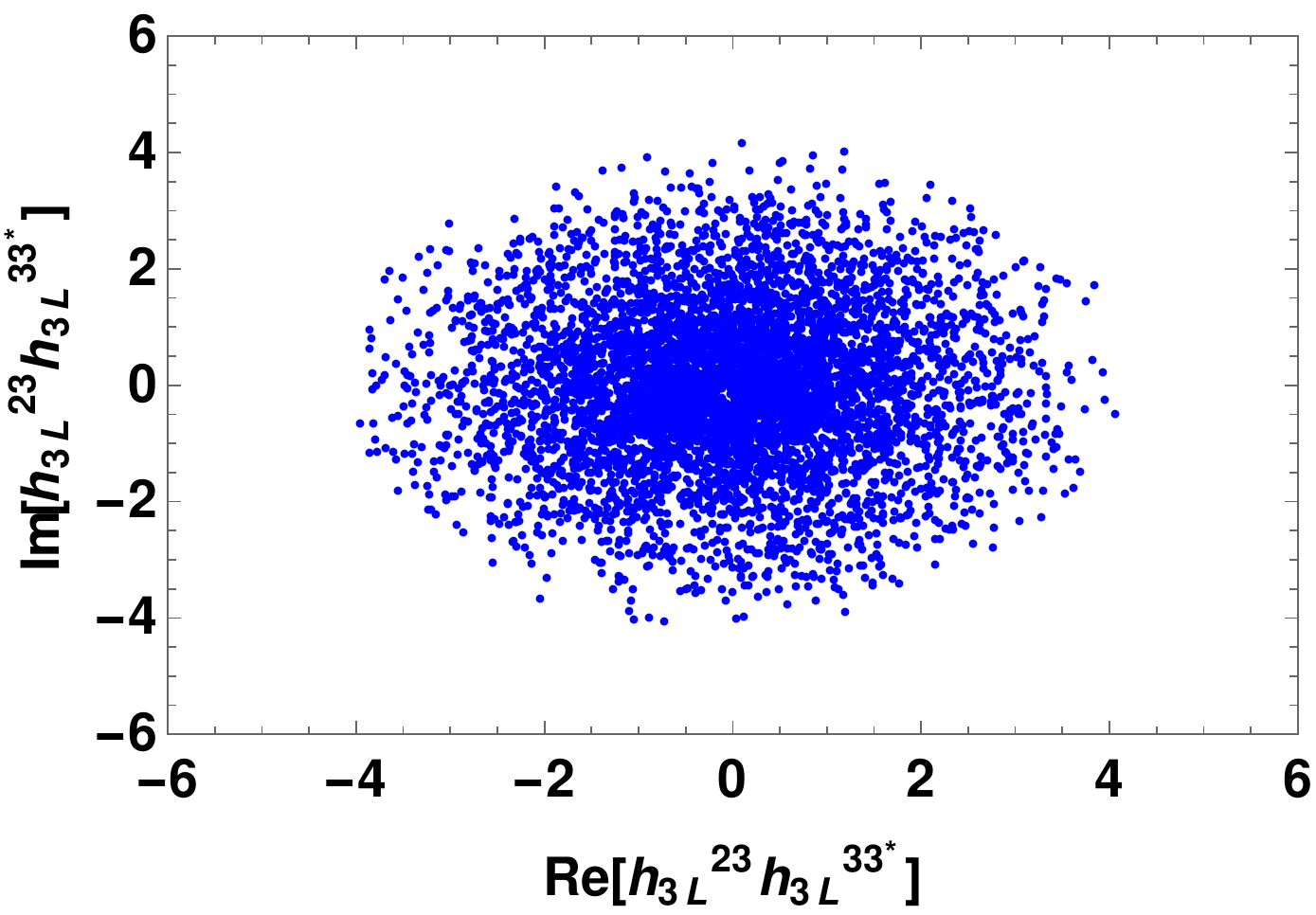}
\caption{Constraints on the  real and imaginary parts of the leptoquark couplings from $\bar B \to X_s \tau^+ \tau^-$ process  in the $U(3,3,2/3)$ leptoquark model}
\end{figure}
%%%%%%%%%%%%%%%%%%%%%%%%%%%%%%%%%%%%%%%%%%%%%%%%%%%%%%%%%%%%%
%%%%%%%%%%%%%%%%%
%%%%%%%%%%%%%%%%%%%%%%%%%%%%%%%%%%%%%%%%%%%%%%%%%%%%%%%%%%%%%%%%%%%%%%%%%%%%%
\begin{table}[h]
\caption{Constraints on the real and imaginary parts of the leptoquark coupling for low and high $q^2$ region  from $\bar B \to X_s l^+ l^-$ process, where $l=e, \mu, \tau$}
\begin{center}
\begin{tabular}{| c | c | c | c |}
\hline
~$q^2$ bin~ &~Leptoquark Couplings ~&~ Real part~ &~ Imaginary Part~  \\
\hline
\hline
low $q^2$ &$h_{1(3)L}^{21} {h_{1(3)L}^{{31}^*}}$ & $-0.01 \to 0.01$ & $-0.01 \to 0.01$ \\
 &$h_{1(3)L}^{22} {h_{1(3)L}^{{32}^*}}$   &   $-0.008 \to 0.008$ &$-0.008 \to 0.008$  \\
   \hline
&$h_{1(3)L}^{21} {h_{1(3)L}^{{31}^*}}$ & $-0.022 \to 0.022$ & $-0.022 \to 0.022$ \\
high $q^2$ &$h_{1(3)L}^{22} {h_{1(3)L}^{{32}^*}}$   &   $-0.018 \to 0.018$ &$-0.018 \to 0.018$  \\
&$h_{1(3)L}^{23} {h_{1(3)L}^{{33}^*}}$   &   $-3.8 \to 3.8$ &$-3.8 \to 3.8$  \\
 \hline
\end{tabular}
\end{center}
\end{table}
%%%%%%%%%%%%%%%%%%%%%%%%%%%%%%%%%%%%%%%%%%%%%%%%%%%%%%%%%
\subsection{$\bar B \to X_s \nu \bar \nu$ process}
%%%%%%%%%%%%%%%%%%%%%%%%%%%%%%%%%%%%%%%%%%%%%%%%%%%%%%%
The study of the  processes involving $b \to s \nu \bar \nu$ transitions are quite important, as they are  related to $b \to s l^+ l^-$ processes by $SU(2)_L$ and are also very sensitive to the  search for new physics beyond  the SM. 
The inclusive decay $\bar B \to X_s \nu \bar{\nu}$ is theoretically very clean since both the perturbative   and the non-perturbative corrections are small. Thus, these decays do not suffer  from the form factor uncertainties.

The effective Hamiltonian for $b \to s \nu \bar \nu$ process is given by \cite{nu-nubar}
\begin{equation}
\mathcal{H}_{eff} = \frac{-4G_F}{\sqrt{2}} V_{tb}V_{ts}^*\big(C^\nu_L \mathcal{O}^\nu_L +C^\nu_R \mathcal{O}^\nu_R \big) + h.c., \label{nu}
\end{equation}
where  the six-dimensional operators are 
\begin{equation}
\mathcal{O}^\nu_L = \frac{\alpha}{4 \pi} \left(\bar{s}\gamma_\mu L b \right) \big(\bar{\nu} \gamma^\mu \left(1-\gamma_5\right)\nu \big),
 \hspace{1cm} \mathcal{O}^\nu_R = \frac{\alpha}{4 \pi} \left(\bar{s}\gamma_\mu R b \right) \big(\bar{\nu}  \gamma^\mu \left(1-\gamma_5\right)\nu \big) .
\label{nu-op}
\end{equation}
In the SM,   the $C_L^\nu$  coefficient is  computed  using the loop functions \cite{Misiak} and is  given by
\bea
 C^\nu_L = -X(x_t)/\sin^2\theta_w\;, \label{cl}
\eea
whereas the $C_R^\nu$  coefficient is negligible. The branching ratio of $\bar B \to X_s \nu \bar \nu$ process is
\begin{eqnarray} \label{BR-bsnu}
\frac {d\Gamma}{d s_b} &= & m_b^5 \frac{\alpha^2 G_F^2}{128 \pi^5} |V_{ts}^* V_{tb}|^2 \kappa(0) \big (|C_L^\nu |^2 + |C_R^\nu|^2\big )  
\lambda^{1/2} (1,\tilde{m}_s^2, s_b)\nn \\ &  \times&
\left[3s_b \left (1+ \tilde{m}_s^2 - s_b-4\tilde{m}_s \frac{Re(C_L^\nu C_R^{\nu *})}{|C_L^\nu|^2 + |C_R^\nu|^2}\right ) + \lambda \big(1,\tilde{m}_s^2, s_b \big )  \right],
\end{eqnarray}
where $\tilde{m}_s = m_s/m_b$, $s_b = s/m_b^2$ and $\kappa(0) = 0.83$ is the QCD correction to the $b \to s \nu \bar \nu$  matrix element \cite{Grossman}. For numerical analysis, we have used the quark masses as $m_s=0.1$ GeV and $m_b=4.8$ GeV.
  It should be noted from (\ref{Lagrangian}) that  $U_3$ leptoquark has  additional Wilson coefficient contribution to $b \to s \nu_i \bar \nu_i$ process, which is given by 
\bea
C_L^{ LQ} =\frac{2 \pi}{\sqrt{2}G_F \alpha V_{tb}V_{ts}^*} \sum_{m,n=1}^3 V_{m3}V_{n2}^* \frac{h_{3L}^{ni} h_{3L}^{mi^*}}{M_{U_3^{-1/3}}^2}. 
\eea
In the presence of LQ the total decay rate of $\bar B \to X_s \nu \bar \nu$ process can be obtained from (\ref{BR-bsnu}) by replacing the Wilson coefficient $C_L^\nu \to C_L^\nu + C_L^{ LQ}$.  Using  all the particle masses and the lifetime of $B$ meson  from \cite{pdg}, the  branching ratio in the SM is found to be
\bea
{\rm Br}(\bar B \to X_s \nu \bar \nu) = (2.74 \pm 0.16) \times 10^{-5},
\eea
and the corresponding  experimental upper limit measured by the ALEPH collaboration is given by \cite{Aleph}
\bea
{\rm Br}(\bar B \to X_s \nu \bar \nu) < 6.4 \times 10^{-4}.
\eea
Since $U_3^{2/3}$ and $U_3^{-1/3}$ LQs are coming from the same $SU(2)$
triplet, one can constrain $h_{3L}^{2l} h_{3L}^{3l*}$ couplings by assuming that both the LQs have the same mass.
 Now comparing the theoretical and experimental branching ratio, we show the constraints on $U(3,3,2/3)$ leptoquark couplings in Fig. 6.  From the figure, the allowed ranges of real and imaginary part of the couplings are found as
\bea
-0.02 \leq {\rm Re}[h_{3L}^{2i} h_{3L}^{3i^*}] \leq 0.02, ~~~~~-0.02 \leq {\rm Im}[h_{3L}^{2i} h_{3L}^{3i^*}] \leq 0.02.
\eea
%%%%%%%%%%%%%%%%%%%%%%%%%%%%%%%%%%%%%%%%%%%%%%%%%%%%%%%%%%%%%%%%%%%%%%%%%%%%%%
\begin{figure}[h]
\centering
\includegraphics[scale=0.6]{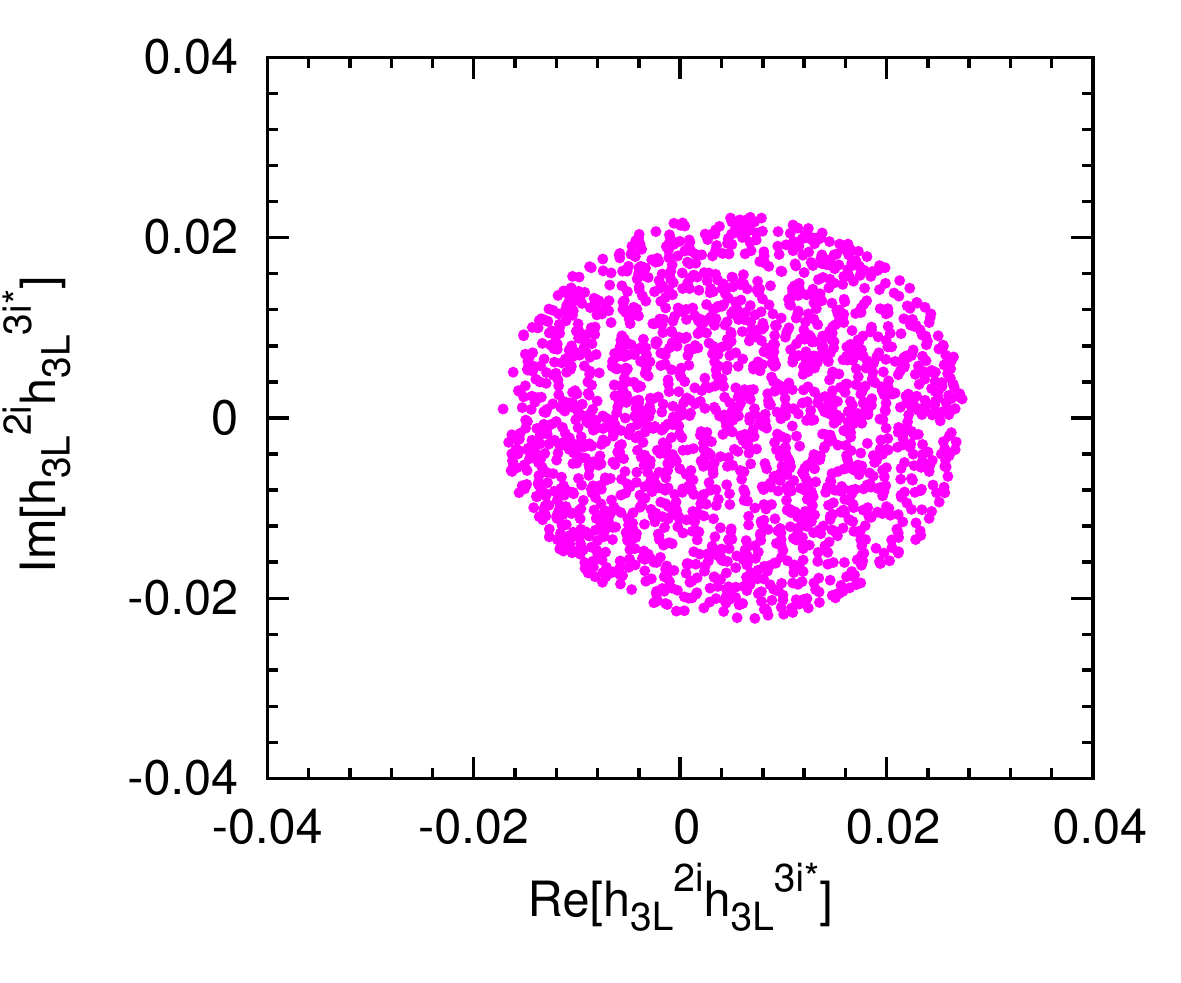}
\caption{Constraints on the  real and imaginary parts of the leptoquark couplings from $\bar B \to X_s \nu \bar \nu$ process  in the $U(3,3,2/3)$ leptoquark model.}
\end{figure}
%%%%%%%%%%%%%%%%%%%%%%%%%%%%%%%%%%%%%%%%%%%%%%%%%%%%%%%%%
\subsection{$B_{u}^+ \to l^+ \nu_l$ processes}
%%%%%%%%%%%%%%%%%%%%%%%%%%%%%%%%%%%%%%%%%%%%%%%%%%%%%%%%%%
The rare leptonic  $B_{u}^+ \to l^+  \nu_l$ decay modes, where $l=e, \mu, \tau$  mediated by $ b \to u l \nu$ transitions  can  provide significant constraints on models of new physics. Neglecting the electromagnetic radiative corrections, the branching ratios of the  $B_{u}^+ \to l^+  \nu_l$ processes in the $U_{1, 3}$ leptoquark model are given by \cite{fazio},
\bea \label{BR-Bulnu}
{\rm Br} (B_{u}^+ \to l^+  \nu_l)&=&\frac{G_F^2 M_{B_u} m_l^2}{8 \pi} \Big( 1-\frac{m_l^2}{M_{B_u}^2}\Big)^2 f_{B_u}^2 \left | V_{ub} \right |^2 \tau_{B^+} \nn \\ && \times \Big | \left(1 + C_{V_1}-C_{V_2} \right)+\frac{M_{B_u}^2}{m_l (m_b +m_u)}  C_{S_1} \Big |^2,
\eea
where $C_{V_{1,2}}$ and $C_{S_{1}}$  Wilson coefficients arise due to $U_{1, 3}$ leptoquark exchange and are  negligible in the SM. Using  the particle masses and life time of $B_u^+$ meson from \cite{pdg}, the decay constants $f_{B_{u, d}}=190.5(4.2)$ MeV \cite{fB} and $|V_{ub}| = 4.13 (49) \times 10^{-3}$ \cite{pdg}, the  branching ratios  in the SM are found to be
\bea
&&{\rm Br} (B_u^+ \to e^+ \nu_e)= (8.9 \pm 0.23) \times 10^{-12}, \nn \\
&&{\rm Br} (B_u^+ \to \mu^+ \nu_\mu)= (3.83 \pm 0.1)  \times 10^{-7}, \nn \\
&&{\rm Br} (B_u^+ \to \tau^+ \nu_\tau)= (8.48 \pm 0.28) \times 10^{-5},
\label{Bu-SM}
\eea
and the corresponding averaged experimental values are \cite{pdg}
\bea
&&{\rm Br} (B_u^+ \to e^+ \nu_e) ~\textless ~9.8 \times 10^{-7},\nn \\
&& {\rm Br} (B_u^+ \to \mu^+ \nu_\mu)~ \textless ~1.0 \times 10^{-6}, \nn  \\ 
&&{\rm Br} (B_u^+ \to \tau^+ \nu_\tau) =(1.14 \pm 0.27) \times 10^{-4}.
\label{Bu-Exp}
\eea
If we apply chirality on LQ, then only $C_{V_1}$ Wilson coefficient will contribute to the branching ratios. 
 Now comparing the theoretical (\ref{Bu-SM}) and experimental (\ref{Bu-Exp}) values, the allowed region of real and imaginary part of LQ couplings  from $B_u^+ \to e^+ \nu_e$ (left panel), $B_u^+ \to \mu^+ \nu_\mu$ (right panel) and $B_u^+ \to \tau^+ \nu_\tau$ (bottom panel) processes are shown in Fig. 7 and  the constrained values are given in Table IV.  From (\ref{BR-Bulnu}), it should be noted that the contribution of $C_{S_1}$ Wilson coefficient  is enhanced by the factor $M_{B_u}^2/m_l$, so  we will neglect the NP in $C_{V_1}$  for simplicity.  Then the branching ratio is only sensitive to the $C_{S_1}$ Wilson coefficient. In Fig. 8, we show the constraint on  $U(3,1,2/3)$ LQ couplings from $B_u^+ \to e^+ \nu_e$ (left panel), $B_u^+ \to \mu^+ \nu_\mu$ (right panel) and $B_u^+ \to \tau^+ \nu_\tau$ (bottom panel) processes and the allowed ranges are given in Table IV.

%%%%%%%%%%%%%%%%%%%%%%%%%%%%%%%%%%%%%%%%%%%%%%%%%%%%%%%%%%%%%%%%%%%%%%%%%%%%%%
\begin{figure}[h]
\centering
\includegraphics[scale=0.6]{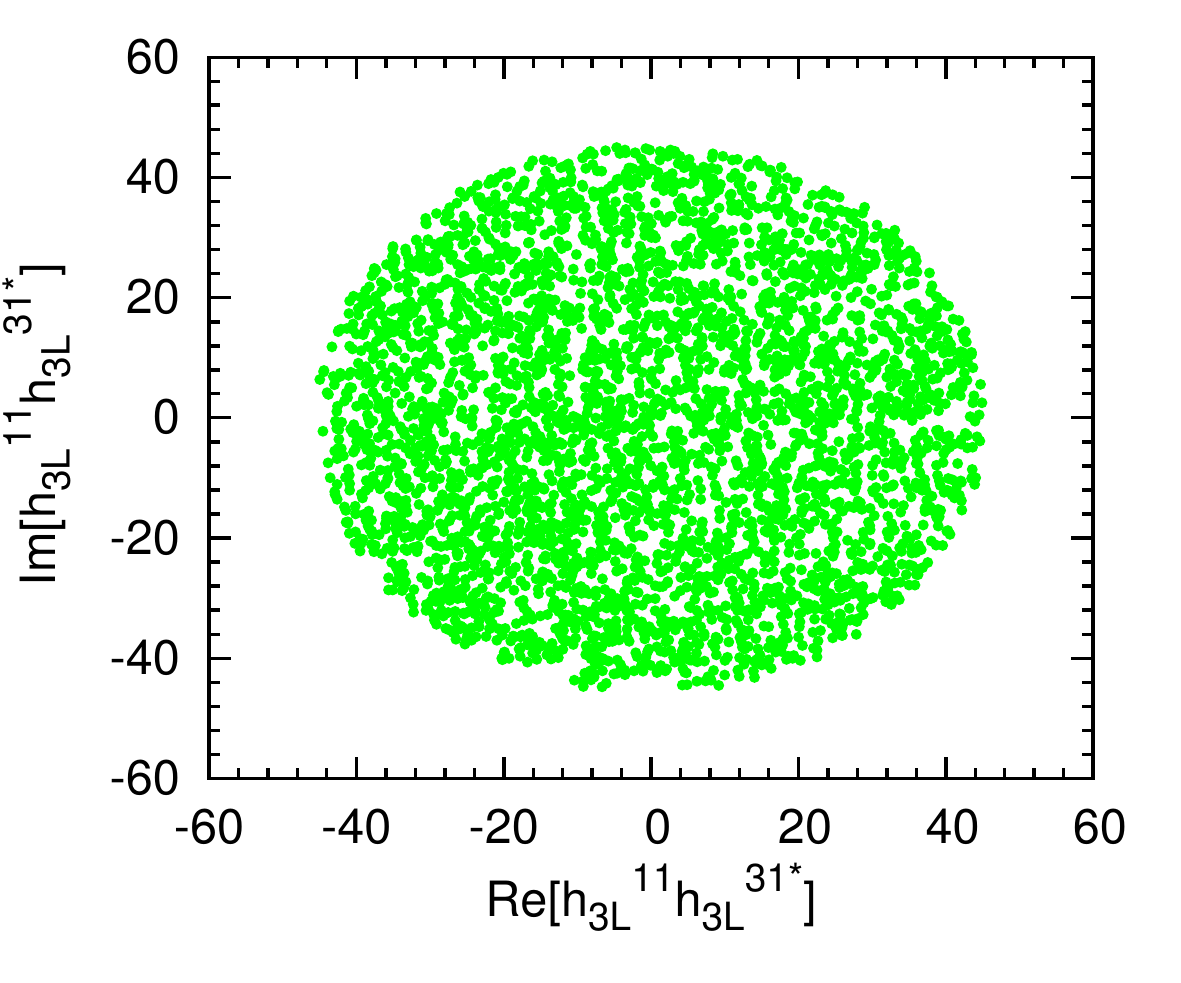}
\quad
\includegraphics[scale=0.6]{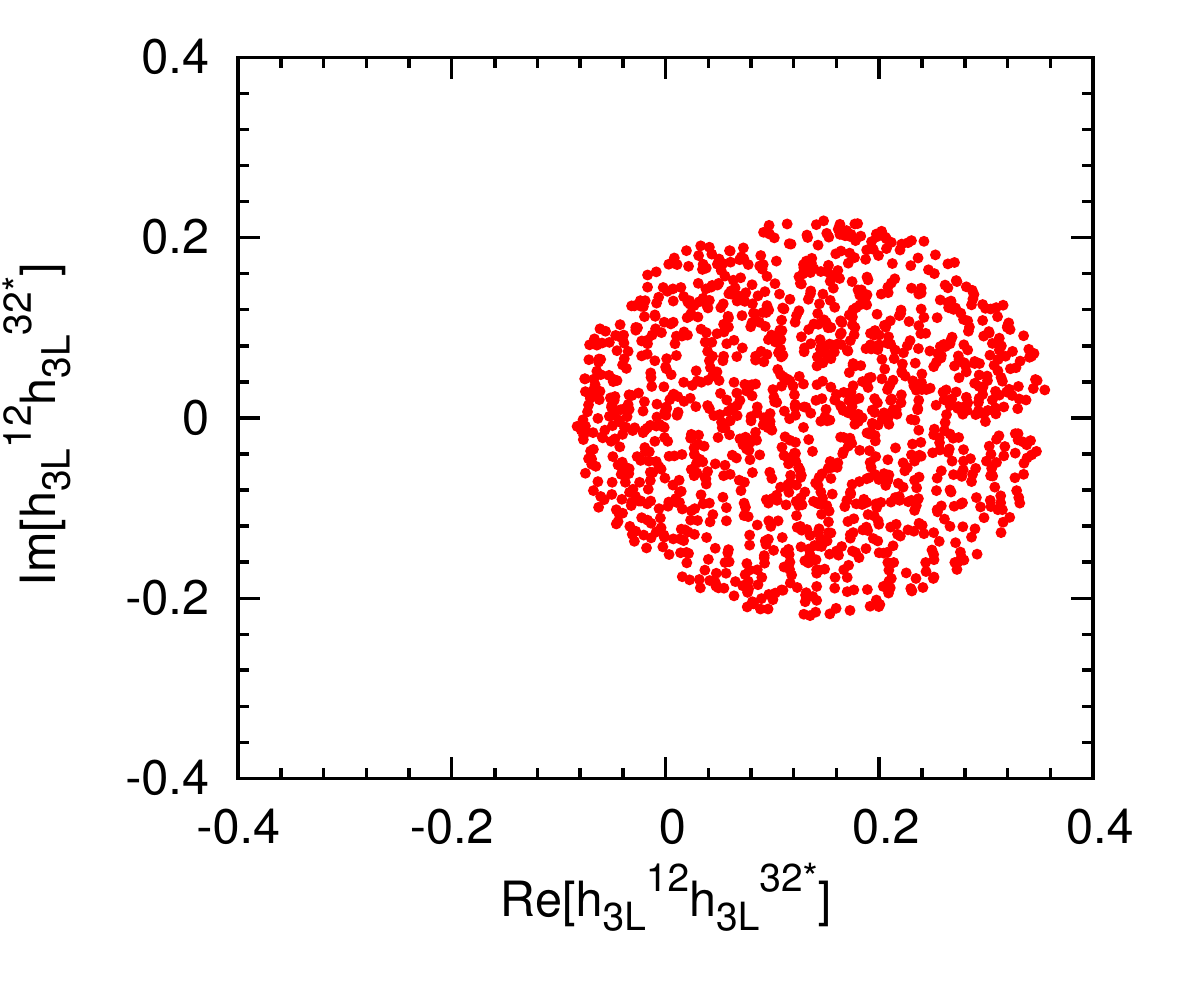}
\quad
\includegraphics[scale=0.6]{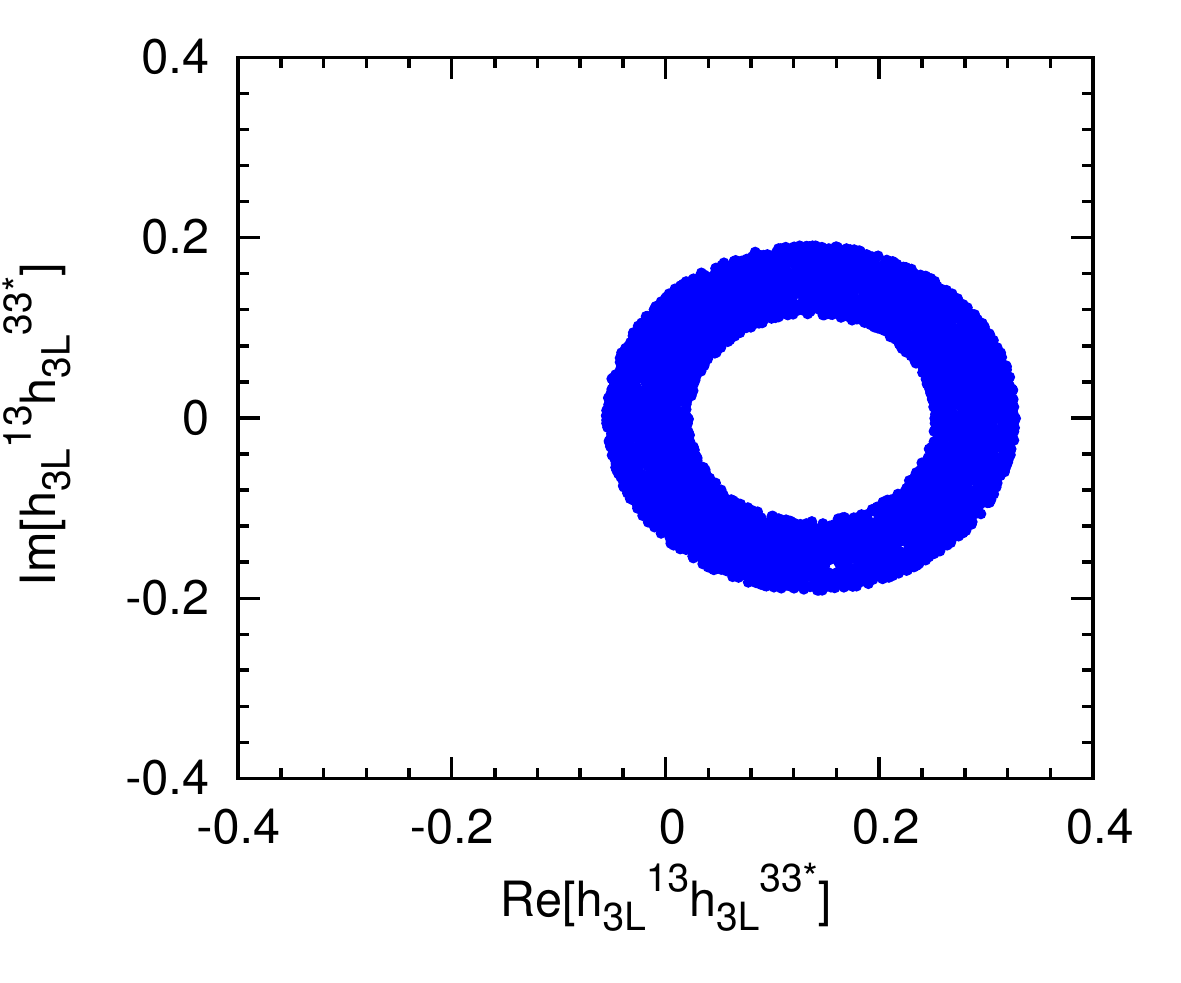}
\caption{Constraints  on the real and imaginary parts of the leptoquark couplings from $B_u^+ \to e^+ \nu_e$ (left panel), $B_u^+ \to \mu^+ \nu_\mu$ (right panel) and $B_u^+ \to \tau^+ \nu_\tau$ (bottom panel) processes in $U(3,3,2/3)$ leptoquark model.}
\end{figure}
%%%%%%%%%%%%%%%%%%%%%%%%%%%%%%%%%%%%%%%%%%%%%%%%%%%%%%%%%%%%%
%%%%%%%%%%%%%%%%%%%%%%%
\begin{figure}[h]
\centering
\includegraphics[scale=0.6]{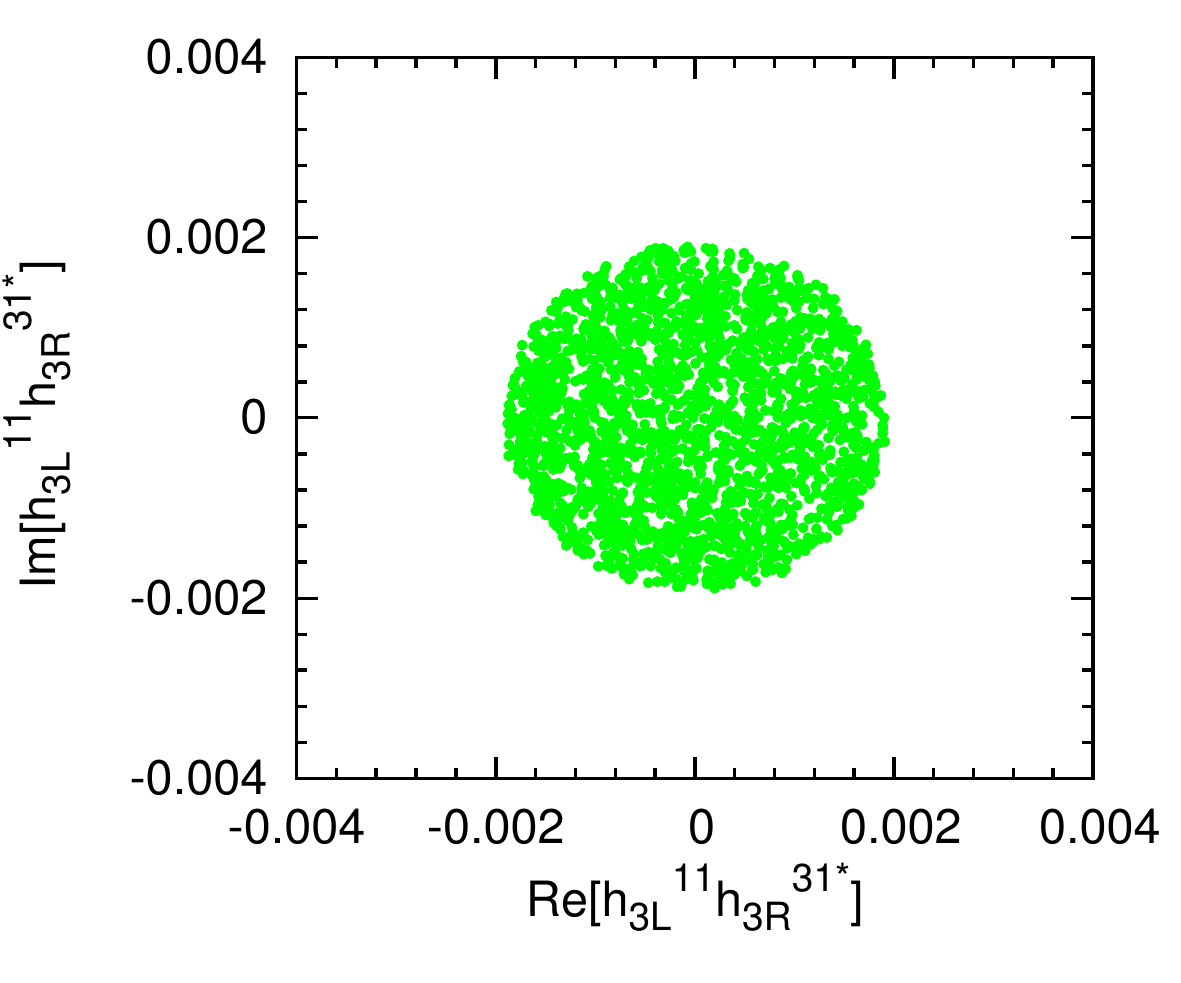}
\quad
\includegraphics[scale=0.6]{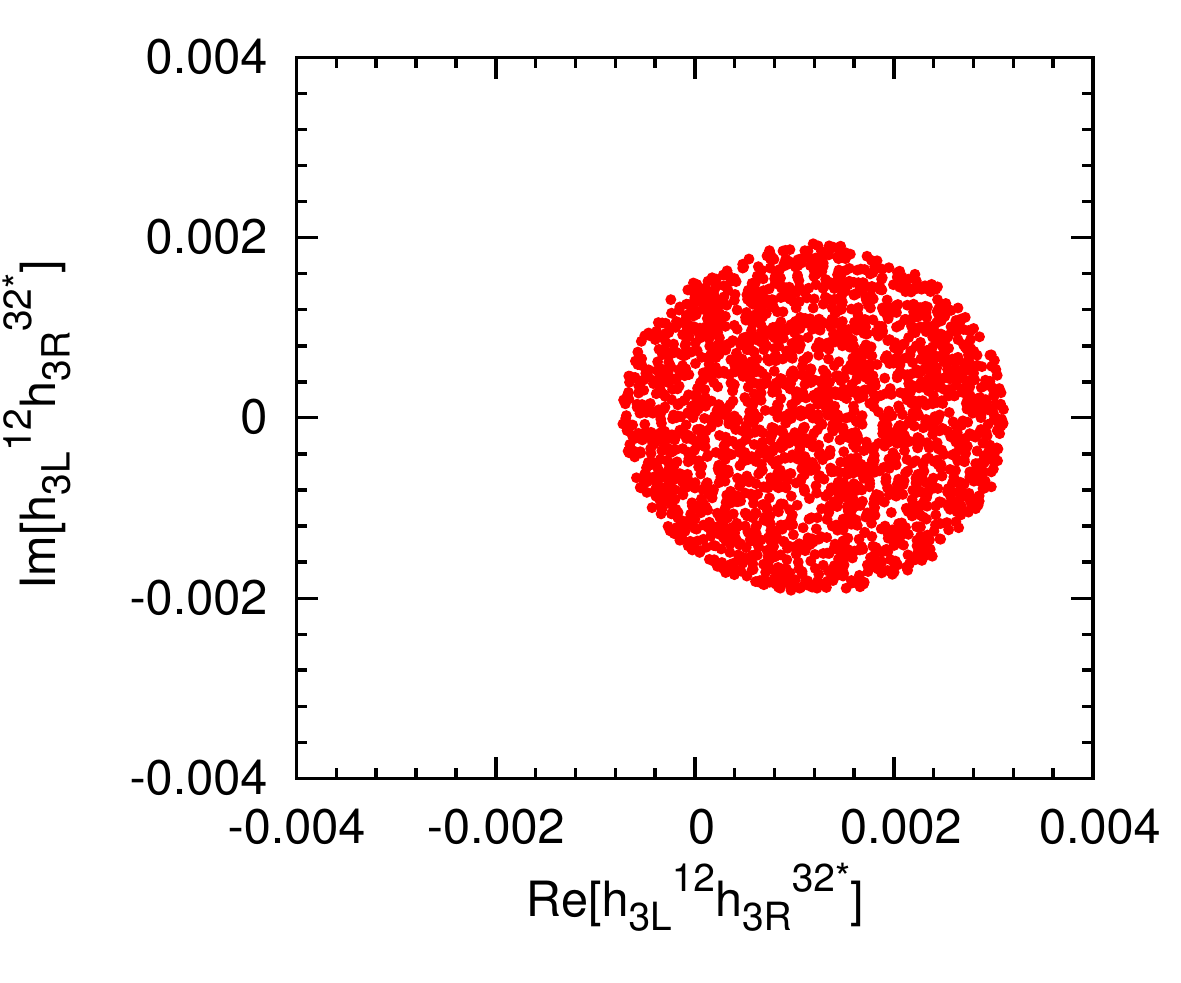}
\quad
\includegraphics[scale=0.6]{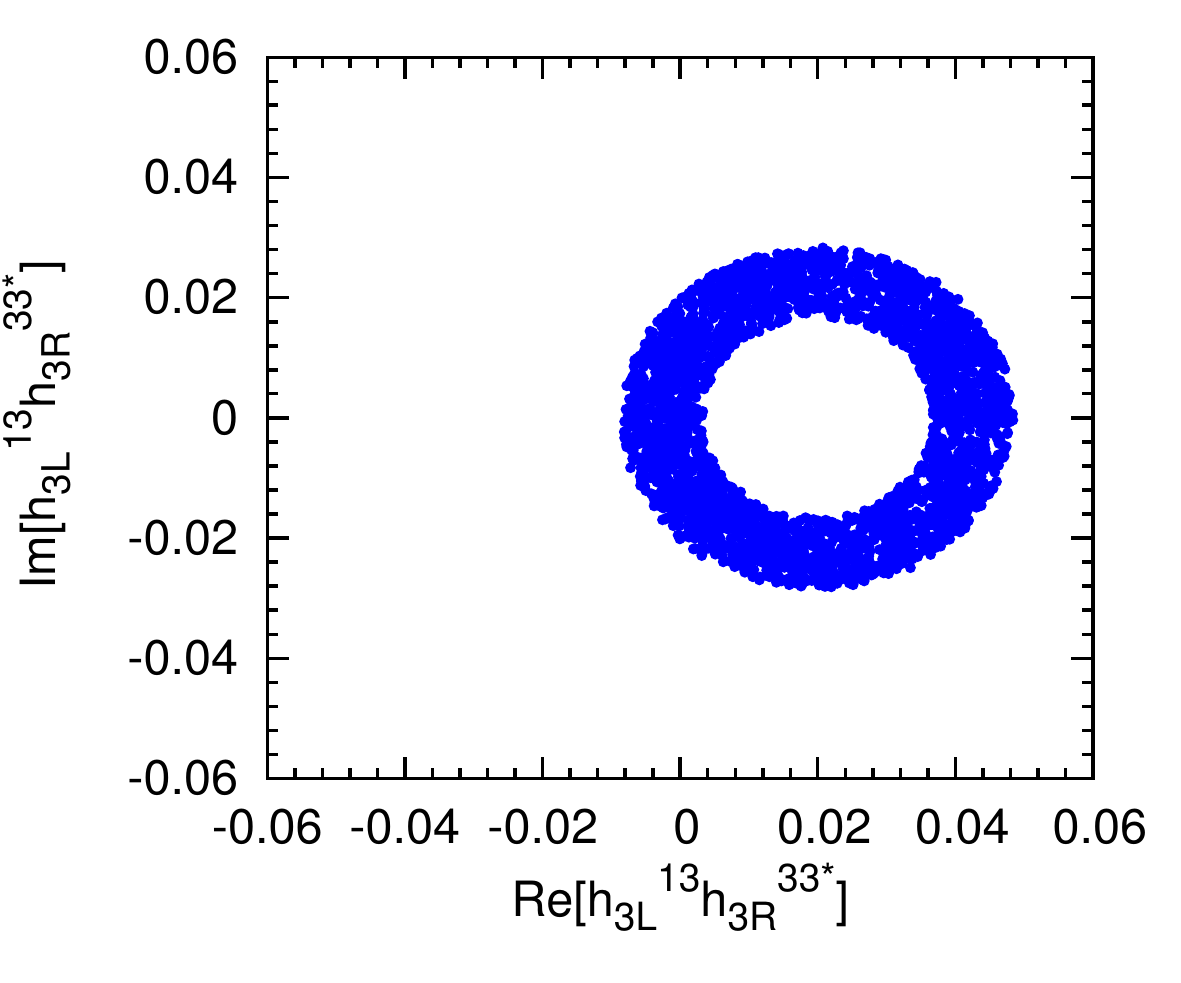}
\caption{Constraints  on the real and imaginary parts of the leptoquark couplings from $B_u^+ \to e^+ \nu_e$ (left panel), $B_u^+ \to \mu^+ \nu_\mu$ (right panel) and $B_u^+ \to \tau^+ \nu_\tau$ (bottom panel) processes in $U(3,1,2/3)$ leptoquark model.}
\end{figure}
%%%%%%%%%%%%%%%%%%%%%%%%%%%%%%%%%%%%%%%%%%%%%%%%%%%%%%%%%%%%%%%%%%%%%%%%%%%%%
\begin{table}[h]
\caption{Constraint on real and imaginary part of the leptoquark couplings  from $B_u^+ \to l^+ \nu_l$ processes, where $l=e, \mu, \tau$}
\begin{center}
\begin{tabular}{ | c | c | c |}
\hline
~Leptoquark Couplings~ & ~~Real part~~ &~~ Imaginary Part~~  \\
\hline
\hline
$h_{1(3)L}^{11} {h_{1(3)L}^{{31}^*}}$ & $-40.0 \to 40.0$ & $-40.0 \to 40.0$ \\
$h_{1(3)L}^{12} {h_{1(3)L}^{{32}^*}}$   &   $-0.08 \to 0.32$ &$-0.2 \to 0.2$  \\
$h_{1(3)L}^{13} {h_{1(3)L}^{{33}^*}}$   &  $0.24 \to 0.32 $ &$-0.2 \to 0.2$  \\
 \hline
 $h_{1(3)L}^{11} {h_{1(3)R}^{{31}^*}}$ & $-0.002 \to 0.002$ & $-0.002 \to 0.002$ \\
$h_{1(3)L}^{12} {h_{1(3)R}^{{32}^*}}$ &$-0.0008 \to 0.0032$  &   $-0.002 \to 0.002$   \\
$h_{1(3)L}^{13} {h_{1(3)R}^{{33}^*}}$   &  $-0.034 \to 0.046 $ &$-0.028 \to 0.028$  \\
 \hline
\end{tabular}
\end{center}
\end{table}
%%%%%%%%%%%%%%%%%%%%%%%%%%%%%%%%%%%%%%%%%%%%%%%%%%%%%%%%%%%%%%%%%%
\section{$B \to D^{(*)} l \bar \nu$ process}
%%%%%%%%%%%%%%%%%%%%%%%%%%%%%%%%%%%%%%%%%%%%%%%%%%%%%%%%%
In this section, we discuss the theoretical framework to compute the branching ratios and other physical observables in  $B \to D^{(*)} l \bar \nu$ processes. The hadronic matrix elements  between the initial $B$ meson and final $D$ meson can be parameterized in terms of the form factors $F_0 (q^2)$, $F_1 (q^2)$ and $F_T (q^2)$ as  \cite{sakaki}
\bea
\Big \langle D (k) | \bar{c} \gamma_\mu  b| \bar{B} (p) \Big \rangle & =& \Big[ \left(p+k\right)_\mu -\frac{M^2_B - M^2_D}{q^2}q_\mu \Big]  F_1\left(q^2\right) + q_\mu \frac{M^2_B - M^2_D}{q^2}  F_0\left(q^2\right),\nn\\
\Big \langle D (k) | \bar{c}\sigma_{\mu \nu} b| \bar{B} (p) \Big \rangle &=& -i \left(p_\mu k_\nu -k_\mu p_\nu \right) \frac{2F_T \left(q^2 \right)}{M_B + M_D}, 
\label{B-D-form-factor}
\eea
where $p$, $k$ are  the 4-momenta of the  $B$  and $D$ mesons respectively and $q^2=(p-k)^2$ is the momentum transfer to the dilepton system. The expression for $F_{1,0, T} (q^2)$ form factors in terms of heavy quark effective theory (HQET) form factors $(h_{\pm, T} (q^2))$   are given in Appendix A \cite{sakaki, Tanaka}. Using Eqn. (\ref{B-D-form-factor}) the differential decay rate of $B \to D \tau \bar \nu_l$ process with respect to $q^2$ is given by \cite{sakaki, Tanaka}
\bea
\frac{d \Gamma  \left(\bar{B} \to D \tau \bar{\nu}_l \right)}{dq^2} &=& \frac{G_F^2|V_{cb}|^2}{192 \pi^3 M_B^3} q^2 \sqrt{\lambda_D \left(q^2 \right)} \Big( 1-\frac{m_\tau^2}{q^2} \Big) ^2  \nn \\ & \times &
\Bigg [ \Big | \delta_{l\tau}+C_{V_1}^l \Big|^2 \Big ( \Big( 1+\frac{m_\tau^2}{2q^2} \Big) {H_{V, 0}^s}^2 + \frac{3}{2}\frac{m_\tau^2}{q^2} {H_{V, t}^s}^2 \Big ) \nn \\ &  + & \frac{3}{2} \Big | C_{S_1}^l \Big|^2 {H_S^s}^2 + 3 {\rm Re} \left[ \left(\delta_{l\tau}+C_{V_1}^l \right) C_{S_1}^{l *}\right] \frac{m_\tau}{\sqrt{q^2}} H_S^s H_{V, t}^s \Bigg ], \hspace{0.7cm}
\eea
where $\lambda_{D}\left(q^2 \right) = \big[(M_B -M_{D})^2-q^2 \big]\big[(M_B +M_{D})^2-q^2 \big]$ and the  hadronic amplitudes ($H^s_{V, (0, t)}$ and $H_S^s$)  are given in Appendix A.

The matrix element in the $B \to D^* \tau \bar \nu_l$ process can be parametrized as  \cite{sakaki}
\bea
\Big \langle D^* (k, \varepsilon) | \bar{c} \gamma_\mu b| \bar{B} (p) \Big \rangle &=& -i \epsilon_{\mu \nu \rho \sigma} {\varepsilon^\nu}^* p^\rho k^\sigma \frac{2V(q^2)}{M_B+M_{D^*}}\;, \nn \\
\Big \langle D^* (k, \varepsilon) | \bar{c} \gamma_\mu \gamma_5 b| \bar{B} (p) \Big \rangle &=& {\varepsilon^{\mu }}^* \left( M_B + M_{D^*}\right) A_1(q^2) -\left(p+k \right)_\mu (\varepsilon^* \cdot q) \frac{A_2 (q^2)}{M_B + M_{D^*}} \nn \\ &-& q_\mu  (\varepsilon^* \cdot q) \frac{2M_{D^*}}{q^2} \left[A_3(q^2) - A_0 (q^2) \right],
\eea
where 
\bea
A_3 (q^2) = \frac{M_B + M_{D^*}}{2 M_{D^*}} A_1 (q^2) - \frac{M_B - M_{D^*}}{2 M_{D^*}} A_2 (q^2)\;,
\eea
and the $V(q^2)$ and $A_{0,1,2} (q^2)$ in terms of HQET form factors are presented in Appendix B.  The differential decay distribution  with respect to $q^2$ is given as \cite{sakaki, Tanaka}
\bea
\frac{d \Gamma  \left(\bar{B} \to D^* \tau \bar{\nu}_l \right)}{dq^2} &=& \frac{G_F^2|V_{cb}|^2}{192 \pi^3 M_B^3} q^2 \sqrt{\lambda_{D^*} \left(q^2 \right)} \Big( 1-\frac{m_\tau^2}{q^2} \Big) ^2  \nn \\ & \times & \Bigg [ \Big | \delta_{l\tau}+C_{V_1}^l\Big|^2 \Bigg ( \Big( 1+\frac{m_\tau^2}{2q^2} \Big) \left(H_{V, +}^2 + H_{V, -}^2 + H_{V, 0}^2 \right)  + \frac{3}{2}\frac{m_\tau^2}{q^2} H_{V, t}^2 \Bigg ) \nn \\  &+& \frac{3}{2} \Big | C_{S_1}^l \Big|^2 H_S^2  + 3 {\rm Re} \left[ \left(\delta_{l\tau}+C_{V_1}^l\right) C_{S_1}^{l *} \right] \frac{m_\tau}{\sqrt{q^2}} H_S H_{V, t} \Bigg ], 
\eea
where $H_{V, \pm}$, $H_{V, 0}$, $H_{V, t}$  and $H_{S}$ are the  hadronic amplitudes described in Appendix B.  Another interesting observable, i.e., the lepton non-universality parameter,  is  the ratio of branching fractions of $B \to D^{(*)} \tau \bar{\nu}_\tau$  to $B \to D^{(*)} l \bar{\nu}_l$  processes, defined as \cite{Bauer, sakaki, Tanaka, fazio, kosnik}
\bea
R_{D^{(*)}} = \frac{{\rm Br}\left(\bar{B} \to D^{(*)} \tau \bar{\nu}_\tau \right)}{{\rm Br}\left(\bar{B} \to D^{(*)} l \bar{\nu}_l \right)},
\eea
which probes  lepton flavour dependent term in and beyond SM. Similarly in  the $b \to s l^+ l^-$ transition, the lepton non-universality is given by \cite{mohanta1, Isidori}
\bea
R_{K^{(*)}} = \frac{{\rm Br}\left(\bar{B} \to K^{(*)} \mu^+ \mu^- \right)}{{\rm Br}\left(\bar{B} \to K^{(*)} e^+ e^- \right)}.
\eea
The decay rate expressions for $\bar{B} \to K^{(*)} \mu^+ \mu^-$ are taken from \cite{hiller, mohanta2}.
One can also see the $q^2$ variation of these parameters using the relations 
\bea
R_{D^{(*)}}(q^2) =\frac{{d \Gamma }\left(\bar{B} \to D^{(*)} \tau \bar{\nu}_\tau \right)/d q^2}{{ d \Gamma}\left(\bar{B} \to D^{(*)} l \bar{\nu}_l \right)/d q^2},~~~~
R_{K^{(*)}}(q^2) = \frac{{d \Gamma}\left(\bar{B} \to K^{(*)} \mu^+ \mu^- \right)/d q^2}{{ d \Gamma}\left(\bar{B} \to K^{(*)} e^+ e^- \right)/dq^2}.
\eea 
Besides the branching ratios and lepton non-universality parameters, the following  interesting observables  could be sensitive to new physics. 
\begin{itemize}
\item
The $\tau$ forward-backward asymmetry  in the $B \to D^{(*)} \tau \bar{\nu}_\tau$ processes is defined as \cite{sakaki, fazio}
\bea
A_{FB} (q^2) = \frac{\int_0^1 \frac{d\Gamma}{d\cos \theta}d\cos \theta-\int_{-1}^0 \frac{d\Gamma}{d\cos \theta}d\cos \theta}{\int_{-1}^1 \frac{d\Gamma}{d\cos \theta}d\cos \theta} = \frac{ b_{\theta}(q^2) }{d\Gamma / dq^2},
\eea
where $\theta$ is the angle between the direction of the charged lepton and the $D^{(*)}$ meson in the $\tau \bar{\nu}$ rest frame. The  expression for $b_\theta (q^2)$ can be found in \cite{sakaki}. 
\item
$\tau$ polarization parameter is defined as  \cite{sakaki}
\bea
P_\tau (q^2) = \frac{ d\Gamma (\lambda_\tau = 1/2)/dq^2 - d\Gamma (\lambda_\tau = -1/2)/dq^2}{d\Gamma (\lambda_\tau = 1/2)/dq^2 + d\Gamma (\lambda_\tau = -1/2)/dq^2},
\eea
where the decay distribution $d\Gamma(\lambda=\pm 1/2)/dq^2$ is given in Appendix C . 
\item
The longitudinal and transverse  polarization of $D^*$ can be defined as  \cite{fazio}
\bea
F_{L, T}^{D^*}(q^2) = \frac{d\Gamma_{L, T} \left(B \to D^* \tau \bar{\nu} \right)/ dq^2}{d\Gamma \left(B \to D^* \tau \bar{\nu} \right) / dq^2},
\eea
where the subscripts $L, T$ denote the longitudinal and  transverse components respectively, and $d\Gamma_{T}/dq^2 = d\Gamma_{+} / dq^2+ d\Gamma_{-} / dq^2$. The complete expression for $d\Gamma_{\pm} / dq^2$ is presented in Appendix C.

\item
Analogous to $R_{D^*}$, one can also define the ratio of   longitudinal and transverse $D^*$ polarization distribution of $B \to D^* \tau \bar{\nu}_\tau$ to the corresponding $B \to D^* l \bar{\nu}_l$  process as \cite{fazio}
\bea
R_{L, T}^{D^*}(q^2) = \frac{d\Gamma_{L, T} \left(B \to D^* \tau \bar{\nu} \right) / dq^2 }{d\Gamma_{L, T} \left(B \to D^* l \bar{\nu} \right) / dq^2}.
\eea
\end{itemize}

After getting familiar with the expressions for branching ratios and different physical observables of  $B \to D^{(*)} l \bar \nu_l$ processes, we now proceed for numerical estimation. All the particle masses and the life time of $B$ meson are taken from \cite{pdg} and the CKM matrix element $|V_{cb}|=0.0424(9)$ \cite{Vcb}.   Now using the constrained leptoquark parameter space as discussed in section III and the Eqns. (\ref{CV1}, \ref{CV2}, \ref{CS1}), we calculate bound on the new Wilson coefficients $C_{V_1} (C_{S_1})$.  If we apply chirality on  vector LQs, then $C_{V_1}$ is the only additional Wilson coefficient to the SM. As the constraint on $C_{V_1}$ is found to be same for both $U_{1, 3}$ leptoquark (with only a sign difference),  we  present the effects of only $U_3$ leptoquark  in our analysis. We show in Fig. 9, the branching ratio of $B \to D e \bar \nu$ (top-left panel), $B \to D \mu \bar \nu$ (top-right panel) and $B \to D \tau \bar \nu$ processes (bottom panel) with respect to $q^2$ in $U_{3}$  vector LQ model. Here darker blue dashed lines represent the SM contribution and the  orange bands are due to new physics contribution from  LQ model. The lighter blue bands  correspond to the uncertainties arising in the SM due to the uncertainties associated with the CKM matrix elements and the hadronic form factors. Similarly the $q^2$ variation of branching ratio of $B \to D^* e \bar \nu$ (top-left panel),  $B \to D^* \mu \bar \nu$ (top-right panel) and $B \to D^* \tau \bar \nu$ (right panel)  processes in the LQ model are presented in Fig. 10. The branching ratios of $B \to D^{(*)} \tau \bar \nu$ process has significant deviation from its SM value whereas  the deviation in $B \to D^{(*)} l \bar \nu_l$ process is negligible.  The integrated values of branching ratios of these processes in SM and LQ model are given in Table V. 
In Fig. 11, we present the plot for the $D^*$ polarization distributions in $B \to D^* l \nu_l$. The left panel of the figure is for $R_L^{D^*}$ and right panel for $R_T^{D^*}$. The predicted numerical values are given in Table V.   Since the LQ contribution  does not affect some observables like forward-backward asymmetry, $\tau$ polarization and $F_{L, T}^{D^*}(q^2)$,  we don't provide the corresponding  results.

 In Fig. 12, we show the variation of lepton non-universality parameters, $R_D (q^2)$ (left panel) and $R_D^* (q^2)$ (right panel) with respect to $q^2$ and the corresponding numerical values are presented in Table VI. Now using the constraints on real and imaginary part of the LQ couplings as given in  Table II and III,  and the Eqns. (\ref{C9-bs}, \ref{c9p-bs}, \ref{cs-bs}, \ref{csp-bs}), we compute the constraint on the new $C_{9, 10, S, P}^{(\prime) {\rm NP}}$ Wilson coefficients. Using the constrained parameters, the plot for  $R_K^{\mu e} (q^2)$  in low $q^2$ (left panel) and in high $q^2$ (right panel) are presented in Fig. 13. Fig. 14 shows the  $R_{K^*}^{\mu e} (q^2)$ anomaly plots in low $q^2$ (left panel) and high $q^2$ (right panel) in the LQ model. The predicted numerical values of lepton non-universality ($R_{K^{(*)}}$) are given in Table VI.  From Table VI, one can see that the predicted values of lepton non-universality parameters in the LQ model have significant deviation from the SM and  are within the $1\sigma$ range of experimental limit. We observe that the addition of new    vector LQ  can explain  both the $R_{D^{(*)}}$ and $R_{K^{(*)}}$ anomalies very well. 
 
%%%%%%%%%%%%%%%%%%%%%%%%%%%%%%%%%%%%%%%%%%%%%%%%%%%%%%%%%%%%%%%%%%%%%%%%%%%%%%
\begin{figure}[h]
\centering
\includegraphics[scale=0.4]{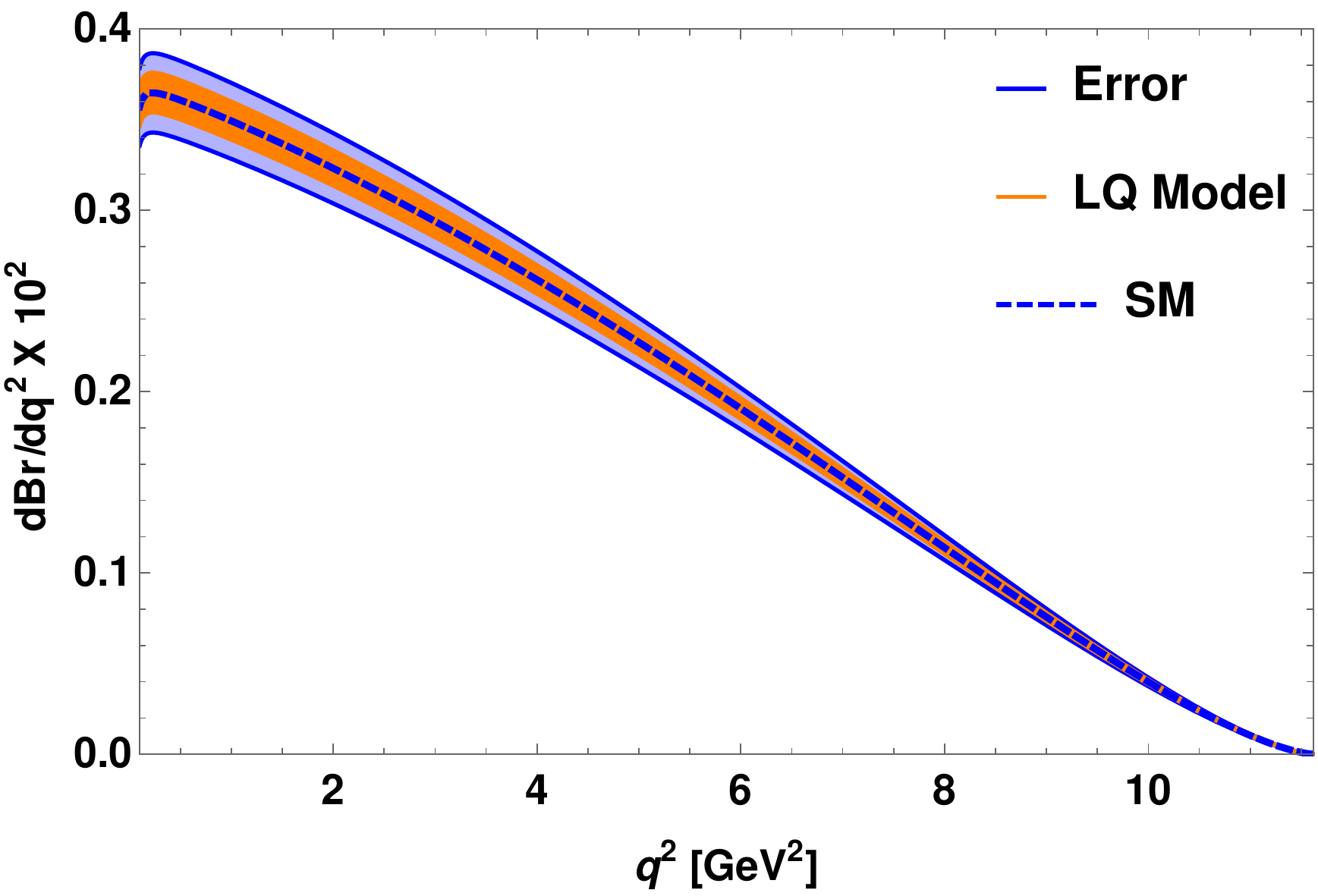}
\quad
\includegraphics[scale=0.4]{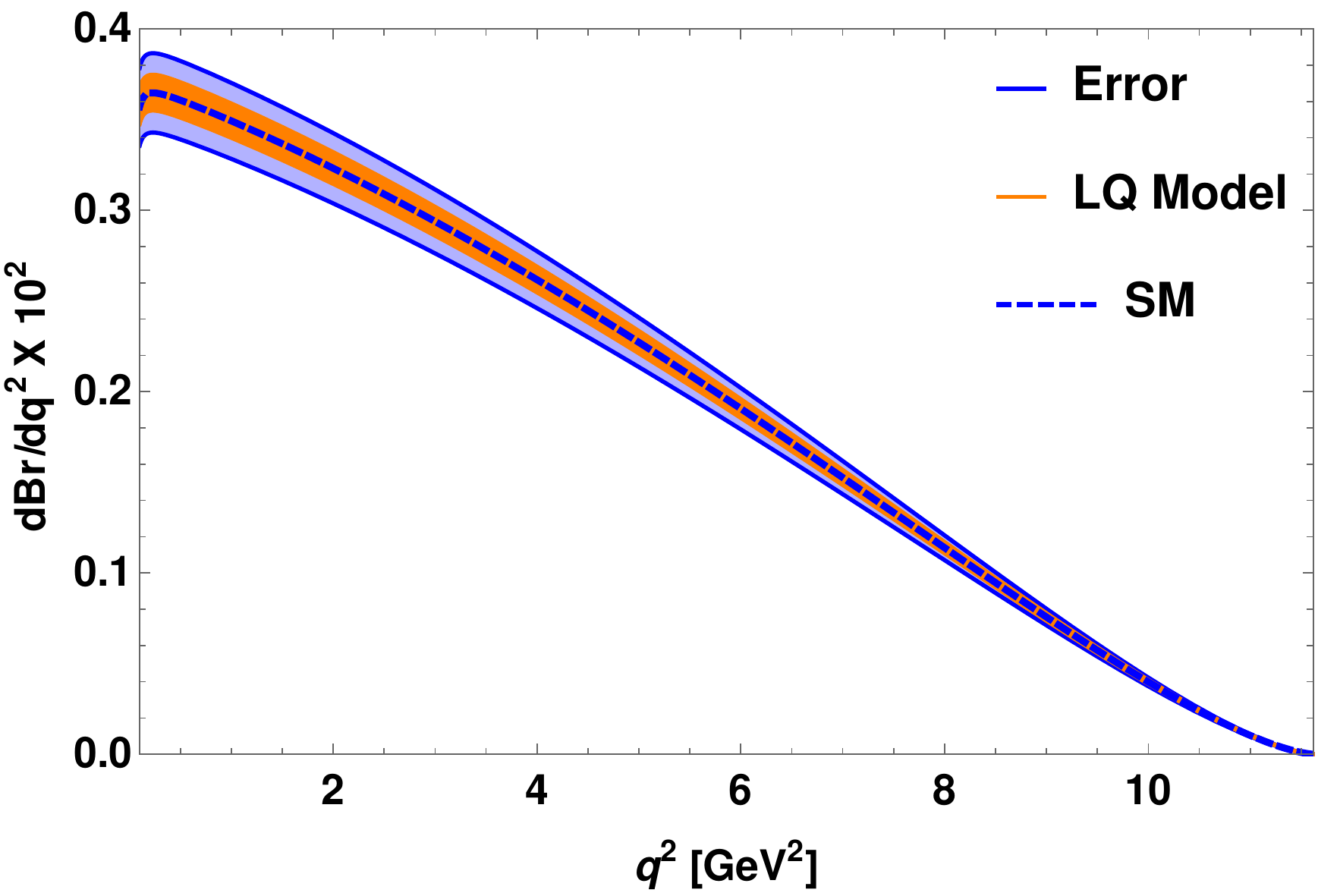}
\quad
\includegraphics[scale=0.4]{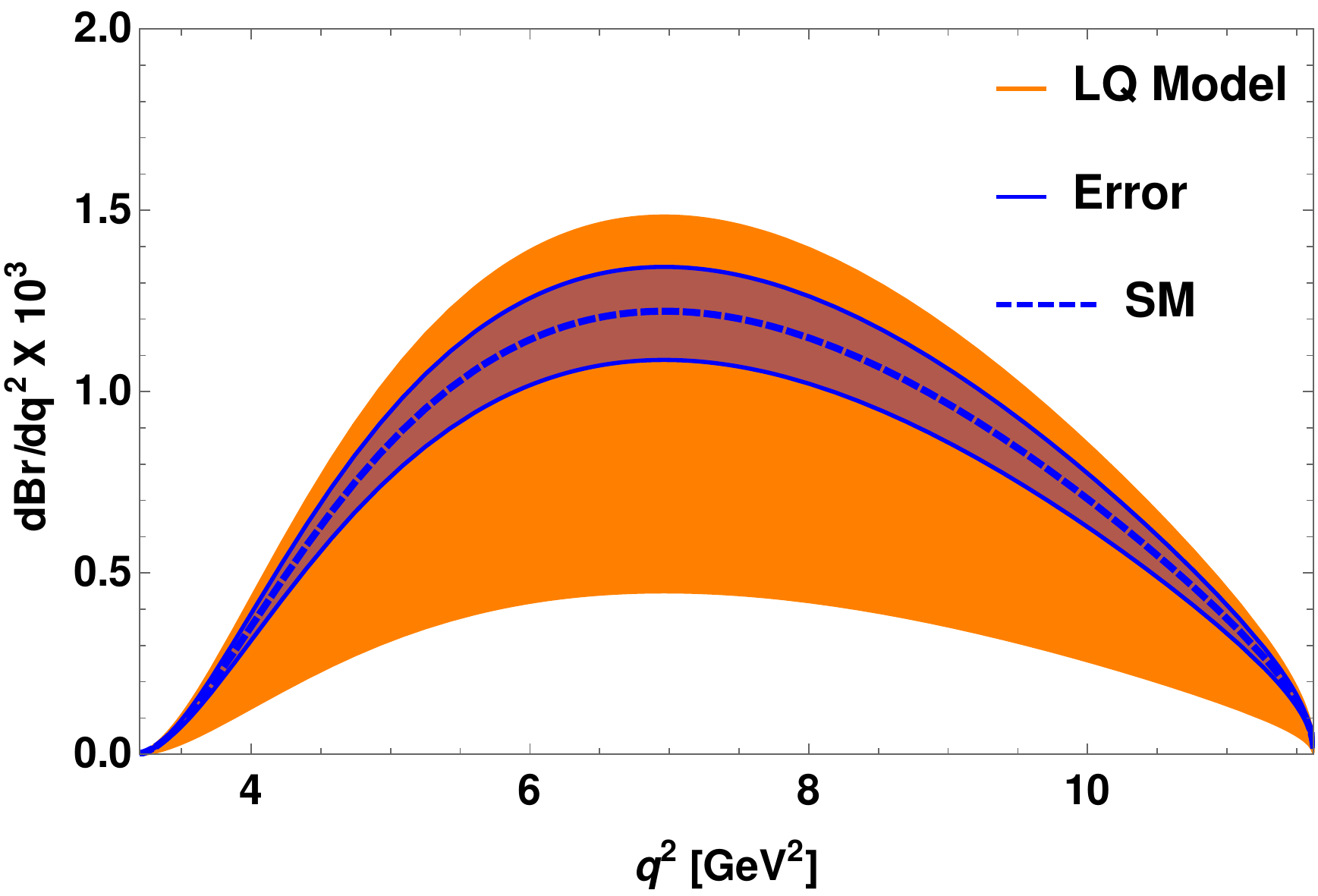}
\caption{The variation of  branching ratios of $B \to D e \bar{\nu}$  (left panel), $B \to D \mu \bar{\nu}$  (right panel) and $B \to D \tau \bar{\nu}$ (bottom panel) processes  with respect to $q^2$ in the leptoquark model. Here darker blue dashed lines are for SM and orange bands represent  leptoquark model. The lighter blue bands stand for the theoretical uncertainties arise due to the  input parameters in the SM.}
\end{figure}
%%%%%%%%%%%%%%%%%%%%%%%%%%%%%%%%%%%%%%%%%%%%%%%%%%%%%%%%%%%%%%%%%%%%%%%%%%%%%%
\begin{figure}[h]
\centering
\includegraphics[scale=0.4]{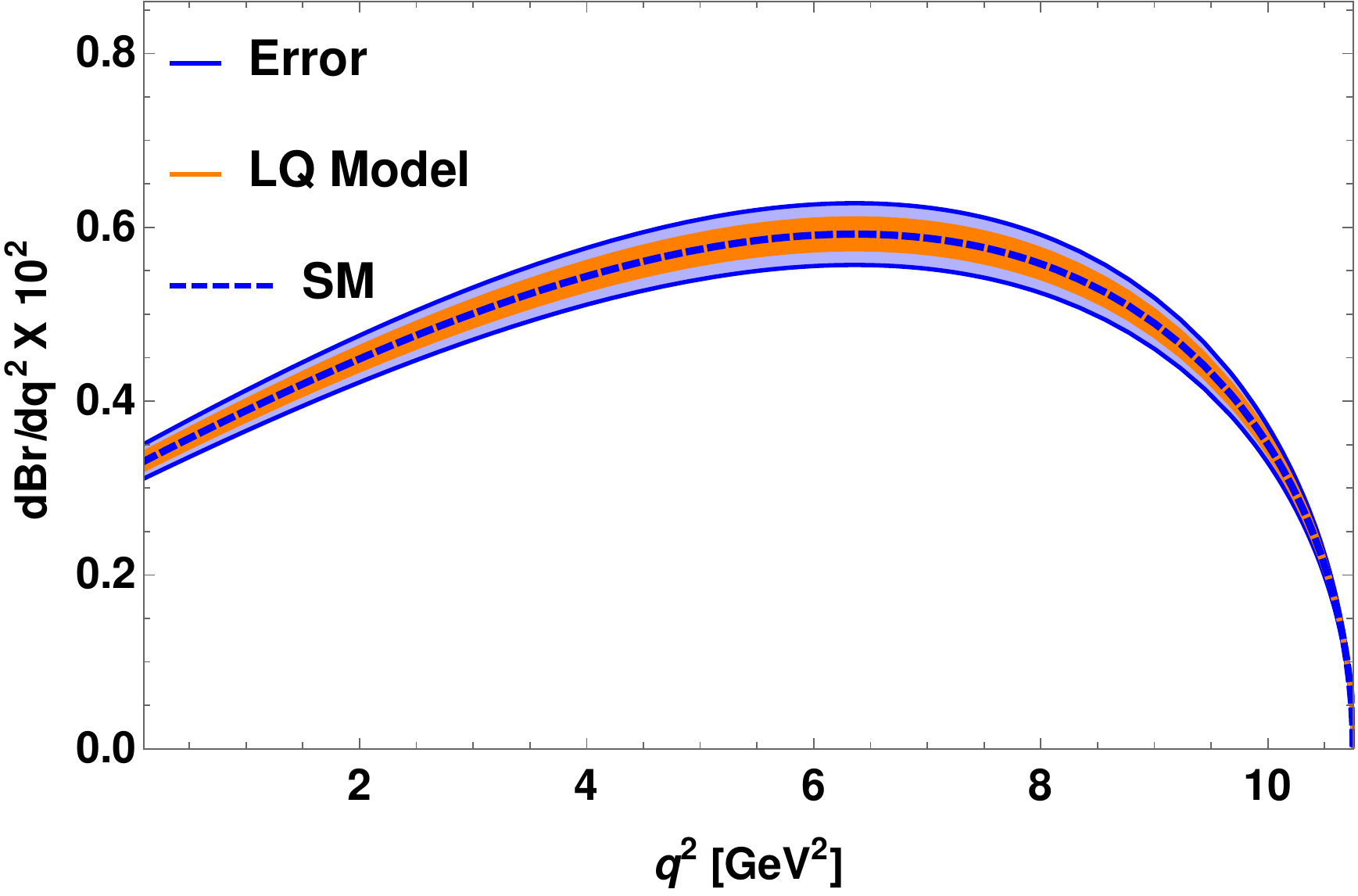}
\quad
\includegraphics[scale=0.4]{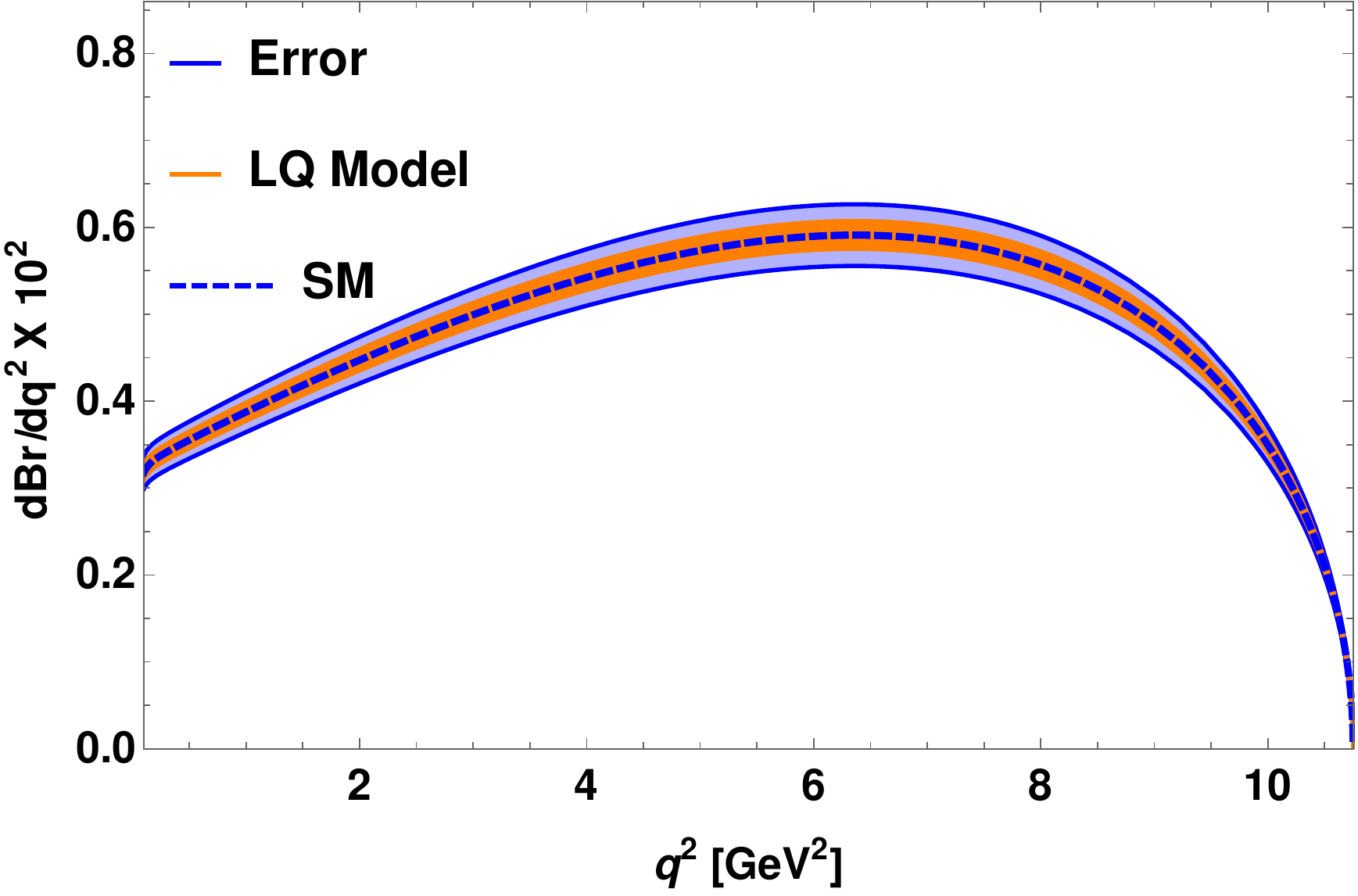}
\quad
\includegraphics[scale=0.4]{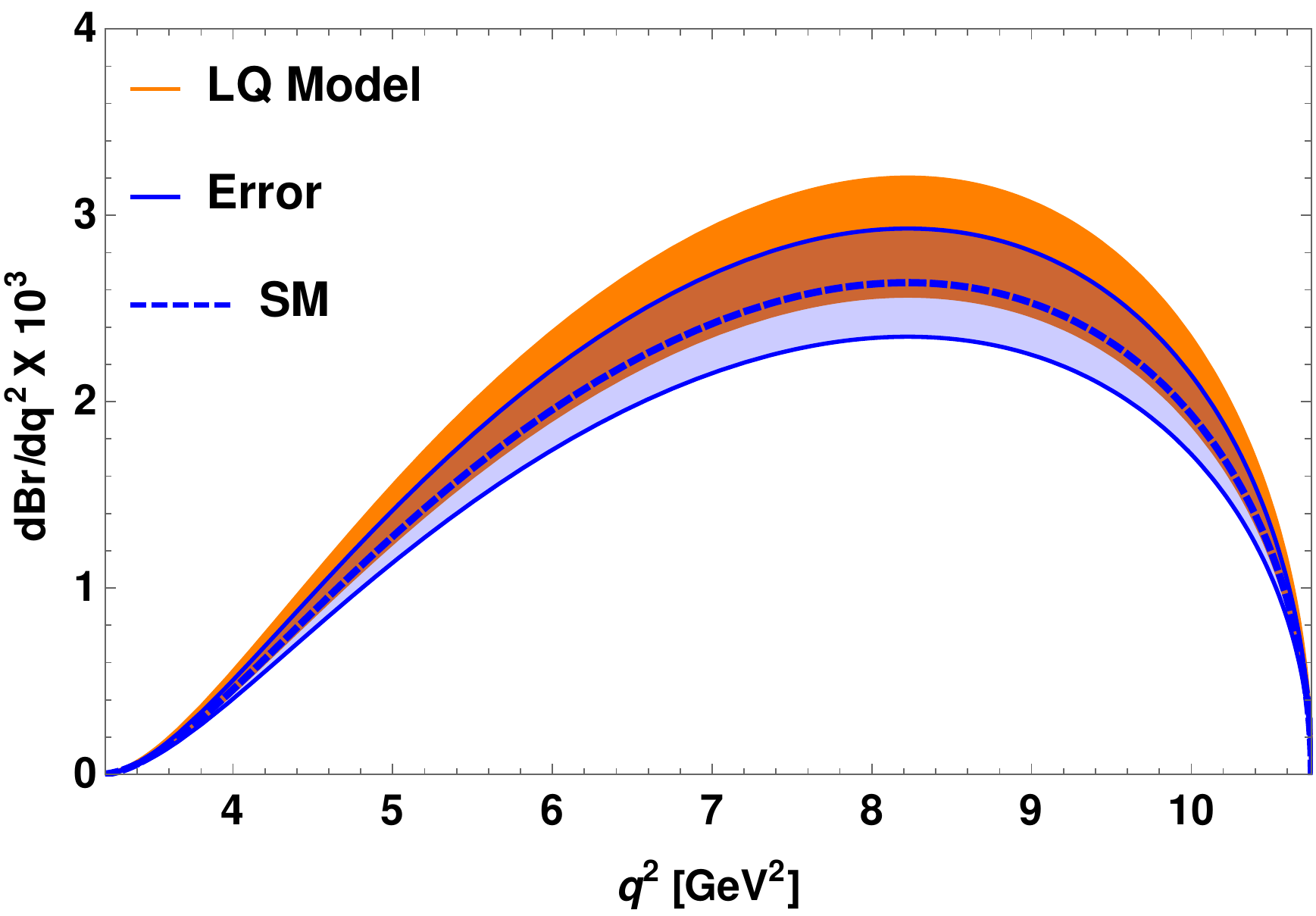}
\caption{The variation of  branching ratios of $B \to D^* e \bar{\nu}$  (left panel), $B \to D^* \mu \bar{\nu}$  (right panel) and $B \to D^* \tau \bar{\nu}$  (bottom panel) processes  with respect to $q^2$ in the leptoquark model.}
\end{figure}
%%%%%%%%%%%%%%%%%%%%%%%%%%%%%%%%%%%%%%%%%%%%%%%%%%%%%%%%%%%%%%%%%%%%%%%%%%%%%%
%%%%%%%%%%%%%%%%%%%%%%%%%%%%%%%%%%%%%%%%%%%%%%%%%%%%%%%%%%%%%%%%%%%%%%%%%%%%%%%%%%%%%%
\begin{figure}[h]
\centering
\includegraphics[scale=0.45]{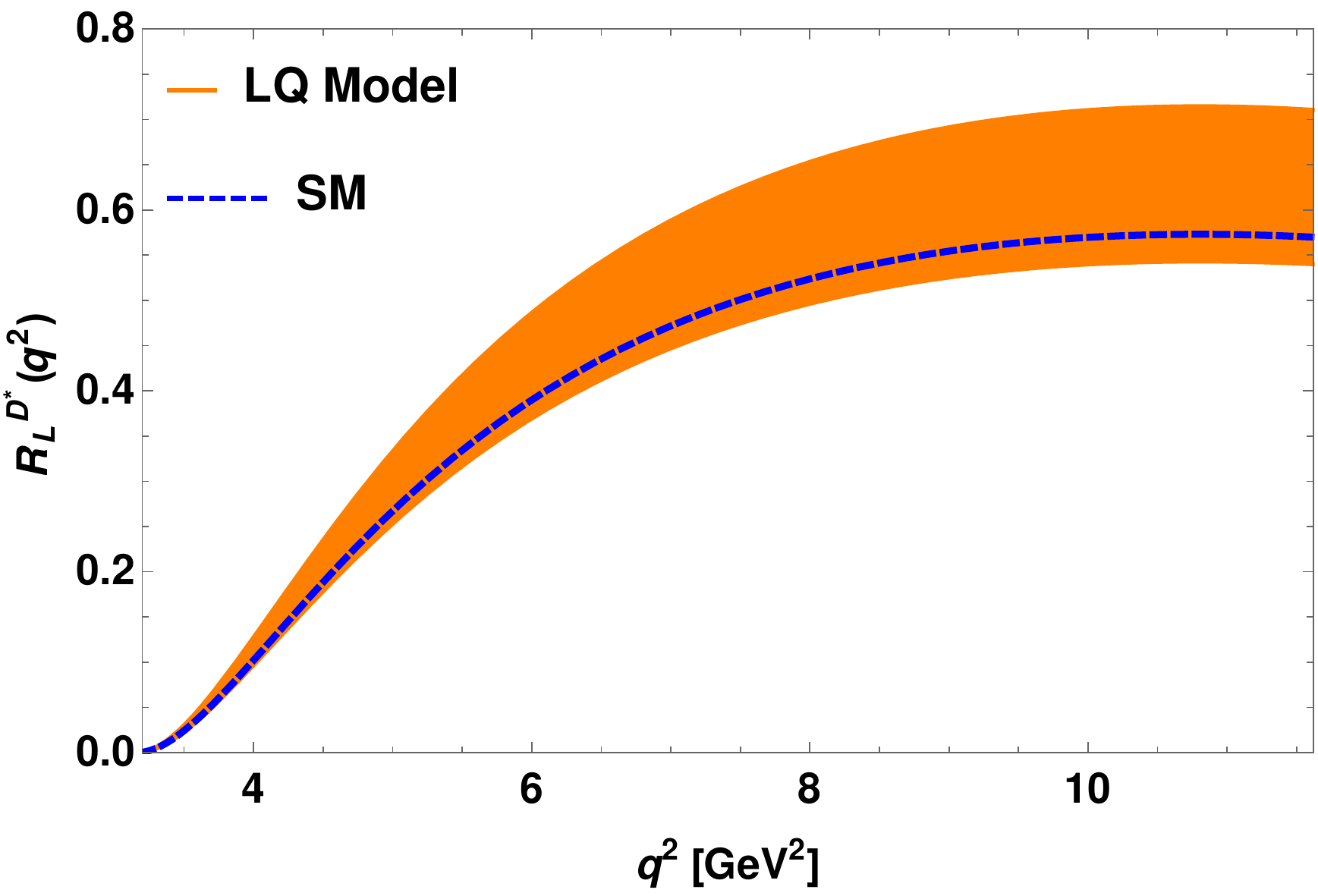}
\quad
\includegraphics[scale=0.45]{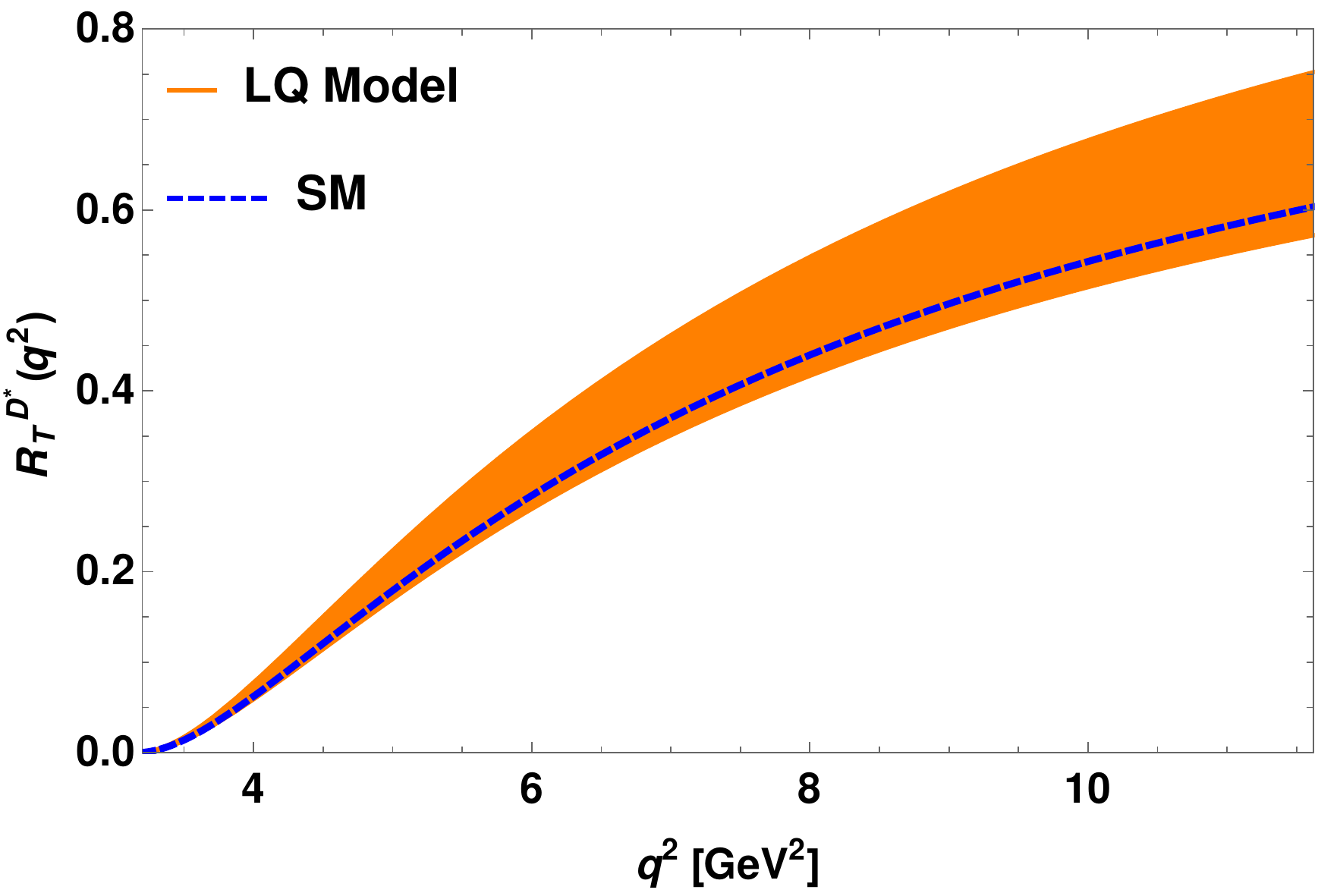}
\caption{The plot for  $R_L^{D^*}$ (left panel) and $R_T^{D^*}$ (right panel) in the leptoquark model}
\end{figure}
%%%%%%%%%%%%%%%%%%%%%%%%%%%%%%%%%%%%%%%%%%%%%%%%%%%%%%%%%%%%%%%%%%%%%%%%%%%%%%%%%%%%%%
\begin{figure}[h]
\centering
\includegraphics[scale=0.45]{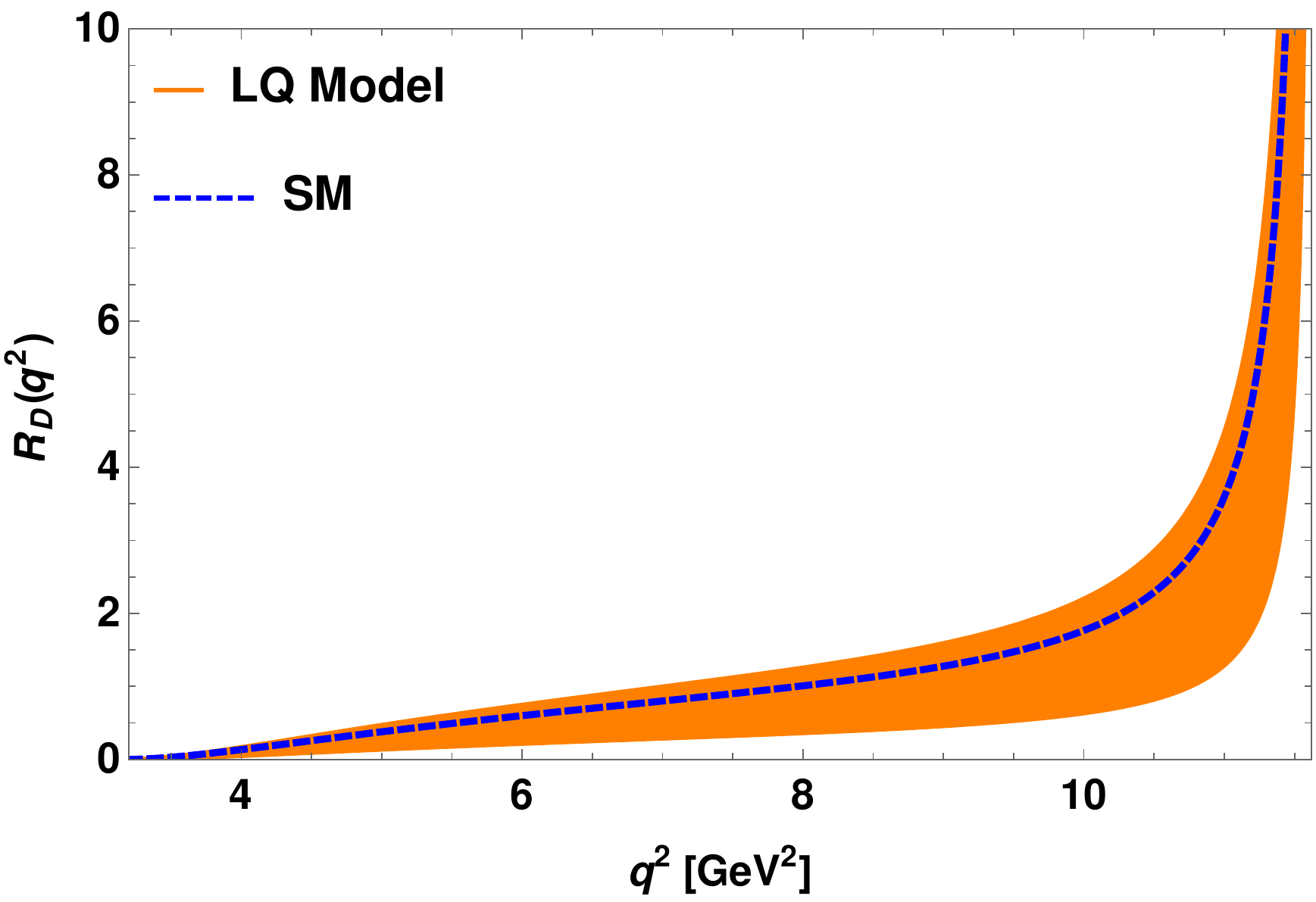}
\quad
\includegraphics[scale=0.45]{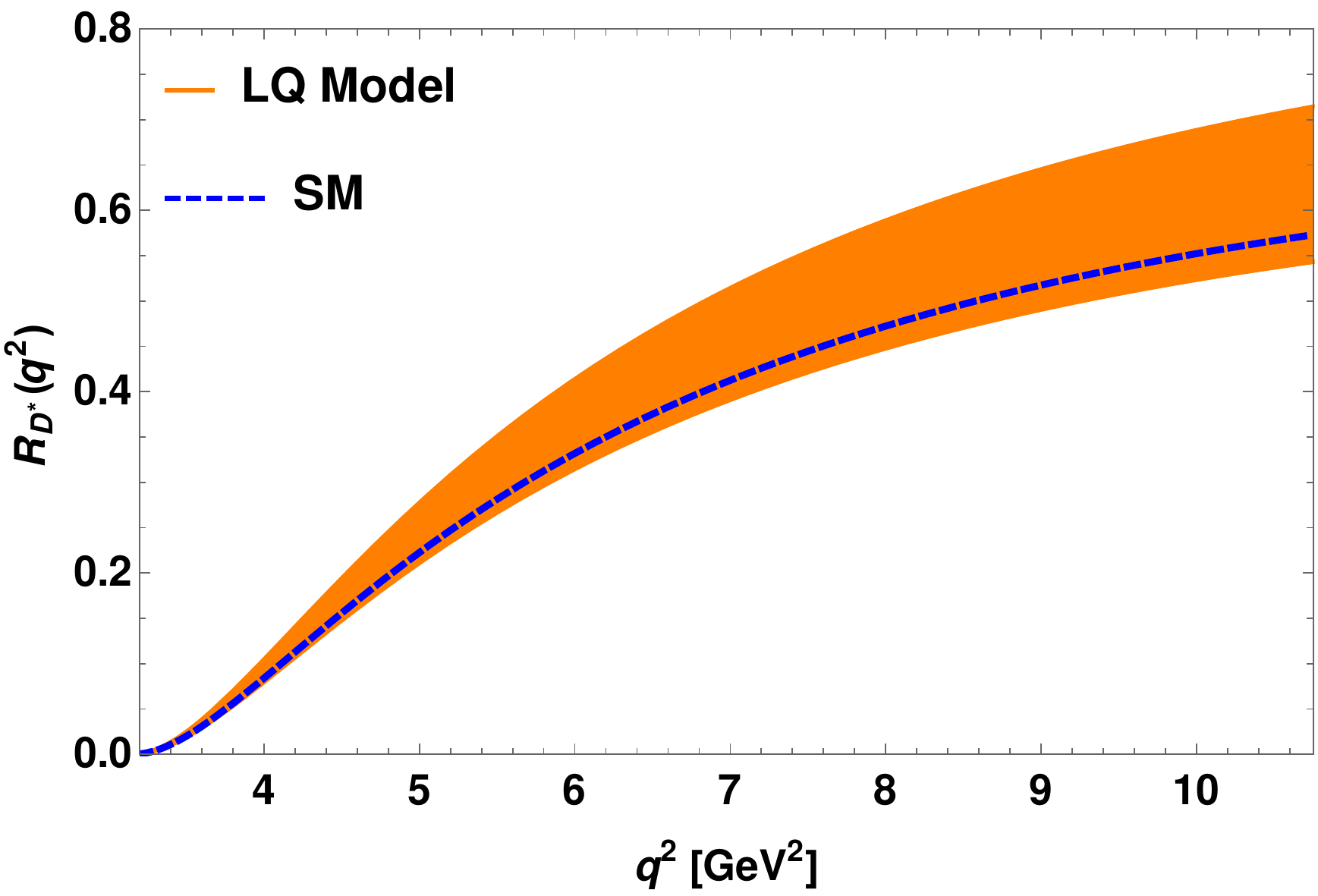}
\caption{The  $q^2$ variation of lepton non-universality $R_D (q^2)$ (left panel) and $R_{D^*}(q^2)$ (right panel) in  leptoquark model.}
\end{figure}
%%%%%%%%%%%%%%%%%%%%%%%%%%%%%%%%%%%%%%%%%%%%%%%%%%%%%%%%%%%%%%%%%%%%%%%%%%%%%%%%%%%%%%
%%%%%%%%%%%%%%%%%%%%%%%%%%%%%
%%%%%%%%%%%%%%%%%%%%%%%%%%%%%%%%%%%%%%%%%%%%%%%%%%%%%%%%%%%%%%%%%%%%%%%%%%%%%%
\begin{figure}[h]
\centering
\includegraphics[scale=0.65]{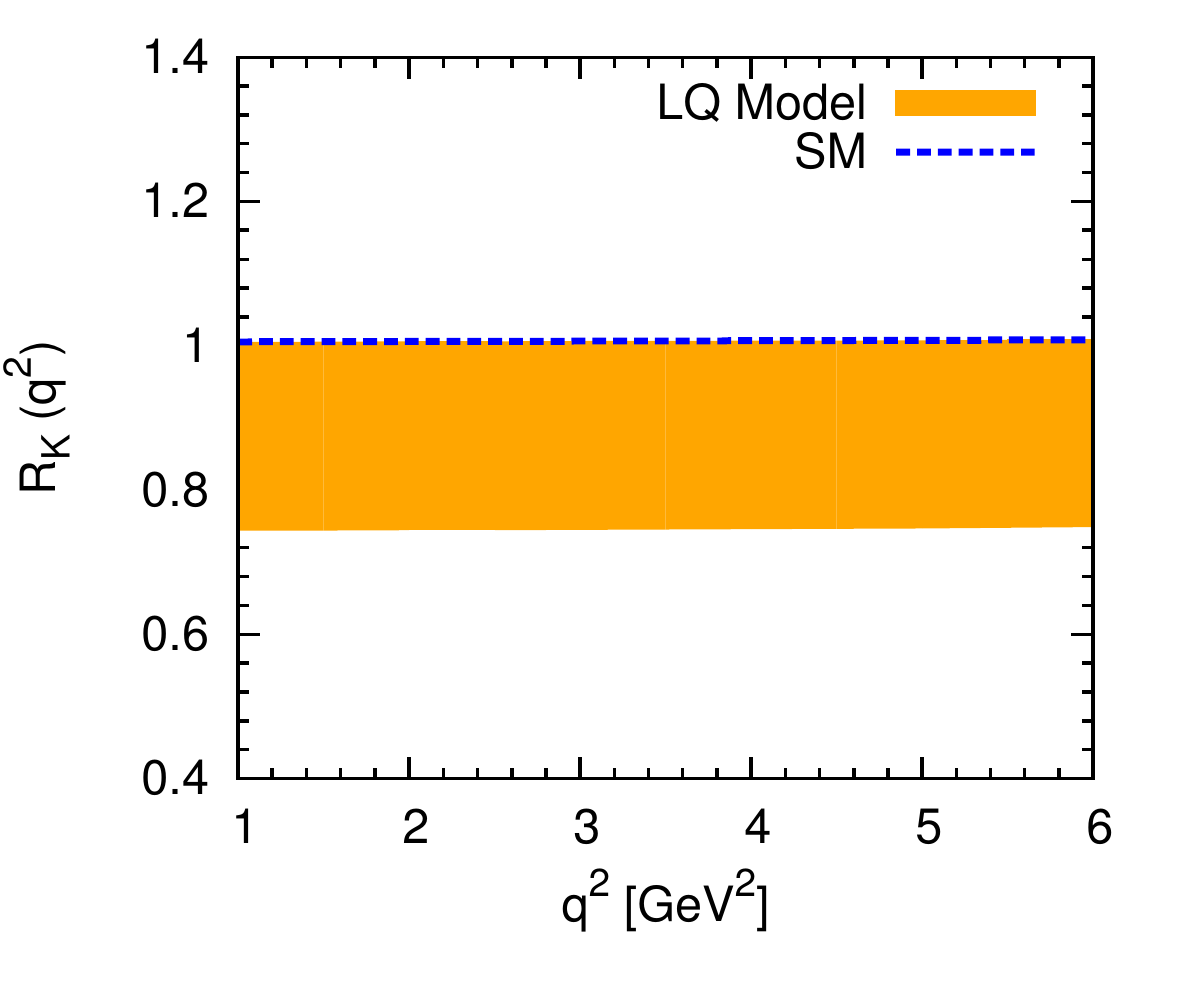}
\quad
\includegraphics[scale=0.65]{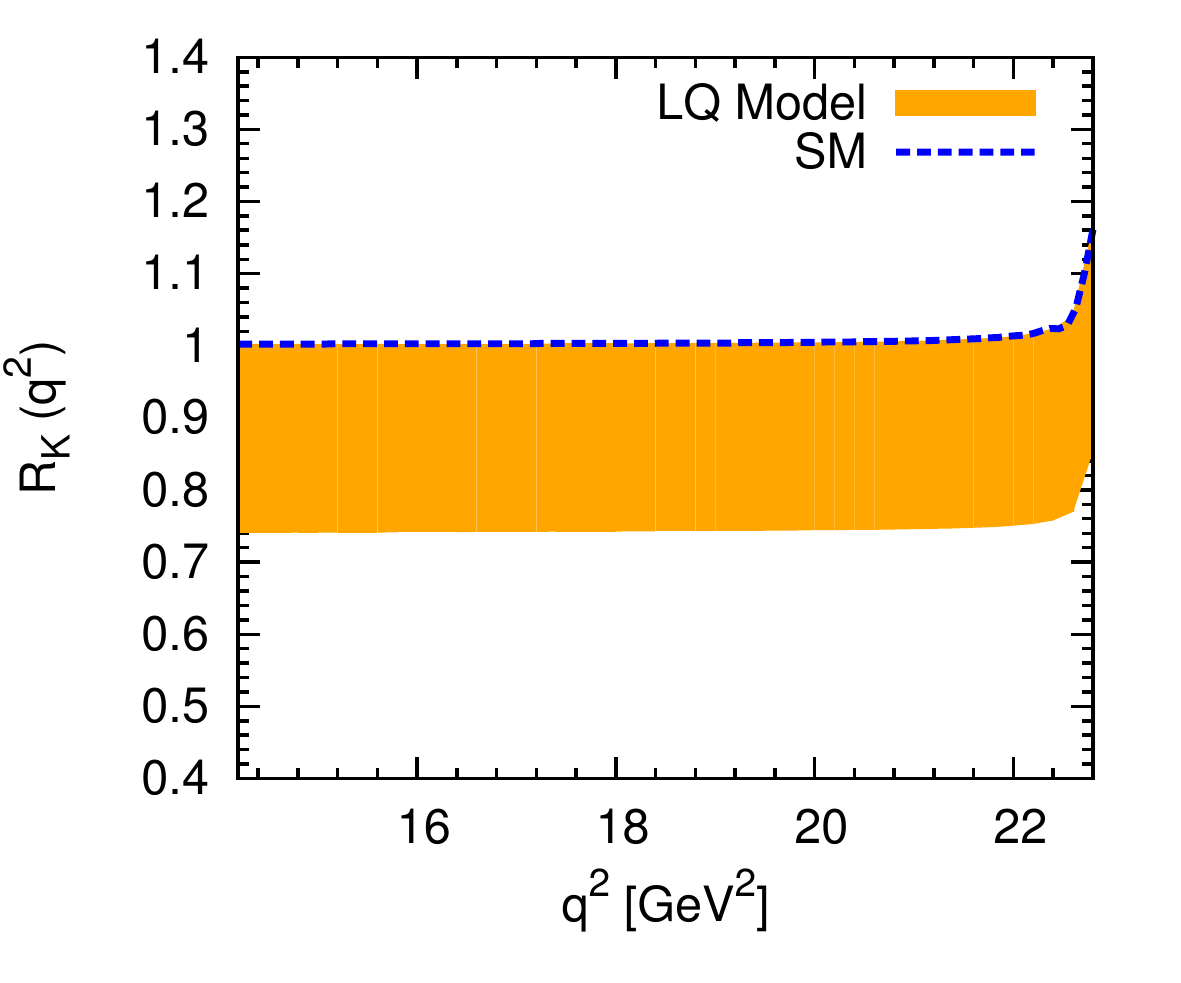}
\caption{The plot for  $R_K (q^2)$ in low $q^2$ (left panel)and  high $q^2$ (right panel)  in the leptoquark model.}
\end{figure}
%%%%%%%%%%%%%%%%%%%%%%%%%%%%%
%%%%%%%%%%%%%%%%%%%%%%%%%%%%%%%%%%%%%%%%%%%%%%%%%%%%%%%%%%%%%%%%%%%%%%%%%%%%%%
\begin{figure}[h]
\centering
\includegraphics[scale=0.65]{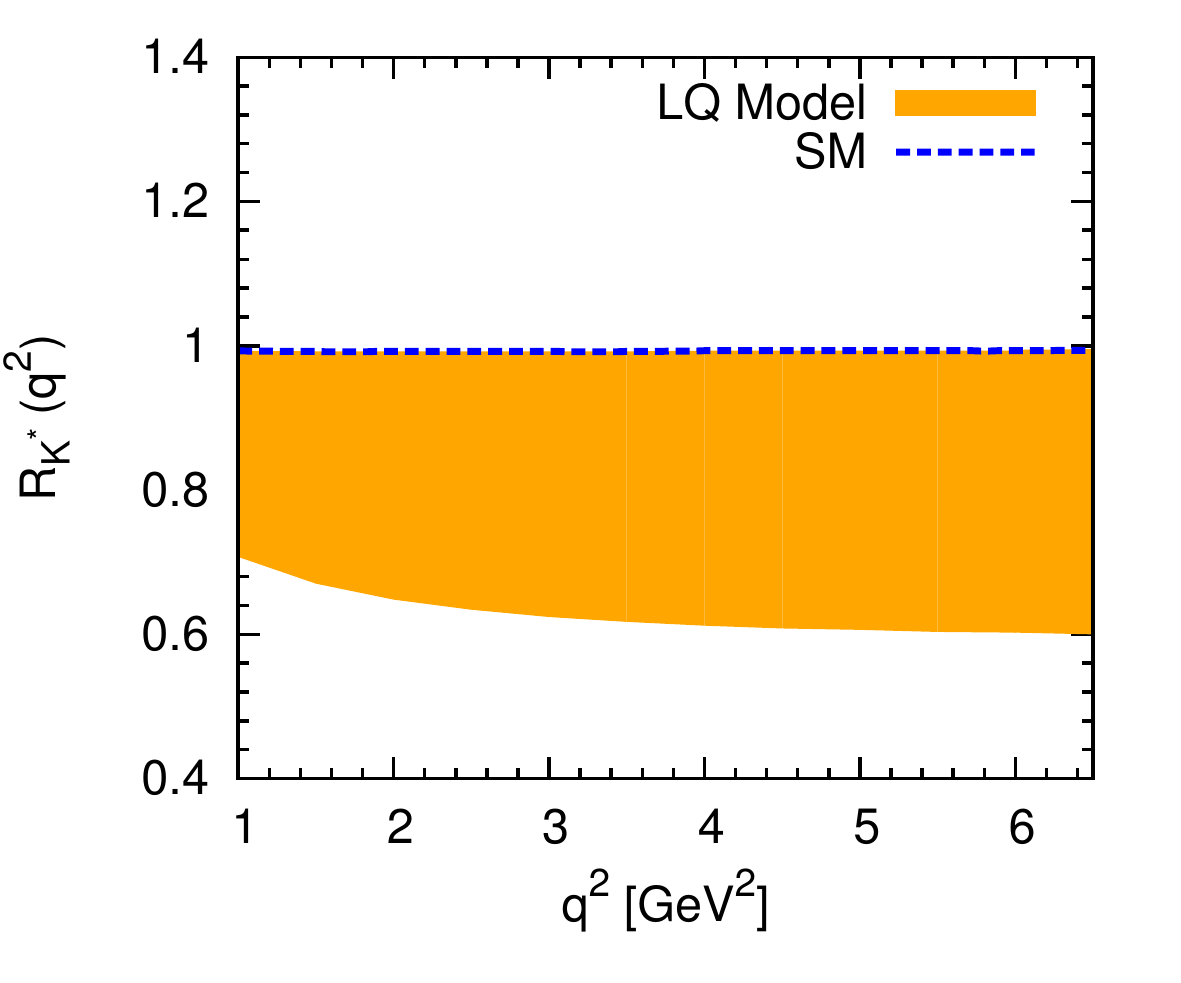}
\quad
\includegraphics[scale=0.65]{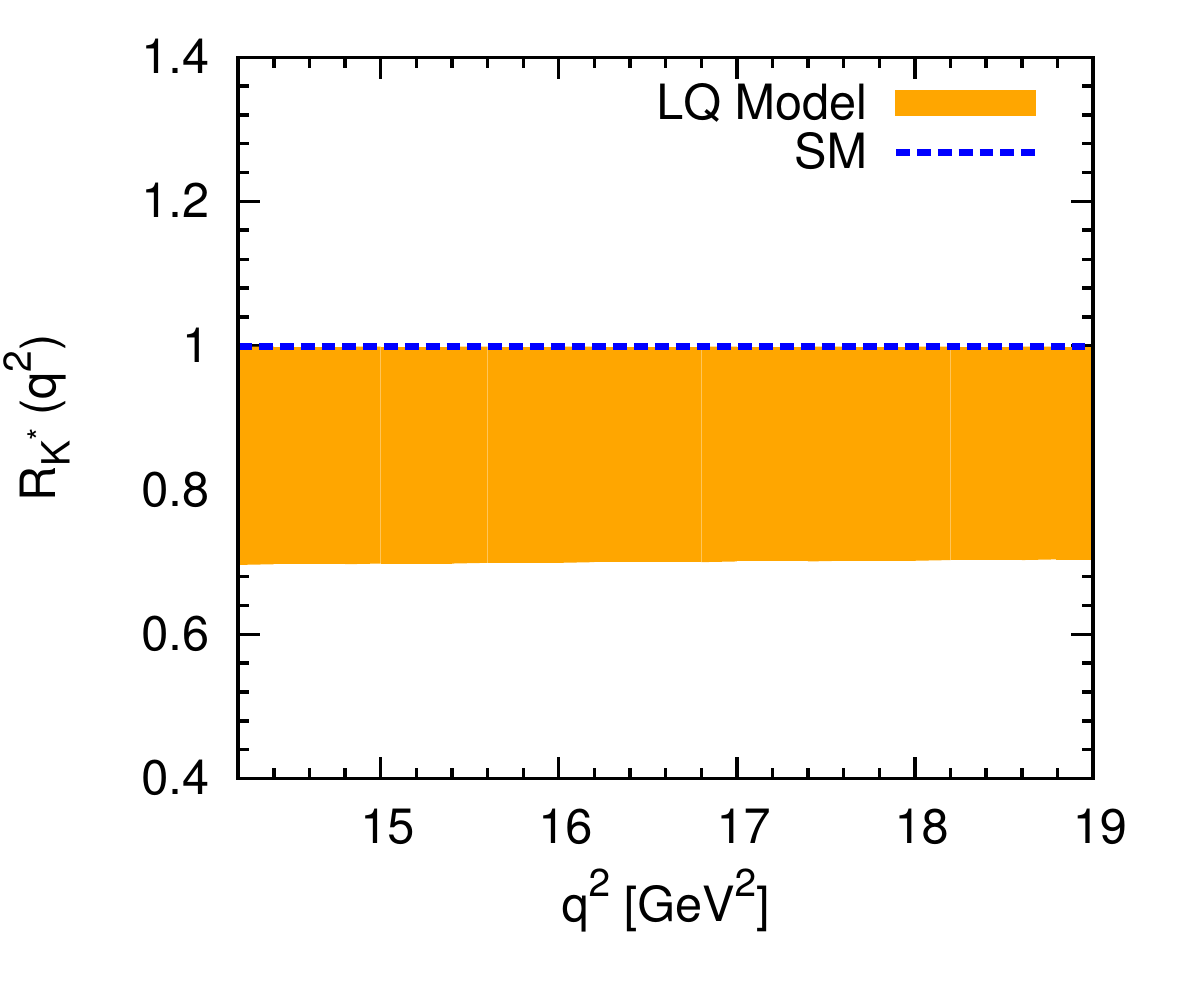}
\caption{The plot for  $R_{K^*} (q^2)$ in low $q^2$ (left panel) and high $q^2$ (right panel) in the leptoquark model.}
\end{figure}
%%%%%%%%%%%%%%%%%%%%%%%%%%%%%%%%%%%%%%%%%%%%%%%%%%%%%%%%%%%%%%%%%%%%%%%%%%%%%%%%%%%%%%%%%%%%%%
%%%%%%%%%%%%%%%%%%%%%%%%%%%%%%%%%%%%%%%%%%%%%%%%%%%%%%%%%%%%%%%%%%%%%%%%%%%%%%
%\begin{figure}[h]
%\centering
%\includegraphics[scale=0.4]{fbDtau.pdf}
%\quad
%\includegraphics[scale=0.4]{fbDstartau.pdf}
%\caption{The  variation of forward-backward asymmetry of $B \to D \tau \bar{\nu}$  (left panel) and $B \to D^* \tau \bar{\nu}$ process (right panel) with respect to $q^2$ in the leptoquark model}
%\end{figure}
%%%%%%%%%%%%%%%%%%%%%%%%%%%%%%%%%%%%%%%%%%%%%%%%%%%%%%%%%%%%%%%%%%%%%%%%%%%%%%%%%%%%%%
%\begin{figure}[h]
%\centering
%\includegraphics[scale=0.4]{ptauDtau.pdf}
%\quad
%\includegraphics[scale=0.4]{ptauDstartau.pdf}
%\caption{The $\tau$-polarization  of $B \to D \tau \bar{\nu}$  (left panel) and $B \to D^* \tau \bar{\nu}$ process (right panel) in the leptoquark model}
%\end{figure}

%%%%%%%%%%%%%%%%%%%%%%%%%%%%%%%%%%%%%%%%%%%%%%%%%%%%%%%%%%%%%%%%%%%%%%%%%%%%%%
%\begin{figure}[h]
%\centering
%\includegraphics[scale=0.4]{flDstartau.pdf}
%\quad
%\includegraphics[scale=0.4]{ftDstartau.pdf}
%\caption{The plot for  $F_L^{D^*}$ (left panel) and $F_T^{D^*}$ (right panel) in the leptoquark model}
%\end{figure}
%%%%%%%%%%%%%%%%%%%%%%%%%%%%%%%%%%%%%%%%%%%%%%%%%%%%%%%%%%%%%%%%%%%%%%%%%%%%%%%%%%%%%%
%%%%%%%%%%%%%%%%%%%%%%%%%%%%%%%%%%%%%%%%%%%%%%%%%%%%%%%%%%%%%%%%%%%%%%%%%%%%%
\begin{table}[h]
\caption{The predicted values of  branching ratios and $D^*$ polarizations of $\bar{B} \to D^{(*)} \tau \bar{\nu}$ processes in the vector leptoquark model.}
\begin{center}
\begin{tabular}{| c | c | c |  }
\hline
~Observables~ &~ SM Predictions~ & ~ Values in LQ Model~ \\
\hline
\hline

Br$\left(\bar{B} \to D l \bar{\nu} \right)$ & $(2.18 \pm 0.13) \times 10^{-2}$  & $(2.13-2.25) \times 10^{-2}$ \\

Br$\left(\bar{B} \to D \tau \bar{\nu} \right)$ & $(6.75\pm 0.08) \times 10^{-3}$  & $(2.48-8.2) \times 10^{-3}$ \\

%$A_{FB}^D$ & $0.36$ & $0.36$ \\

%$P_{\tau}^D$ & $0.327$  & $0.327$  \\
\hline

Br$\left(\bar{B} \to D^* l \bar{\nu} \right)$ & $(5.18 \pm 0.31) \times 10^{-2}$ & $(5.04-5.32) \times 10^{-2}$\\

Br$\left(\bar{B} \to D^* \tau \bar{\nu} \right)$ & $(1.33\pm 0.14) \times 10^{-2}$  & $(1.3-1.6) \times 10^{-2}$ \\

%$A_{FB}^{D^*}$ & $0.2$ & $0.2$\\

%$P_{\tau}^{D^*}$ & $-0.5$ & $-0.5$ \\

$R_{L}^{D^*}$ & $0.227$  & $0.215 - 0.283$ \\

$R_{T}^{D^*}$ & $0.29$  & $0.274-0.36$  \\

%$F_{L}^{D^*}$ & $0.46$  & $0.46$\\

%$F_{T}^{D^*}$ & $0.54$ & $0.54$ \\

 \hline
\end{tabular}
\end{center}
\end{table}

%%%%%%%%%%%%%%%%%%%%%%%%%%%%%%%%%%%%%%%%%%%%%%%%%%%%%%%%%%%%%%%%%%%%%%%%%%%%%%%%%%%%%%
%%%%%%%%%%%%%%%%%%%%%%%%%%%%%%%%%%%%%%%%%%%%%%%%%%%%%%%%%%%%%%%%%%%%%%%%%%%%%
\begin{table}[h]
\caption{The predicted values of $R_{D^{(*)}} $ and $R_{K^{(*)}}$  in the vector leptoquark model.}
\begin{center}
\begin{tabular}{| c | c | c | c |}
\hline
~Observables~ & ~SM Predictions ~ & ~Values in LQ Model~ & ~Experimental Limit~ \\
\hline
\hline
$R_D$ & $0.31$ & $0.11-0.386$ & $0.397 \pm 0.040 \pm 0.028$  \\

$R_{D^*}$ & $0.26$ & $0.243-0.32$ & $0.316 \pm 0.016 \pm 0.010$\\

${R_{K}}_{q^2 \in[1,6]}^{\mu e}$ & $1.006$ & $0.75-1.006$ & $0.745^{+0.090}_{-0.074}\pm 0.036$\\

${R_{K}}_{q^2 \geq 14.18}^{\mu e}$ & $1.004$ & $0.74-1.004$ & $\cdots$\\

${R_{K^*}}_{q^2 \in[1,6]}^{\mu e}$ & $0.996$ & $0.725-0.996$ & $\cdots$\\

${R_{K^*}}_{q^2 \geq 14.18}^{\mu e}$ & $0.999$ & $0.816-0.999$  & $\cdots$\\

 \hline
\end{tabular}
\end{center}
\end{table}

%%%%%%%%%%%%%%%%%%%%%%%%%%%%%%%%%%%%%%%%%%%%%%%%%%%%%%%%%%%%%%%%%%%%%%%%%%%%%%%%%%%%%%
\section{$B_{u,c}^{* +} \to l^+ \nu$ process}
%%%%%%%%%%%%%%%%%%%%%%%%%%%%%%%%%%%%%%%%%%%%%%%%%%%%%%%%%%%%
The rare leptonic $B_{u,c}^{* +} \to l^+ \nu_l$ processes of unstable $B_{u,c}^{* +}$ mesons mediated by $b \to u l \nu$ and $b \to c l \nu$ transitions are studied in this section. Unlike their pseudoscalar partners these  decays are not helicity suppressed, but their shorter lifetimes make the branching ratios to be small.
%In the SM, the $B_{u,c}^{* +} \to l^+ \nu_l$ processes are extremely rare due to  the presence of light neutrino  in the final states and are suppressed by the dominant radiative decays of $B_{u, c}^*$ mesons.
The interaction Lagrangian of charged-current leptonic decays of $B_{u, c}^{(*)}$ mesons are given by \cite{Bs-star}
\bea
\mathcal{L} = -\frac{4G_F}{\sqrt{2}} V_{q^\prime b} \Big [ (1+C_{V_1}) \left(\bar q^\prime \gamma^\mu L b \right) \left(\bar{l} \gamma_\mu L \nu \right) + C_{V_2} \left(\bar q^\prime \gamma^\mu R b \right) \left(\bar{l} \gamma_\mu L \nu \right) \Big],
\eea
where  $q^\prime=u,c$ and $C_{V_{1,2}}$ are the new Wilson coefficients arising due to the exchange of  vector LQ.  The transition amplitudes can be expressed in terms of the decay constants,  defined as
\bea
\langle 0 | \bar{q^\prime}\gamma^\mu  \gamma_5 b | B_{q^\prime} (p_{B_{q^\prime}}) \rangle = -if_{B_{q^\prime}} p_{B_{q^\prime}}^\mu, \nn\\
\langle 0 | \bar{q^\prime}\gamma^\mu   b | B^{*}_{q^\prime} (p_{B^*_{q^\prime}},\epsilon) \rangle = f_{B^{*}_{q^\prime}}  M_{B^{*}_{q^\prime}} \epsilon^\mu,
\label{VIM}
\eea
where $f_{B^{(*)}_{q^\prime}}$ are the decay constant of $B_{q^\prime}^{(*)}$ mesons and $\epsilon$ is the polarization
vector of $B_{q'}^*$. Using Eqn. (\ref{VIM}), the differential decay distribution of $B_{u,c}^{(*) \mp} \to l^\mp \bar{\nu}_l$ processes in the LQ model  are 
\bea
\Gamma(B_{q'}^+ \to {l^+ \nu }) = \frac{G_F^2}{8\pi} |V_{q^\prime b}|^2 \left(1+C_{V_1}-C_{V_2} \right)^2 M_{B_{q^\prime}} f_{B_{q^\prime}}^2 m_l^2,
\eea
and
\bea
\Gamma(B_{q'}^{+*} \to {l^+ \nu }) = \frac{G_F^2}{12\pi} |V_{q^\prime b}|^2 \left(1+C_{V_1}+C_{V_2} \right)^2 M_{B^*_{q^\prime}}^3 f_{B^*_{q^\prime}}^2,
\eea
respectively. 
The input values of masses of  $B_{u, c}^{(*)}$ mesons are taken from \cite{pdg} and the  decay constants  of $B_{u, c}^{(*)}$ mesons are  $f_{B^*}/f_B = 0.941(26)$ \cite{fB-star}, $f_{B_c} = 489$ MeV  \cite{fBc} and $f_{B_c^*}/f_{B_c} = 1$  \cite{Bs-star}.  The branching ratios of $B_c \to l \nu_l$ processes in the SM are 
\bea
&&{\rm Br} (B_c^+ \to e^+ \nu_e)|_{\rm SM} = (2.94 \pm 0.12) \times 10^{-9},  \\
&&{\rm Br} (B_c^+ \to \mu^+ \nu_\mu)|_{\rm SM} = (1.26 \pm 0.05) \times 10^{-4}, \\
&&{\rm Br} (B_c^+ \to \tau^+ \nu_\tau)|_{\rm SM} = (3.6 \pm 0.14) \times 10^{-2}.
\eea
  The decay width of ${B^*_{u, c}}^+ \to l^+ \nu_l$ processes in the SM are
\bea
&&\Gamma (B_u^* \to l \nu_l) = (2.98\pm 0.12) \times 10^{-16}~{\rm GeV}, \\ 
&&\Gamma (B_c^* \to l \nu_l) = (3.9 \pm 0.16) \times 10^{-13}~{\rm GeV}.
\eea
The decay width of ${B^*_{u, c}} \to l \nu_l$ processes are independent of the mass of the final leptons, hence same for all generation in  the SM. 
In order to calculate the branching ratios we need the values of lifetime or the total decay width of $B_{u, c}^*$ mesons.  We have taken the decay width  of $B_{u, c}^*$ meson as $\Gamma_{B_u^*}=0.50(25)$ KeV and $\Gamma_{B_c^*} = 0.03(7)$ KeV respectively, which are computed in Ref. \cite{Bs-star}. 
The predicted branching ratios in the  vector LQ model are presented in Table VII. The branching ratios of ${B_c^*}^+ \to l^+ \nu_l$ processes are of the order of $\sim 10^{-5}$, which are not very suppressed,  and
they could be observed in the LHCb experiment.
However, the branching ratios of ${B_u^*}^+ \to l^+ \nu_l$ are found to be rather small.  We do not find much deviation from the SM in the  $B_{u,c}^* \to l \nu$ processes in the LQ  model. 
%However $B_u^* \to \mu \nu_\mu (\tau \nu_\tau)$ process has one order deviation from the SM. 

%%%%%%%%%%%%%%%%%%%%%%%%%%%%%%%%%%%%%%%%%%%%%%%%%%%%%%%%%%%%%%%%%%%%%%%%%%%%%
\begin{table}[h]
\caption{The predicted values of branching ratios of $B_{u, c}^{*~+} \to l^+ \nu$ processes in the  leptoquark model.}
\begin{center}
\begin{tabular}{| c | c | c |  }
\hline
~Observables~ & ~SM predictions ~ & ~Values in LQ Model~ \\
\hline
\hline
Br ($B_u^* \to e \nu$) & $(5.97 \pm 0.24) \times 10^{-10}$ & $ (0.94 -1.01) \times 10^{-6}$    \\

Br ($B_u^* \to \mu \nu$) &  $(5.97 \pm 0.24) \times 10^{-10}$   & $ (3.67-6.793) \times 10^{-10}$     \\

Br ($B_u^* \to \tau \nu$) & $(5.97 \pm 0.24) \times 10^{-10}$ & $(4.2-3.67) \times 10^{-10}$  \\

\hline
Br ($B_c^* \to e \nu$) & $(1.3\pm 0.052) \times 10^{-5}$ & $ (1.27-1.34)\times 10^{-5}$ \\

Br ($B_c^* \to \mu \nu$) & $(1.3\pm 0.052) \times 10^{-5}$ & $(1.26-1.34)\times 10^{-5}$  \\

Br ($B_c^* \to \tau \nu$) & $(1.3 \pm 0.052) \times 10^{-5}$ & $(1.26-1.58)\times 10^{-5}$   \\

 \hline
\end{tabular}
\end{center}
\end{table}

%%%%%%%%%%%%%%%%%%%%%%%%%%%%%%%%%%%%%%%%%%%%%%%%%%%%%%%%%%%%%%
\section{conclusion}
%%%%%%%%%%%%%%%%%%%%%%%%%%%%%%%%%%%%%%%%%%%%%%%%%%%%%%%%%

In this work we considered the vector leptoquark model to explain the anomalies observed in semileptonic $\bar B \to D^{(*)} \tau \bar{\nu}$ decay process in light of recent $B$-factories result, especially the deviation of  $R_{D^{(*)}}$ observables from the  SM predictions. There are two relevant vector leptoquark $(U_1, U_3)$  states which conserve baryon and lepton numbers and can  simultaneously explain the processes mediated by quark level transitions $b \to c l \bar{\nu}_l$ and $b \to s l^+ l^-$.  We constrained the leptoquark couplings by using the branching ratios of $B_s \to l^+ l^-$, $\bar B \to X_s l^+ l^-$, $\bar B \to X_s \nu \bar \nu$ and $B_u^+ \to l^+ \nu_l $ processes, where $l$ is any charged lepton. We estimated the branching ratios, forward backward asymmetries, lepton non-universality, $\tau$ and $D^*$ polarization parameters in the $\bar B \to D^{(*)} l \bar \nu_l$ processes. We looked into the lepton non-universality parameters in both $\bar B \to D^{(*)} l \bar \nu_l$ and $\bar B \to K^{(*)} l^+ l^-$ processes and found that both the $R_{K^{(*)}}$ and $R_{D^{(*)}}$ anomalies could be explained by $U_{1, 3}$ vector leptoquarks. We also studied the rare $B_{u, c}^* \to l \nu$ decay processes of $B_{u, c}^*$ vector mesons.  The branching ratios of the decay modes $B_c^* \to l \nu$ are not very suppressed, i.e., ${\cal O}(10^{-5})$, which could be observed in the LHCb experiment.

%%%%%%%%%%%%%%%%%%%%%%%%%%%%%%%%%%%%%%%%%%%%%%%%%%%%%%%%%%%%%%%%%%%%%%%%%%%%%%%%
\appendix

\section{$B \to D \tau \bar \nu_l$ form factors}
%%%%%%%%%%%%%%%%%%%%%%%%%%%%%%%%%%%%%%%%%%%%%%%%%%%%%%%%%%%%
The nonzero hadronic amplitudes for $B \to D \tau \bar \nu_l$ process are
\bea
&&H_{V, 0}^s (q^2) \equiv H_{V_{1, 0}}^s (q^2)\equiv H_{V_{2, 0}}^s (q^2) = \sqrt{\frac{\lambda_D \left(q^2 \right)}{q^2}} F_1 \left(q^2 \right), \nn \\
&&H_{V, t}^s (q^2) \equiv H_{V_{1, t}}^s (q^2) \equiv H_{V_{2, t}}^s (q^2)  = \frac{M_B^2-M_D^2}{\sqrt{ q^2}} F_0 \left(q^2 \right), \nn \\
&&H_S^s (q^2) \equiv H_{S_{1}}^s (q^2) = H_{S_{2}}^s (q^2) \simeq \frac{M_B^2-M_D^2}{m_b-m_c} F_0 \left(q^2 \right),
\eea
where the form factors $F_{0, 1}$  are defined as
\bea
&&F_1 (q^2) = \frac{1}{2\sqrt{M_B M_D}} \Bigg [ \left(M_B + M_D \right) h_+\left(\omega (q^2) \right)-\left(M_B - M_D \right) h_- \left(\omega (q^2) \right) \Bigg ], \nn \\
&&F_0 (q^2) = \frac{1}{2\sqrt{M_B M_D}} \Bigg [ \frac{\left(M_B + M_D \right)^2-q^2}{M_B + M_D }  h_+\left(\omega (q^2) \right)-\frac{\left(M_B - M_D \right)^2-q^2}{M_B - M_D } h_- \left(\omega (q^2) \right)  \Bigg].\hspace{0.4cm}
%&& F_T (q^2) = \frac{M_B + M_D}{2\sqrt{M_B M_D}}h_T \left(\omega (q^2) \right).
\eea
Here $h_\pm \left(\omega (q^2) \right) $ are the HQET form factors taken from the Ref. \cite{sakaki, Neubert}.

%%%%%%%%%%%%%%%%%%%%%%%%%%%%%%%%%%%%%%%%%%%%%%%%%%%%%%%%%%%
\section{$B \to D^* l \bar \nu$ form factors}
%%%%%%%%%%%%%%%%%%%%%%%%%%%%%%%%%%%%%%%%%%%%%%%%%%%%%%%%%%%

The hadronic amplitude for $B \to D^* l \bar \nu$ process are 
\bea
&&H_{V, \pm} (q^2) \equiv H_{V_1,\pm}^\pm (q^2) =- H_{V_2,\mp}^\mp (q^2) = \left(M_B + M_{D^*} \right) A_1(q^2) \mp \frac{\sqrt{\lambda_{D^*}(q^2)}}{M_B + M_{D^*}} V(q^2), \nn \\
&&H_{V, 0} (q^2) \equiv H_{V_1,0}^0 (q^2) =- H_{V_2,0}^0 (q^2) = \frac{M_B + M_{D^*}}{2M_{D^*} \sqrt{q^2}} \Big [ - \left(M_B^2 - M_{D^*}^2-q^2 \right) A_1(q^2) \nn \\ && \hspace{6.5cm} + \frac{\lambda_{D^*}(q^2)}{( M_B + M_{D^*})^2} A_2(q^2) \Bigg ], \nn \\
&&H_{V, t} (q^2) \equiv H_{V_1,t}^0 (q^2) =- H_{V_2,t}^0 (q^2) = - \sqrt{ \frac{\lambda_{D^*} (q^2)}{q^2}} A_0 (q^2), \nn \\
&&H_{S} (q^2) \equiv H_{S_1}^0 (q^2) =- H_{S_2}^0 (q^2) \simeq -\frac{\sqrt{\lambda_{D^*} (q^2)}}{m_b + m_c} A_0 (q^2),
\eea
where the form factors are defined as 
\bea
V(q^2) &=& \frac{M_B +M_{D^*}}{2\sqrt{M_B M_{D^*}}} h_V \left(\omega (q^2) \right), \nn \\
A_1(q^2) &=& \frac{(M_B +M_{D^*})^2 - q^2}{2\sqrt{M_B M_{D^*}} (M_B +M_{D^*})} h_{A_1} \left(\omega (q^2) \right), \nn \\
A_2 (q^2) &=& \frac{M_B +M_{D^*}}{2\sqrt{M_B M_{D^*}}} \Big [ h_{A_3} \left(\omega (q^2) \right)+\frac{M_{D^*}}{M_B}h_{A_2} \left(\omega (q^2) \right) \Big ], \nn \\
A_0 (q^2) &=& \frac{1}{2\sqrt{M_B M_{D^*}}} \Bigg [  \frac{(M_B +M_{D^*})^2 - q^2}{2M_{D^*}} h_{A_1} \left(\omega (q^2) \right) \nn \\ &-& \frac{M_B^2 -M_{D^*}^2 + q^2}{2M_B} h_{A_2} \left(\omega (q^2) \right)-  \frac{M_B^2 -M_{D^*}^2 - q^2}{2M_{D^*}} h_{A_3} \left(\omega (q^2) \right) \Bigg ].
\eea
The complete expression for HQET form factors $h_i, i=V, A_{1, 2, 3}$ are given in \cite{sakaki, Neubert}.

%%%%%%%%%%%%%%%%%%%%%%%%%%%%%%%%%%%%%%%%%%%%%%%%%%%%%%%%%%%%%%%%%
\section{$\tau$ and $D^*$ polarizations}
%%%%%%%%%%%%%%%%%%%%%%%%%%%%%%%%%%%%%%%%%%%%%%%%%%%%%%%%%%

For a fixed polarization of $\tau$,  the decay distribution of $B \to D \tau \bar{\nu}_l$ process with respect to $q^2$ are given as 
\bea
\frac{d\Gamma^{\lambda_{\tau = 1/2}} \left(\bar{B} \to D \tau \bar{\nu}_l \right)}{dq^2} & =& \frac{G_F^2 |V_{cb}|^2}{192 \pi^3 M_B^3} q^2 \sqrt{\lambda_D (q^2)} \Big( 1-\frac{m_\tau^2}{q^2} \Big)^2 \times \nn \\  &&\Bigg[ \frac{1}{2} \Big | \delta_{l\tau} + C_{V_1}^l  \Big |^2 \frac{m_\tau^2}{q^2} \left( {H_{V, 0}^s}^2 + 3 {H_{V, t}^s}^2 \right) + \frac{3}{2} \Big | C_{S_1}^l \Big | ^2 {H_S^s}^2 \nn \\ && +3{\rm Re} \Big[\left( \delta_{l\tau} + C_{V_1}^l \right)  {C_{S_1}^{l*}} \Big]\frac{m_\tau}{\sqrt{q^2}} H_S^s H_{V, t}^s \Bigg],
\eea
\bea
\frac{d\Gamma^{\lambda_{\tau = -1/2}} \left(\bar{B} \to D \tau \bar{\nu}_l \right)}{dq^2} & =& \frac{G_F^2 |V_{cb}|^2}{192 \pi^3 M_B^3} q^2 \sqrt{\lambda_D (q^2)} \Big( 1-\frac{m_\tau^2}{q^2} \Big)^2  \Big | \delta_{l\tau} + C_{V_1}^l \Big |^2 {H_{V, 0}^s}^2,
\eea
and for $B \to D^* \tau \bar{\nu}_l$ process
\bea
\frac{d\Gamma^{\lambda_{\tau = 1/2}} \left(\bar{B} \to D^* \tau \bar{\nu}_l \right)}{dq^2} & =& \frac{G_F^2 |V_{cb}|^2}{192 \pi^3 M_B^3} q^2 \sqrt{\lambda_ {D^*} (q^2)} \Big( 1-\frac{m_\tau^2}{q^2} \Big)^2 \times \nn \\  &&\Bigg[ \frac{1}{2} \Big | \delta_{l\tau} + C_{V_1}^l  \Big |^2 \frac{m_\tau^2}{q^2} \left(H_{V, +}^2 + H_{V, -}^2 + H_{V, 0}^2 + 3 H_{V, t}^2 \right)   \nn \\ &&
+ \frac{3}{2} \Big | C_{S_1}^l \Big | ^2 {H_S}^2  +3{\rm Re} \Big[\left( \delta_{l\tau} + C_{V_1}^l  \right) {C_{S_1}^{l*}} \Big]  \frac{m_\tau}{\sqrt{q^2}} H_S H_{V, t} \Bigg], \hspace{0.3cm}
\eea
\bea
\frac{d\Gamma^{\lambda_{\tau = -1/2}} \left(\bar{B} \to D^* \tau \bar{\nu}_l \right)}{dq^2} & =& \frac{G_F^2 |V_{cb}|^2}{192 \pi^3 M_B^3} q^2 \sqrt{\lambda_ {D^*} (q^2)} \Big( 1-\frac{m_\tau^2}{q^2} \Big)^2  \nn \\  &&   \times \Bigg[ \Big | \delta_{l\tau} + C_{V_1}^l  \Big |^2  \left(H_{V, +}^2 + H_{V, -}^2 + H_{V, 0}^2  \right)   \Bigg ].
\eea
The $q^2$ distributions of $B \to D^* \tau \bar{\nu}_l$ process for a given polarization of $D^*$ are given as
\bea
\frac{d\Gamma^{\lambda_{D^* = \pm 1}} \left(\bar{B} \to D^* \tau \bar{\nu}_l \right)}{dq^2} & =& \frac{G_F^2 |V_{cb}|^2}{192 \pi^3 M_B^3} q^2 \sqrt{\lambda_ {D^*} (q^2)} \Big( 1-\frac{m_\tau^2}{q^2} \Big)^2  \nn \\  && \times \Bigg[ \Big( 1+\frac{m_\tau^2}{2 q^2} \Big) \Big | \delta_{l\tau} + C_{V_1}^l  \Big |^2 H_{V, \pm}^2  \Bigg ], 
\eea
\bea
\frac{d\Gamma^{\lambda_{D^* = 0}} \left(\bar{B} \to D^* \tau \bar{\nu}_l \right)}{dq^2} & =& \frac{G_F^2 |V_{cb}|^2}{192 \pi^3 M_B^3} q^2 \sqrt{\lambda_ {D^*} (q^2)} \Big( 1-\frac{m_\tau^2}{q^2} \Big)^2 \times \nn \\  &&\Bigg[\Big | \delta_{l\tau} + C_{V_1}^l  \Big |^2 \Big[ \Big( 1+\frac{m_\tau^2}{2 q^2} \Big) H_{V, 0}^2 + \frac{3}{2} \frac{m_\tau^2}{q^2} H_{V, t}^2 \Big ] \nn \\ && + \frac{3}{2} \Big | C_{S_1}^l \Big | ^2 H_S^2  +3{\rm Re} \Big[\left( \delta_{l\tau} + C_{V_1}^l \right)  {C_{S_1}^{l*}}\Big] \frac{m_\tau}{\sqrt{q^2}} H_S H_{V, t} \Bigg].
\eea

%%%%%%%%%%%%%%%%%%%%%%%%%%%%%%%%%%%%%%%%%%%%%%%%%%%%%%%%%%%%%%%%%%%%%%%%%%%%%%%
{\bf Acknowledgments}
%%%%%%%%%%%%%%%%%%%%%%%%%%%%%%%%%%%%%%%%%%%%%%%%%%%%%%%

SS and RM  would like to thank Science and Engineering Research Board (SERB),
Government of India for financial support through grant No. SB/S2/HEP-017/2013.

%%%%%%%%%%%%%%%%%%%%%%%%%%%%%%%%%%%%%%%%%%%%%%%%%%%%%%%%%%%%%%%%%%%%%%%%%%%%%%%%%%%%%

\end{document}